\documentclass[11pt]{article}
\pdfoutput=1 
\usepackage{./jheppub}
\usepackage[toc,page]{appendix}

\usepackage{etoolbox}
\appto\appendix{\addtocontents{toc}{\protect\setcounter{tocdepth}{1}}}
\appto\listoffigures{\addtocontents{lof}{\protect\setcounter{tocdepth}{1}}}
\appto\listoftables{\addtocontents{lot}{\protect\setcounter{tocdepth}{1}}}

\usepackage[usenames,dvipsnames,table]{xcolor}
\usepackage[utf8]{inputenc}

\definecolor{shadecolor}{rgb}{0.90,0.90,0.90}

\usepackage{subfigure}
\usepackage{graphicx}
\usepackage{amsmath, amssymb}

\usepackage{framed}

\newcommand{\be}{\begin{eqnarray}}
\newcommand{\ee}{\end{eqnarray}}
\newcommand{\bega}{\begin{align}}
\newcommand{\ena}{\end{align}}
\numberwithin{equation}{section}

\def\cA{\mathcal{A}}

\def\cN{ {\mathcal N}}
\def\ta{\tilde{a}}

\newcommand{\bea}{\begin{eqnarray}}
\newcommand{\eea}{\end{eqnarray}}
\newcommand{\nn}{\nonumber}

\newcommand{\NN}{\mathcal{N}}

\newcommand{\CC}{{\mathcal C}}

 \newcommand{\SU}{\mathrm{SU}}
 \newcommand{\USp}{\mathrm{USp}}
 \newcommand{\U}{\mathrm{U}}
  \newcommand{\SO}{\mathrm{SO}}

 \newcommand{\cI}{ {\mathcal I}}
 \newcommand{\cT}{ {\mathcal T}}
 \newcommand{\tx}{\tilde{x}}
 
 \newcommand{\cO}{ {\mathcal O}}

  \newcommand{\tth}{\tilde{h}}

\def\ttt{\tilde{t}}
\def\tz{\tilde{z}}

\def\p{\eta_p}

\def\({\left(}
\def\){\right)}

\def\p{\partial}

\newcommand{\thp}[1]{\theta_p\left(#1\right)}
\newcommand{\thq}[1]{\theta_q\left(#1\right)}

\newcommand{\qPoc}[2]{\left(#1;#2\right)_\infty}
\newcommand{\GF}[1]{\Gamma_e\left(#1\right)}
\newcommand{\thf}[1]{\theta_p\left(#1\right)}
\newcommand{\pq}[2]{(pq)^{\frac{#1}{#2}}}
\newcommand{\pqm}[2]{(pq)^{-\frac{#1}{#2}}}

\title{$C_2$ generalization of the van Diejen model from the minimal $(D_5,D_5)$ conformal matter.}

\preprint{}

\author[a]{Belal Nazzal,}

\author[b]{Anton Nedelin,}

\affiliation[a]{Department of Physics, Technion, Haifa, 32000, Israel}

\affiliation[b]{Section de Math\'ematiques, Universit\'e de Gen\`eve, 1211 Gen\`eve 4, Switzerland}

\emailAdd{sban@campus.technion.ac.il, anton.nedelin@gmail.com}

\abstract{We study superconformal indices of $4d$ compactifications of the $6d$ minimal $(D_{N+3},D_{N+3})$ conformal matter
	theories on a punctured Riemann surface. Introduction of  supersymmetric surface defect in these theories is done at the
	level of the index by the action of the finite difference operators on the corresponding indices. There exist at least three
different types of such operators according to three types of punctures with $A_N, C_N$ and $\left(A_1\right)^N$ global symmetries. We mainly concentrate
on $C_2$ case and derive explicit expression for an infinite tower  of difference operators generalizing the van Diejen model. We check various properties of these
operators originating from the geometry of compactifications. We also provide an expression for the kernel function of both our $C_2$ operator and previously derived
$A_2$ generalization of van Diejen model. Finally we also consider compactifications with $A_N$-type punctures and derive the full tower of commuting difference
operators corresponding to this root system generalizing the result of our previous paper.}

\begin{document}

\maketitle
\flushbottom


\section{Introduction}
\label{sec:intro}

Intricate connection between supersymmetric gauge theories and integrable systems plays an important
role in modern theoretical physics and mathematics. From physics point of view whenever some sector of the
gauge theory is related to an integrable system, observables in this sector can be computed using
rich variety of integrability techniques. Canonical example of this situation is the integrability
of $\NN=4$ Super Yang-Mills (SYM) theory \cite{Minahan:2002ve,Beisert:2010jr}. In particular it connects calculation of the planar scaling
dimensions of $\NN=4$ SYM to the integrable system of spin chains and allows to compute these scaling dimensions at arbitrary coupling.
On the other hand exploration of such connections can
also shed light on some questions about integrable systems and even lead to construction of new classes of such systems making
this kind of studies interesting from mathematics point of view.

In our work we are exploring a particular class of connections between six dimensional $(1,0)$ superconformal field theories and
elliptic quantum mechanics Hamiltonians in the spirit of Bethe/gauge correspondence of Nekrasov and Shatashvili
\cite{Nekrasov:2009ui,Nekrasov:2009rc,Bullimore:2014awa,Hatsuda:2018lnv}.
In particular we consider four-dimensional theories with four supercharges obtained by compactification of a $6d$ SCFT on a punctured Riemann surface.
In order for an integrable model to emerge in this setting we have to introduce surface defects into our $4d$ theory and study its superconformal index.

This construction was first established in the context of compactifications of $6d$ $(2,0)$ of ADE type \cite{Gaiotto:2012xa,Lemos:2012ph}. In particular the
corresponding $4d$ superconformal indices with the defect were found to be closely related to the Ruijsenaars-Schneider (RS) elliptic analytic finite difference
operators (A$\Delta$Os). Later these results were extended to many other cases: class $S_k$ $4d$ theories \cite{Gaiotto:2015usa},
compactifications of $A_2$ and $D_4$ minimal $6d$ SCFTs \cite{Razamat:2018zel} and compactifications of rank one E-string theories \cite{Nazzal:2018brc}.
In some of these cases the obtained A$\Delta$Os were already known in the literature. For example in E-string compactifications \textit{van Diejen (vD)  model}  \cite{Diejen,MR1275485}
was observed. But in some cases operators were previously unknown as in the case of the minimal $6d$ SCFTs compactifications. Study of such novel
operators constitutes an interesting field of  research with some initial steps already taken in this direction \cite{Ruijsenaars:2020shk,minimal_operators}.

From the point of view of physics the connection of superconformal indices in the presence of defects with the integrable quantum mechanics Hamiltonians allows
one to bootstrap index of an arbitrary theory obtained by compactifying corresponding $6d$ theory on a punctured Riemann surface. In particular if we
know the eigenfunctions of the corresponding A$\Delta$O we can compute the index of any theory obtained in such compactifications including non-Lagrnagian
theories for which there are no other methods of computing the index. As mentioned previously, the first setting where these ideas were tested
are $4d$ class-S theories obtained in the compactifications of $6d$ $(2,0)$ theory \cite{Gaiotto:2012xa}. In this case the corresponding integrable system
was given by  elliptic RS model. Eigenfunctions of the elliptic  RS Hamiltonians are not known in general but some of their limits are well studied in the mathematical
literature. These limits were used in order to prove previously established relations \cite{Gadde:2011uv,Gadde:2009kb,Gadde:2011ik} of superconformal indices
 with the Macdonald and Schur polynomials allowing one to compute indices of class S non-Lagrangian theories.

The construction outlined above relies on the intermediate $5d$ layer. For things to work there should exist an effective $5d$ gauge theory obtained by
compactifying original $6d$ SCFT on a circle with a choice of holonomies for its global symmetries. In particular different
$5d$ compactifications lead to different types of punctures on the Riemann surfaces used to obtain $4d$ theories.
One of the interesting problems related to this fact is the compilation of the dictionary between known  compactifications of various
$6d$ SCFTs and elliptic  integrable systems.
In our previous paper \cite{Nazzal:2021tiu}  we have considered compactifactions of the $6d$ minimal $(D_{N+3},D_{N+3})$ conformal
matter theories \cite{Razamat:2019ukg,Razamat:2020bix}. In particular we have derived A$\Delta$Os corresponding
to the intermediate $5d$ $\SU(N+1)$ gauge theory or equivalently the $A_N$ type puncture on the compactification surface.
These elliptic A$\Delta$Os appeared to be previously unknown $A_N$ generalizations of vD model. On the other hand there are at least two
more $5d$ effective descriptions of $6d$ SCFTs corresponding to $\USp(2N)$ and $\SU(2)^N$ gauge theories giving rise to the punctures with the same
$C_N$ and $(A_1)^{\otimes N }$ global symmetries. These two descriptions should lead to additional higher-rank generalizations of the vD model.

In our present paper we follow this line of research and closely study compactifications of the minimal $(D_{N+3},D_{N+3})$ conformal matter theory
on a Riemann surface with the $C_N$-type punctures. In particular we concentrate on the next to simplest case of $N=2$\footnote{The simplest case of $N=1$, as
well as the case of $(A_1)^1$, were studied in our previous paper \cite{Nazzal:2021tiu}. As a result we observed two alternative parametrizations
of the standard $BC_1$ vD model.} and derive corresponding infinite tower of A$\Delta$Os. We also devote part of the paper to the study and proof of some
remarkable properties of these operators that follow from the geometry of corresponding compactifications. In particular we prove the novel
kernel property for two operators of different type but same rank. We also discuss commutation property of obtained operators.

The paper is organised as follows. In section  \ref{sec:an:oper} we review our previous results for $A_N$ type operators. In addition
to the previous results we also derive full tower of such A$\Delta$Os which was not obtained previously.
In Section \ref{sec:c2:operator} we derive in details novel generalization of vD operators corresponding to $C_2$ root system.\footnote{We would like to stress that the operator we derive is different from the 
standard higher-rank vD operator \cite{Diejen, MR1275485} since the latter one is 
associated with the affine $BC_n$ type root system while we discuss $C_2$ root system 
in our paper. We expect that the canonical $BC_n$ vD model can be obtained in 
a similar manner using compactifications of the rank-Q E string theory. Unfortunatelly so far only compactifications on the  spheres with two punctures are known for 
these theories so we are not able to prove this conjecture. }
In section \ref{sec:c2:properties} we discuss properties of derived $C_2$ A$\Delta$Os. In particular we pay special attention to a new kernel
function for simultaneously $A_2$ and $C_2$ operators and give fully analytic proof of the corresponding kernel equation. We also briefly discuss commutation
relations and check them perturbatively in expansion. In section \ref{sec:discussion} we briefly summarize  our results and  discuss
plans for the future research further developing these results. Finally the paper has a number of Appendices collecting useful formulas as well
as technical details of various calculations in our paper.

\section{$A_N$ operators.}
\label{sec:an:oper}

In this section we will briefly review derivation of the $A_N$ generalization of the van Diejen operator.
So called basic version\footnote{See  the definition of the basic operators and their towers below.}
of this operator has been derived in our previous paper \cite{Nazzal:2021tiu}. Here we will extend this result to the full tower of operators.

\begin{figure}[!btp]
        \centering
        \includegraphics[width=\linewidth]{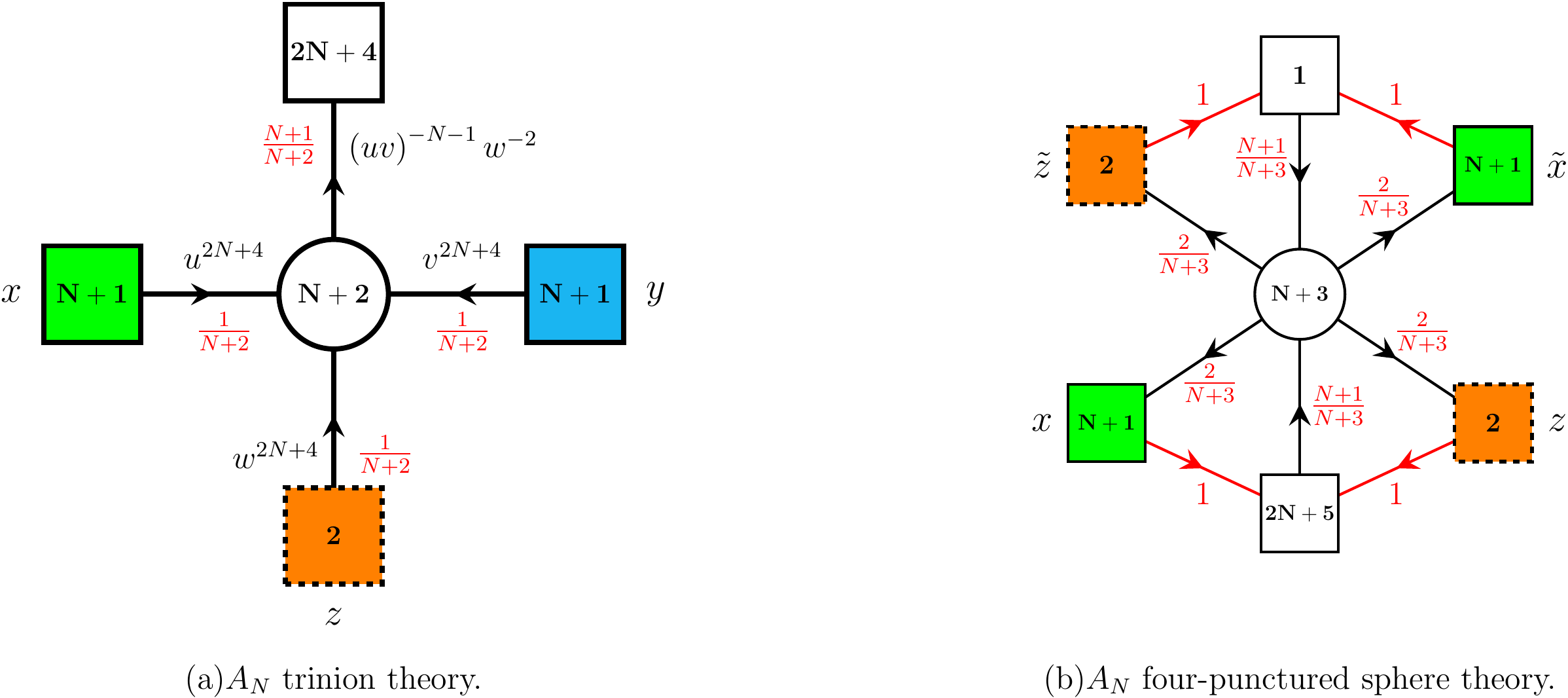}
	\caption{(a) $A_N$ three-punctured sphere with two maximal and one minimal puncture. (b) $A_N$ four-punctured sphere obtained by S-gluing two three-punctured spheres}
    \label{pic:an:trinion}
\end{figure}

Just as in \cite{Nazzal:2021tiu} in order to derive A$\Delta$O we start with the $4d$ three-punctured sphere
theory which was first introduced in \cite{Razamat:2020bix}. This theory is obtained in the compactification of the
$6d$  minimal $(D_{N+3},D_{N+3})$ conformal matter on a sphere with two maximal $\SU(N+1)$ punctures and one
minimal $\SU(2)$ puncture. The quiver of this theory is shown in Figure \ref{pic:an:trinion}.
 In addition to the global symmetry, the punctures are characterized by the moment map operators. In particular
 both $\SU(N+1)$ maximal and $\SU(2)$ minimal punctures are characterized by $(2N+4)$ mesonic and $2$ baryonic moment maps:
\be
&&M_u={\bf N+1}^x\otimes \left( {\bf 2N+4}_{u^{N+3}v^{-N-1}w^{-2}}\oplus {\bf 1}_{\left( uv^{N+1} \right)^{2N+4}}\right)
\oplus \overline{ {\bf N+1} }^x\otimes{\bf 1}_{\left(u^N w^2\right)^{2N+4}}\,,
\nn\\
&&M_v={\bf N+1}^y\otimes \left( {\bf 2N+4}_{v^{N+3}u^{-N-1}w^{-2}}\oplus {\bf 1}_{\left( vu^{N+1} \right)^{2N+4}}\right)
\oplus \overline{ {\bf N+1} }^y\otimes{\bf 1}_{\left(v^N w^2\right)^{2N+4}}\,,
\nn\\
&&M_w={\bf 2}^z\otimes\left( {\bf 2N+4}_{\left(uvw^{-2} \right)^{-N-1}}\oplus {\bf 1}_{\left(wv^{N+1}\right)^{2N+4}}
\oplus {\bf 1}_{\left( wu^{N+1}\right)^{2N+4}}\right)\,,
\label{an:moment:maps}
\ee
where $a_i,\, u,\, v,\, w$ are fugacities of the Cartans of the $6d$ global $\SO(4N+8)$ symmetry. Subscripts of the moment maps written above denote their charges
w.r.t. to these symmetries.

Further $S$-gluing two such trinions along the maximal punctures we can obtain four-punctured sphere with zero flux two maximal and two minimal punctures.
To obtain the corresponding $4d$ gauge theory we should just
take two copies of trinion theories shown in Figure \ref{pic:an:trinion} and then identify
and gauge corresponding global symmetries of the maximal punctures. Here and everywhere else in the paper
all operations with gauge theories are expressed in terms of the superconformal indices. In this langauge, the gluing procedure takes the following form:
\be
K_4^A(x,\tx,z,\tz)=\kappa_N\oint \prod\limits_{i=1}^N\frac{dy_i}{2\pi i y_i}
\prod\limits_{i\neq j}^{N+1}\frac{1}{\GF{\frac{y_i}{y_j}}}\bar{K}_3^{A}(\tx,y,\tz)K_3^{A}(x,y,z)\,.
\ee
where $K^{A}_3(x,y,z)$ is the index of the trinion with $y$
being fugacity of the global $\SU(N+1)$ symmetry of the puncture we glue along and $\bar{K}_3^{A}$ is the index of the conjugated trinion.
Finally $\kappa_N$ is the usual constant given by:
\be
\kappa_N\equiv \frac{\qPoc{q}{q}^{N}\qPoc{p}{p}^{N}}{\left(N+1\right)!}\,.
\label{kappa:def}
\ee

Performing this $S$-gluing operation we obtain the four-punctured sphere theory
shown in Figure \ref{pic:an:trinion} with the corresponding superconformal index specified in \eqref{sun:4pt:sphere}. This theory was previously obtained
by us in \cite{Nazzal:2021tiu} to derive basic $A_N$ A$\Delta$Os. Now in order to derive the operator we should close two minimal punctures
of this four-punctured sphere. To do it we should break the global symmetry of the puncture.
This can be achieved by giving a non-trivial vev $\langle \p_{12}^L \p_{34}^K  M\rangle\neq 0$ to the derivatives of
one of the moment map operators. When we close punctures with at least one of $K$ or $L$ not equal to zero,
i.e. vev is space-time dependent, we effectively insert defect into the theory \cite{Gadde:2013dda,Gaiotto:2012xa}.
At the level of the superconformal index, closing the puncture amounts to giving a corresponding weight
to the fugacity of the puncture's global symmetry. Once we do it we hit a pole of the index.
Then computation of the residue of this pole results in the superconformal index of the IR theory
that the UV theory flowed to due to the introduction of the vev. In our case we choose
to close $\SU(2)_z$ minimal puncture with the defect and $\SU(2)_{\tz}$ puncture without, i.e. choosing $L=K=0$ in the vev.
In particular let's say that we are going to compute A$\Delta$O acting on the puncture with the moment maps of charges $\tth_i$
and an overall $\U(1)$ charge
\be
\tth\equiv \prod\limits_{i=1}^{2N+6}\tth_i\,.
\label{total:u1}
\ee
For example the moment maps of the $\SU(N+1)_x$ puncture of the trinion shown in Figure \ref{pic:an:trinion} correspond to $M_u$ operators specified in
\eqref{an:moment:maps} and have the following charges:
\be
\tth_i=u^{N+3}v^{-N-1}w^{-2}a_i,\, \quad i=1,\dots,2N+4\,, \quad \prod_{i=1}^{2N+4}a_i=1\,,
\nn\\
\tth_{2N+5}=\left(uv^{N+1}\right)^{2N+4},\,\quad  \tth_{2N+6}= \left(u^{N} w^{2}\right)^{-2N-4}\,,\quad \tth=\left(uw^{-1}\right)^{8(N+2)}\,,
\label{a2:moment:maps}
\ee
where we have flipped the charge $\tth_{2N+6}$ of the last moment map since this parametrization will be more natural
for our operators and in this case all of the moment maps of the maximal puncture
transform in the fundamental representation of $\SU(N+1)_x$. We can now notice from
\eqref{an:moment:maps} that charges of the minimal puncture moment maps are related to the charges of the
moment maps of the maximal punctures by simple relation
\be
\tth^{\SU(2)}_i=\left(\tth^{\SU(N+1)}_i\right)^{-1}\left(\tth^{\SU(N+1)}\right)^{\frac{1}{4}}\,,
\ee

So we can express everything in terms of only the moment maps of the maximal  puncture we act on. Now assume we give
vevs to the moment maps (mesonic or baryonic) with charges $\tth_i^{-1}\tth^{1/4}$ of both $\SU(2)_z$
and $\SU(2)_{\tz}$. Moment
maps we use should be the same in order to keep total flux of the $6d$ global symmetries zero.
At the level of the index calculations it corresponds to computing the residue of the index of the four-punctured sphere theory at the pole
\be
z=Z_{i;L,M}^*=(pq)^{-\frac12} \tth_i\tth^{-\frac{1}{4}} q^{-M} p^{-L} ,
\;\;\;\;\;\;\; \tilde{z}=\tilde{Z}^*_{i;0,0}=(pq)^{-\frac12} \tth_i^{-1}\tth^{\frac{1}{4}} q^{-\tilde{M}} p^{-\tilde{L}}\,,
\label{an:pole}
\ee
where $L,M,\tilde{L},\tilde{M}$ are positive integers corresponding to the powers of derivatives inside the vev.
As we mentioned previously it is enough to introduce
defect only for one of the two punctures. Hence we choose $\tilde{M}=\tilde{L}=0$ and keep
$L\,,M$ general. Then we compute corresponding residues of the index of the four-punctured
sphere theory and obtain theory for the tube with two maximal $\SU(N+1)$ punctures and a codimension-two defect.
Its superconformal index is given by\footnote{Here and further we often omit some overall factors which are irrelevant for
the derivations of A$\Delta$Os. Because of this we use $\sim$ instead of strict equality here.}:
\be
K_{(2;i;L,M)}^A(x,\tx)\sim\mathrm{Res}_{z\to Z^*_{i;L,M}\,,\tz\to \tilde{Z}^*_{i;0,0}}K_4^A(x,\tx,z,\tz)
\ee
Finally in order to obtain desired A$\Delta$O we glue our tube with the defect to an arbitrary Riemmann
surface with maximal $\SU(N+1)_{\tx}$ puncture. As the result of this gluing we expect to obtain action of a finite-difference operator
on the index $\cI(\tx)$ of this $4d$ $\NN=1$ theory:
\be
\cO^{(A_N;h_{k};L,M)}_x\cdot\cI(x)=\kappa_N\oint\prod\limits_{j=1}^N\frac{d\tx_j}{2\pi i\tx_{j}}\prod\limits_{i\neq j}^{N+1}\frac{1}{\GF{\frac{\tx_i}{\tx_j}}}K_{(2;i;L,M)}^A(x,\,\tx)\cI(\tx)
\label{c2:final:gluing}
\ee
Details of all the calculations summarized above are given in Appendix \ref{app:an:operator}. They result in the following operator:
\begin{shaded}
\begin{align}
&\cO_x^{(A_N;\tth_k;M,L)} =\sum\limits_{(\sum_{i=1}^{N+2}{m_i}=M)}\sum\limits_{(\sum_{i=1}^{N+2}{l_i}=L)}\sum\limits_{(\sum_{i=1}^{N+1}{s_i}=M-m_{N+2})}
\sum\limits_{(\sum_{i=1}^{N+1}{r_i}=L-l_{N+2})} \,C_{L,M}^{\vec l,\vec m,\vec r,\vec s}
\times\nn\\
&
\prod\limits_{i=1}^{N+1}
\frac{
\prod_{b\neq k}^{2N+6}\prod_{n=0}^{s_i-1} \thp{ (pq)^\frac12 \tth_b^{-1}   x_i^{-1} q^{n-m_i} p^{-l_i} } \prod_{n=0}^{r_i-1} \thq{(pq)^\frac12  \tth_b^{-1} x_i^{-1} q^{s_i-m_i} p^{n-l_i}}
}
{
\prod_{j\neq i}^{N+1}\prod_{n=0}^{m_j-1}\thp{ q^{n-m_i }p^{ -l_i } x_j/x_i} \prod_{n=0}^{l_j-1}\thq{ q^{m_j-m_i }p^{n -l_i } x_j/x_i}
}
\nn\\&
\frac{\prod_{n=0}^{M-s_i-1}\thp{(pq)^\frac12 \tth_k^{-1} x_i^{-1} p^{r_i-l_i} q^{n+s_i-m_i} }\prod_{n=0}^{L-r_i-1}\thq{(pq)^\frac12 \tth_k^{-1} x_i^{-1} p^{n+r_i-l_i} q^{M-m_i} }}
{
 \prod_{n=0}^{m_{N+2}-1}\thp{(pq)^{-\frac12} \tth_k\tth^{-\frac12}   x_i^{-1}  p^{-L-l_i} q^{n-M-m_{i} } }\prod_{n=0}^{l_{N+2}-1}\thq{(pq)^{-\frac12} \tth_k\tth^{-\frac12} x_i^{-1}  p^{n-L-l_i } q^{m_{N+2}-M-m_i } }
}
\nn\\&
\frac
{  \prod_{n=0}^{M-m_{N+2}-s_i-1}\thp{(pq)^\frac12 \tth_k^{-1}\tth^{\frac12}   x_i  p^{l_i} q^{n+m_i } }\prod_{n=0}^{L-l_{N+2}-r_i-1}\thq{(pq)^\frac12 \tth_k^{-1}\tth^{\frac12} x_i  p^{n+l_i } q^{M-m_{N+2}-s_i } }  }
{  \prod_{n=0}^{m_i-1}\thp{(pq)^\frac12 \tth_k^{-1}\tth^{\frac12}   x_i  p^{L-l_{N+2}+l_i} q^{n+M-m_{N+2} } }\prod_{n=0}^{l_i-1}\thq{(pq)^\frac12 \tth_k^{-1}\tth^{\frac12} x_i  p^{n+L-l_{N+2} } q^{M-m_{N+2}+m_i } }  }
\nn\\&
\frac{
1}{
\prod_{j\neq i}^{N+1}\prod_{n=0}^{r_i-1}\thq{ q^{m_j-m_i+s_i-s_j}p^{n+l_j-l_i-r_j} x_j/x_i}\prod_{n=0}^{s_i-1}\thp{ q^{n+m_j-m_i-s_j}p^{l_j-l_i-r_j} x_j/x_i}
}
\times\nn\\
&\hspace{10cm}\Delta_q^{m_i-s_i}(x_i)\Delta_p^{l_i-r_i}(x_i)\,,
\label{AN-HT}
\end{align}
\end{shaded}
where $C_{L,M}^{\vec l,\vec m,\vec r,\vec s}$ are $x$-independent constant factors given by,

\begin{shaded}
\be
&
C_{L,M}^{\vec l,\vec m,\vec r,\vec s}=
\frac{
\prod_{n=1}^{M-m_{N+2}} \thp{ q^{-n} } \prod_{n=1}^{L-l_{N+2} } \thq{ q^{m_{N+2}-M} p^{-n} }
}
{
\prod_{n=1}^{s_i} \thp{ q^{-n} p^{-r_i} } \prod_{n=1}^{r_i } \thq{   p^{-n} }
\prod_{n=1}^{m_i} \thp{ q^{-n} p^{-l_i} } \prod_{n=1}^{l_i } \thq{   p^{-n} }
}
\nn
\\&
\frac{\prod_{b\neq k}^{2N+6} \prod_{n=0}^{m_{N+2}-1} \thp{ \tilde{h}_k \tilde{h}^{-\frac12} \tilde{h}_b q^{n-M} p^{-L} }
\prod_{n=0}^{l_{N+2}-1} \thq{ \tilde{h}_k \tilde{h}^{-\frac12} \tilde{h}_b q^{m_{N+2}-M} p^{n-L} }
}{
\prod_{n=1}^{m_{N+2}} \thp{ pq\, \tilde{h}_k^{-2} \tilde{h}^\frac12 q^{2M-n} p^{2L-l_{N+2}} }
\prod_{n=1}^{l_{N+2}} \thq{ pq\, \tilde{h}_k^{-2} \tilde{h}^\frac12 q^{2M} p^{2L-n } }
}
\nn\\
\ee
\end{shaded}

Also $\Delta$s are shift operators defined as follows:
\be
\Delta^m_a(x_i)f(x)\equiv f\left(x_i\to a^mx_i\right)\,, \quad a=q,p\,.
\label{delta:shift}
\ee
 The operator contains all shifts of the form $q^{m_i}p^{l_i}x_i$ where $\vec{m}$ and $\vec{l}$ are all possible partitions of length $N+1$ of $M$ and $L$ correspondingly.
  At each level, i.e. fixed $M$ and $L$ there are
 $2N+6$ operators due to $2N+6$ moment maps with the charges $\tth_k\tth^{-\frac{1}{4}}$. There are also $2N+6$ other operators obtained by giving vevs to the flipped
 moment maps of the charge $\tth_k^{-1}\tth^{\frac{1}{4}}$. They have similar form and properties so we do not present them here. All the operators should commute with each other, and we checked this in expansion in $p,q$ for a few of the simplest cases.  Now we will refer to the case $M=1$ and $L=0$,
 or vice versa, as \textit{basic operators}. These basic operators for $A_N$ generalizations of vD model
 were derived by us previously in \cite{Nazzal:2021tiu}. The operators above also reproduce our previous results when we fix $M=1,\, L=0$
 or $M=0,\, L=1$.
Further in our paper we will also need the basic operator obtained by closing flipped moment maps. This operator can be found in \cite{Nazzal:2021tiu} and has the following form:
\begin{shaded}
\be
&&\cO_{x}^{(A_N;\tth_i^{-1};1,0)} \cdot {\cal I}(x)\equiv
\nn\\
&&\hspace{1.5cm}\left(\sum\limits_{l\neq m}^{N+1}A^{(A_N;\tth_i^{-1};1,0)}_{lm}(x)\Delta_q^{-1}(x_l)\Delta_q(x_m)+ W^{(A_N;\tth_i^{-1};1,0)}\left(x,\tth\right)\right) \cI(x)\,,
\label{sun:op:noflip:gen}
\ee
\end{shaded}
where the shift part of this operator  is given by
\be
A^{(A_N;\tth_i^{-1};1,0)}_{lm}(x)=\frac{\prod\limits_{j=1}^{2N+6}\thf{\pq{1}{2} \tth_j^{-1}x_l^{-1}}}
{\thf{\frac{x_m}{x_l}}\thf{q\frac{x_m}{x_l}}}
\prod\limits_{k\neq m\neq l}^{N+1}\frac{\thf{\pq{1}{2}\tth_i^{-1}x_k^{-1}}
\thf{\pq{1}{2}\tth_i^{-1}\tth^{1/2} x_k}}{\thf{\frac{x_k}{x_l}}\thf{\frac{x_m}{x_k}}}\,,
\nn\\
\label{sun:op:noflip:gen:shift}
\ee
and the constant part is given by:
\be
&&W^{(A_N;\tth_i^{-1};1,0)}(x,\tth)=\frac{\prod\limits_{j\neq i}^{2N+6}\thf{q^{-1}\tth_i\tth_j\tth^{-1/2}}}
{\thf{q^{-2}\tth_i^{2}\tth^{-1/2}}}
\prod\limits_{k=1}^{N+1}\frac{\thf{\pq{1}{2}\tth_i^{-1} x_k^{-1}}}
{\thf{\pqm{1}{2}\tth_i\tth^{-1/2}q^{-1}x_k^{-1}}}+
\nn\\
&&\sum\limits_{m=1}^{N+1}\frac{\prod\limits_{j\neq i}^{2N+6}\thf{\pq{1}{2}\tth_jx_m}}{\thf{\pq{1}{2}\tth_i^{-1}\tth^{1/2}q x_m}}
\prod\limits_{k\neq m}^{N+1}\frac{\thf{\pq{1}{2}\tth_{i}^{-1}\tth^{1/2}x_k}
\thf{\pq{1}{2}\tth_i^{-1}x_k^{-1}}}
{\thf{q^{-1}\frac{x_k}{x_m}}\thf{\frac{x_m}{x_k}}}\,.
\label{sun:op:noflip:gen:const}
\ee
This constant part is elliptic function in each $x_i$ variable with periods $1$ and $p$. It has  poles in the fundamental domain at the following positions:
\be
x_i=q^{\pm 1}x_r\,,\quad x_i=sq^{\pm \frac{1}{2}}P_i^{-\frac{1}{2}}\,,\quad x_i=sq^{\pm \frac{1}{2}}p^{\frac{1}{2}}P_i^{-\frac{1}{2}}\,,\quad s=\pm 1\,,
\label{an:poles}
\ee
where we defined
\be
P_i\equiv \prod\limits_{j\neq i}^N x_j\,.
\label{Pi:def}
\ee
From expression \eqref{sun:op:noflip:gen:const} it looks like there are extra poles in the constant part, but careful examination shows that residues at these values of
$x$'s are zero so there are no real poles there. At the poles \eqref{an:poles} we have the following residues:
\begin{align}
\mathrm{Res}_{x_l=qx_r}W^{(A_N;\tth_i^{-1};1,0)}(x,\tth)&=-\frac{qx_r}{\qPoc{p}{p}^2\thf{q^{-1}}}\prod\limits_{j=1}^{2N+6}\thf{\pq{1}{2}\tth_jx_r}
\times\nn\\
&\hspace{0cm} \prod\limits_{k\neq l\neq r}^{N+1}\frac{\thf{\pq{1}{2}\tth_i^{-1}\tth^{1/2}x_k}\thf{\pq{1}{2}\tth_i^{-1}x_k^{-1}}}{\thf{q^{-1}\frac{x_k}{x_r}}\thf{\frac{x_r}{x_k}}}
\nn\\
\mathrm{Res}_{x_l=q^{-1}x_r}W^{(A_N;\tth_i^{-1};1,0)}(x,h)&=\frac{q^{-1}x_r}{\qPoc{p}{p}^2\thf{q^{-1}}}\prod\limits_{j=1}^{2N+6}\thf{\pq{1}{2}\tth_j^{-1}x_r^{-1}}
\times\nn\\
&\hspace{0cm} \prod\limits_{k\neq l\neq r}^{N+1}\frac{\thf{\pq{1}{2}\tth_i^{-1}\tth^{1/2}x_k}\thf{\pq{1}{2}\tth_i^{-1}x_k^{-1}}}{\thf{q^{-1}\frac{x_r}{x_k}}\thf{\frac{x_k}{x_r}}}
\nn\\
\mathrm{Res}_{x_l=sq^{-\frac{1}{2}}P_l^{-\frac{1}{2}}}W^{(A_N;\tth_i^{-1};1,0)}(x,h)&=s\frac{q^{-\frac{1}{2}}P_l^{-\frac{1}{2} }}{2 \qPoc{p}{p}^2\thf{q^{-1}}}
\prod\limits_{j=1}^{2N+6}\thf{sp^{1/2}\tth_jP_l^{-1/2}}
\times\nn\\
&\prod\limits_{k\neq l}^N \frac{\thf{\pq{1}{2}\tth_i^{-1}\tth^{1/2}x_k}\thf{\pq{1}{2}\tth_i^{-1}x_k^{-1}}}{\thf{sq^{-1/2}P_l^{-1/2}x_k^{-1}}\thf{sq^{-1/2}P_l^{1/2}x_k}}\,,
\nn\\
\mathrm{Res}_{x_l=sq^{\frac{1}{2}}P_l^{-\frac{1}{2}}}W^{(A_N;\tth_i^{-1};1,0)}(x,\tth)&=-s\frac{q^{\frac{1}{2}}P_l^{-\frac{1}{2} }}{2 \qPoc{p}{p}^2\thf{q^{-1}}}
\prod\limits_{j=1}^{2N+6}\thf{sp^{1/2}\tth_jP_l^{-1/2}}
\times\nn\\
&\prod\limits_{k\neq l}^N \frac{\thf{\pq{1}{2}\tth_i^{-1}\tth^{1/2}x_k}\thf{\pq{1}{2}\tth_i^{-1}x_k^{-1}}}{\thf{sq^{-1/2}P_l^{-1/2}x_k^{-1}}\thf{sq^{-1/2}P_l^{1/2}x_k}}\,,
\nn\\
\mathrm{Res}_{x_l=sp^{\frac{1}{2}}q^{-\frac{1}{2}}P_l^{-\frac{1}{2}}}W^{(A_N;\tth_i^{-1};1,0)}(x,\tth)&=s\frac{q^{-\frac{1}{2}}p^{\frac{3}{2}}\tth^{-\frac{1}{2}}
P_l^{\frac{1}{2} }}{2 \qPoc{p}{p}^2\thf{q^{-1}}}
\prod\limits_{j=1}^{2N+6}\thf{s\tth_jP_l^{-1/2}}
\times\nn\\
&\prod\limits_{k\neq l}^N \frac{\thf{\pq{1}{2}\tth_i^{-1}\tth^{1/2}x_k}\thf{\pq{1}{2}\tth_i^{-1}x_k^{-1}}}{\thf{sp^{-1/2}q^{-1/2}P_l^{-1/2}x_k^{-1}}\thf{sp^{1/2}q^{-1/2}P_l^{1/2}x_k}}\,,
\nn\\
\mathrm{Res}_{x_l=sp^{\frac{1}{2}}q^{\frac{1}{2}}P_l^{-\frac{1}{2}}}W^{(A_N;\tth_i^{-1};1,0)}(x,\tth)&=-s\frac{q^{\frac{1}{2}}p^{\frac{3}{2}}\tth^{-\frac{1}{2}}
P_l^{\frac{1}{2} }}{2 \qPoc{p}{p}^2\thf{q^{-1}}}\prod\limits_{j=1}^{2N+6}\thf{s\tth_jP_l^{-1/2}}
\times\nn\\
&\prod\limits_{k\neq l}^N \frac{\thf{\pq{1}{2}\tth_i^{-1}\tth^{1/2}x_k}\thf{\pq{1}{2}\tth_i^{-1}x_k^{-1}}}{\thf{sp^{-1/2}q^{-1/2}P_l^{-1/2}x_k^{-1}}\thf{sp^{1/2}q^{-1/2}P_l^{1/2}x_k}}\,,
\label{an:residues}
\end{align}

These are completely general expressions for the whole family of $A_N$ operators.

\section{Derivation of $C_2$ operators.}
\label{sec:c2:operator}

In this  section we will discuss derivation of the $C$-type
rank-$2$ analytic finite difference operators A$\Delta$O. In our previous
paper \cite{Nazzal:2021tiu} we have already derived the basic operator for $A_N$ generalization
of van Diejen model and the calculation of the full tower of these operators
is summarized in Section \ref{sec:an:oper}.
Unfortunately full $C_N$ generalization is still out of reach for us due to technical
complications but we can concentrate on the study of rank-$2$ generalizations with
$C_2$  root system.

We will derive $C_2$ operator corresponding to the insertion
of the codimension-two defect due to the nontrivial vev of the
holomorphic derivatives of the moment maps $\langle \partial_{\pm}^K M \rangle\neq 0$.
To simplify our calculations of A$\Delta$O we will start with the derivation
of four- and  three-punctured spheres with $C_2$ type punctures.

\begin{figure}[!btp]          \centering
           \subfigure[A tube with one $\SU(N+1)$ and one $\USp(2N)$ puncture. ]{\label{pic:cn:tube}\includegraphics[width=0.3\linewidth]{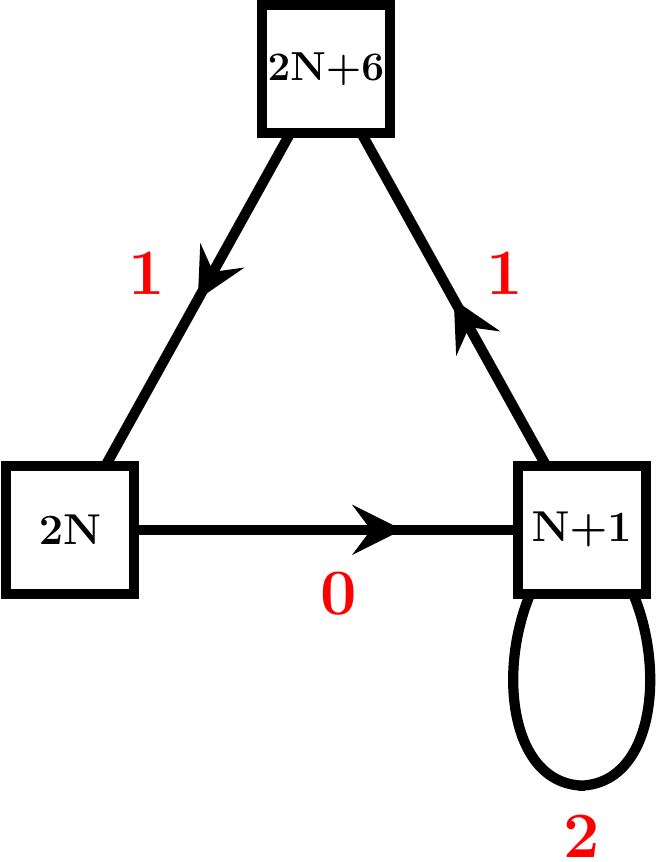}}
           \hspace{27mm}
           \subfigure[Trinion with two maximal $\USp(2N)$ and one minimal $\SU(2)$ puncture.]{\label{pic:cn:trinion}\includegraphics[width=0.5\linewidth]{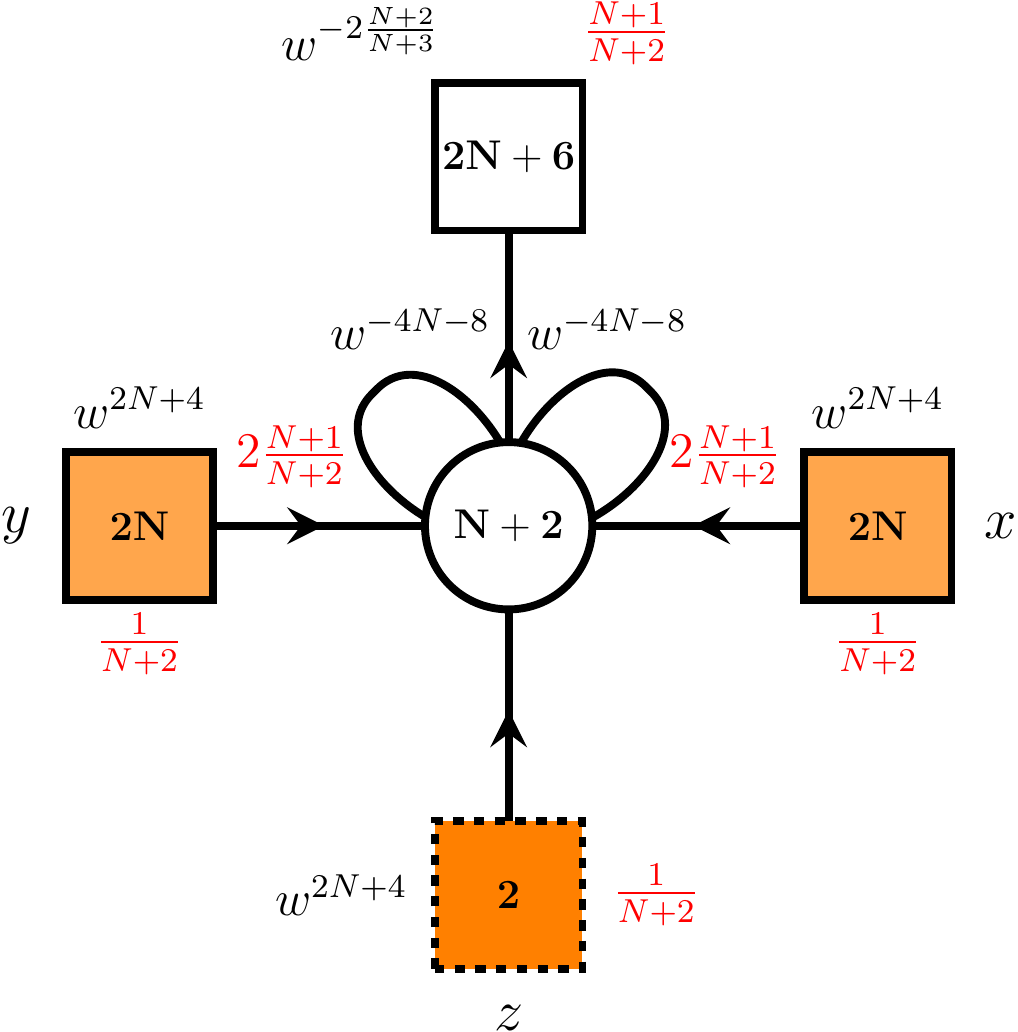}}
           \caption{Tube and trinion with maximal $C_N$ punctures.}
\end{figure}

The basic $C_N$ trinion was already derived by us in Appendix B.5 of
\cite{Nazzal:2021tiu} and is shown in Figure \ref{pic:cn:trinion}. Generalizations of
this trinion to the cases of three maximal punctures and higher numbers of minimal punctures
can also be found in \cite{Kim:2023qbx}.  In order to
derive this trinion we used a tube theory with one $\USp(2N)$ and one $\SU(N+1)$
maximal punctures shown in Figure \ref{pic:cn:tube}. This theory was first discussed
in \cite{Kim:2018bpg}. In order to obtain trinion theory shown in Figure
\ref{pic:cn:trinion} we start with the three-punctured sphere with two maximal
$\SU(N+1)$ and one minimal $\SU(2)$ punctures and glue two $A_NC_N$  tubes to
the maximal punctures. The moment maps of the punctures of the resulting trinion are given by $(2N+6)$ mesons both for maximal and minimal punctures:
\be
&M_x={\bf 2N}_x\otimes \left( {\bf 2N+6}\right)_{w^{2\frac{(N+2)^2}{N+3}}}\,,
\qquad M_y={\bf 2N}_y\otimes \left( {\bf 2N+6}\right)_{w^{2\frac{(N+2)^2}{N+3}}}\,,
\nn\\
&M_z={\bf 2}_z\otimes \left( {\bf 2N+6}\right)_{w^{2\frac{(N+2)^2}{N+3}}}\,,
\label{cn:moment:maps}
\ee
where $w$ is one of the parameters of Cartans of the global $\SO(4N+8)$ symmetry
of the $6d$ minimal conformal matter theory. In general we will use two types of
parametrization of $6d$ global symmetry in this paper. First one is
natural to use in compactifications with $C_N$ type punctures and
it is given by
\be
\ta_i\,,\quad i=1,\dots,2N+6\,,\quad \prod\limits_{i=1}^{2N+6}\ta_i=1\,, \quad w\,.
\label{cn:parametrization}
\ee
Another parametrization is useful when we work with $A_N$ expressions and
is given by:
\be
a_i\,,\quad i=1,\dots,2N+4\,,\quad \prod\limits_{i=1}^{2N+4}a_i=1\,,
\quad u\,,v\,,w\,.
\label{an:parametrization}
\ee
These two parametrizations are related by the following map that we will often need in our calculations:

\be
\tilde{a}_l=\left(uv \right)^{-N-1}w^{-\frac{2}{N+3}}a_l\,,~l=1,\dots,2N+4\,,
\nn\\
\tilde{a}_{2N+6}=u^{2(N+1)(N+2)}w^{2\frac{N+2}{N+3}}\,,\quad \tilde{a}_{2N+5}=v^{2(N+1)(N+2)}w^{2\frac{N+2}{N+3}}\,.
\label{ancn:dict}
\ee

Now in order to derive finite difference operators we should proceed in the same way as in the case
of $A_N$ operators summarized in Section \ref{sec:an:oper}. We start by taking two $C_N$
trinion theories  $\cT^{\cA}_{x,y,z}$ where $\cA$ is the non-zero flux of the trinion and $x,y,z$ in the subscript are fugacities of
the global symmetries of two maximal and one minimal punctures correspondingly. Then we take an arbitrary $\cN=1$ theory
obtained in the compactification of the minimal $(D_{N+3},D_{N+3})$ conformal matter theory on the Riemmann surface of genus $g$
with $s$ punctures denoted as $\CC_{g,s}$ with at least one $\USp(2N)_{\tx}$ maximal puncture. Gluing all three surfaces together along the maximal punctures
results in a Riemmann surface of the same genus and global symmetry fluxes but two extra minimal punctures.
We will be performing all these operations at the level of superconformal indices. There gluing amounts to identifying
global symmetries of the punctures we glue and gauging it. So for the indices we can write down the following
identity after gluing:
\be
\cI\left[\CC_{g,s}(\tx)\oplus\bar{\cT}^{\cA}_{\tx,y,z_1}\oplus\cT^{\cA}_{y,x,z_2} \right]=\frac{\qPoc{q}{q}^4\qPoc{p}{p}^4}{2^4\cdot 2!\cdot 2!}\oint\prod\limits_{i=1}^2\frac{d\tx_i}{2\pi i \tx_i}\frac{dy_i}{2\pi i y_i}
\prod\limits_{i=1}^N\frac{1}{\GF{y_i^{\pm 2}}}
\times\nn\\
\prod\limits_{i<j}^N\frac{1}{\GF{y_i^{\pm 1}y_j^{\pm 1}}}
\prod\limits_{i=1}^N\frac{1}{\GF{\tx_i^{\pm 2}}}\prod\limits_{i<j}^N\frac{1}{\GF{\tx_i^{\pm 1}\tx_j^{\pm 1}}}
K_3^{C}(x,y,z_2)\bar{K}_3^{C}(\tx,y,z_1)\cI\left[\CC_{g,s}(\tx)\right]\,,
\label{trinions:gluing}
\ee

where $\cI\left[\CC_{g,s}\right]$ is the index of a theory obtained by compactifications of $6d$ theory on the Riemmann surface
$\CC_{g,s}$ and $K_3^{C}(x,y,z)$ is the index of the trinion theory shown in Figure \ref{pic:cn:trinion} while $\bar{K}_3^{C}(x,y,z)$
is its conjugate. Definition and
properties of the elliptic $\Gamma$-function $\GF{x}$ are given in Appendix
\ref{app:spec:func}. Then, closing $\SU(2)_z$ and $\SU(2)_{\tz}$ minimal punctures
with or without introduction of defects we can obtain finite difference operators.
As discussed previously in order to close minimal punctures we have to give
certain weights to corresponding global symmetry fuagcities $z_1$ and $z_2$. As a result we get the follwoing
identity
\be
\lim_{z_1\to Z_1^*,z_2\to Z_2^* }\cI\left[\CC_{g,s}(x)\oplus\bar{\cT}^{\cA}_{x,y,z_1}\oplus\cT^{\cA}_{y,\tx,z_2} \right]\sim\cO\cdot\cI\left[\CC_{g,s}(x)\right]\,,
\label{diff:op:def}
\ee
where $\cO$ is some finite difference operator, and in case we close punctures
without introducing defects we just obtain the identity operator.

Here for convenience we will not compute the full expression \eqref{trinions:gluing} directly right away.
Just as for $A_N$ operators derived in Section \ref{sec:an:oper} it appears to be much easier to
perform calculation in a slightly different way. First we derive
the index of the four-punctured sphere with zero flux, two maximal $\USp(2N)$ and two minimal
$\SU(2)$ punctures. For this purpose we perform $S$-gluing of two
trinions $\cT^{AC}_{x,y,z}$ with $\SU(N+1)$ and $\USp(2N)$ maximal punctures and $\SU(2)$ minimal punctures each.
This kind of trinions can be derived by appropriate gluing of
$A_NC_N$ tube theory shown in Figure \ref{pic:cn:tube}, to one of
the maximal punctures of $A_N$ trinion introduced in \cite{Razamat:2020bix}.
Superconformal index of the resulting four-punctured sphere theory is given by:
\be
K_4^C(x,\tx,z,\tz)=\kappa_N\oint
\prod\limits_{i=1}^N\frac{dy_i}{2\pi i y_i}
\prod\limits_{i\neq j}^{N+1}\frac{1}{\GF{\frac{y_i}{y_j}}}
\bar{K}_3^{AC}(\tx,y,\tz)K_3^{AC}(x,y,z)\,.
\label{4puncture:gluing}
\ee
where $K^{AC}_3(x,y,z)$ is the index of the trinion with $y$
being $\SU(N+1)$ fugacity and $\kappa_N$ is the usual constant given in \eqref{kappa:def}.
Geometrically this operation is shown in Figure \ref{pic:gluingmove1}.

Next we close one of the two minimal punctures by giving nonzero
vev to one of the $M_z$ moment maps given in \eqref{cn:moment:maps}.
At the level of the index computations this means that the corresponding
global symmetry fugacity should be consistent with the vev, i.e. in our case
we should fix $\tz$ fugacity to
\be
\tz=\tilde{Z}_{i;K,M}^*\equiv\pqm{1}{2}w^{2\frac{(N+2)^2}{N+3}}\ta_i
q^{-K}p^{-M}\,,
\label{tz:closing}
\ee
where $\ta_i$ can be chosen arbitrary since all the expressions are symmetric
w.r.t. permutations of $\ta_i$. Integers $K$ and $M$ correspond to an
order of the derivative of the moment map in $34$ and $12$ planes correspondingly,
i.e. we give vev to $\langle\partial_{12}^M\partial_{34}^K\widehat{M}_i \rangle\neq 0$.
Physically this corresponds to introducing various
codimension two defects into $4d$ theory. For the first $\SU(2)_{\tz}$
minimal puncture we choose to close it without introducing any defect,
which in turn corresponds to $K=M=0$ choice in \eqref{tz:closing}.

At this value of $\tz$, the superconformal index
\eqref{4puncture:gluing} of the four-punctured sphere has
a pole. Computing the residue at this pole we obtain
the index of the three-punctured sphere:
\be
K_{(3;i,0)}^C(x\,,y\,,z)\sim\mathrm{Res}_{\tz\to \tilde{Z}_{i;0,0 }^*}K_4^C(x\,,y\,,z\,,\tz)
\label{cn:new:trinion:def}
\ee
Here subscript $(3;i,0)$ refers to the fact that we obtain three-punctured
sphere by closing minimal puncture of the four-punctured sphere
choosing $\ta_i$ for the vev in \eqref{tz:closing} and not introducing a defect,
which corresponds to the choice $K=M=0$ in the same equation.

Calculation of \eqref{4puncture:gluing} and \eqref{cn:new:trinion:def}
results in the theory shown in Figure \ref{pic:cn:3pt:new}.
Detailed derivations of this section can be found in Appendix \ref{app:punctured:spheres}.

\begin{figure}[!btp]
          \centering
           \includegraphics[width=0.7\linewidth]{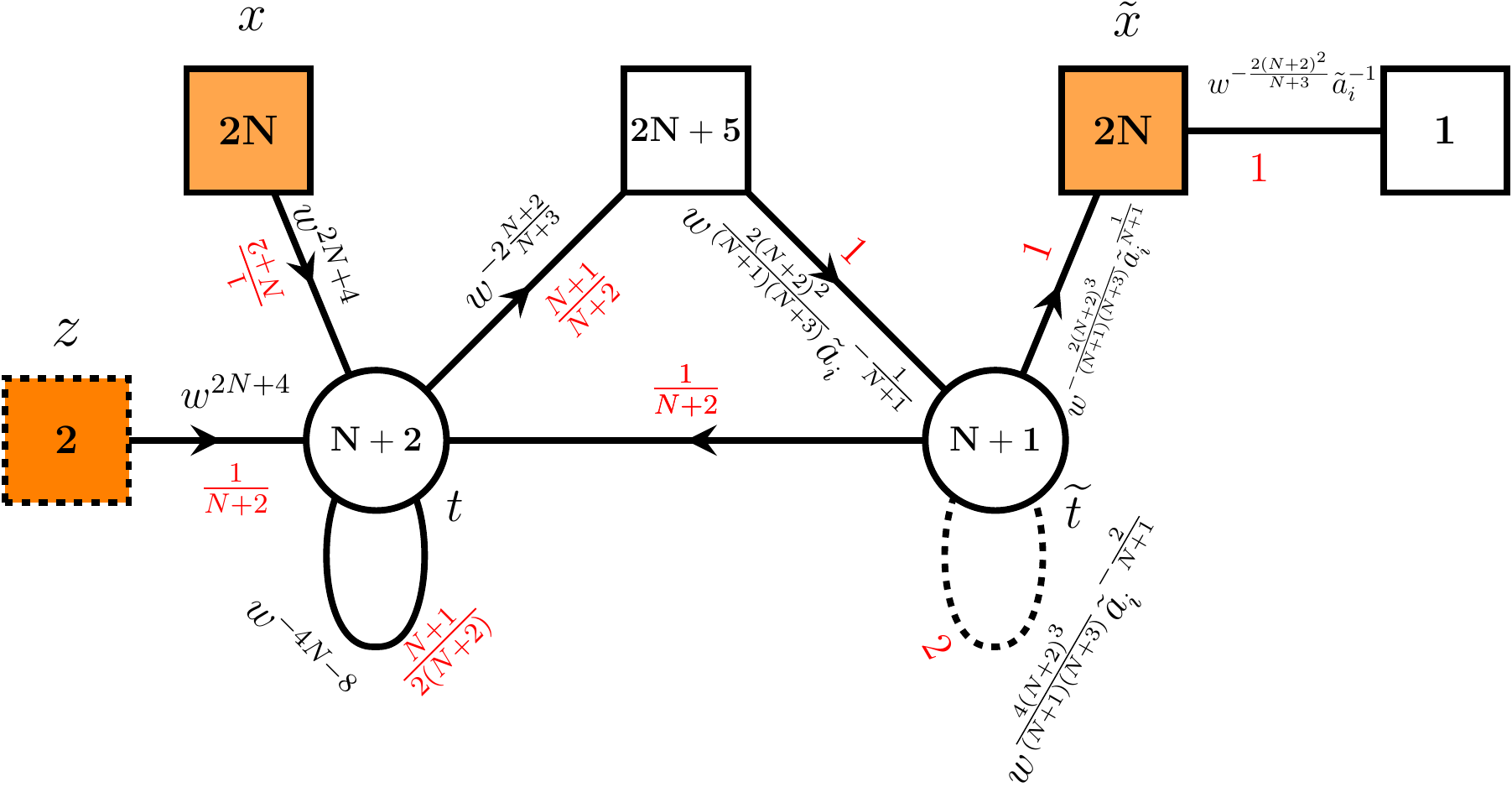}
           \caption{Three-punctured sphere theory with two maximal $\USp(2N)$ punctures and one minimal $\SU(2)$ puncture obtained
           after gluing \eqref{4puncture:gluing} and closing one minimal $\SU(2)_{\tilde{z}}$ puncture without defect. Solid and dashed
           line attached to the gauge node denotes multiplet in $AS$ and $\overline{AS}$ representation correspondingly.}
           \label{pic:cn:3pt:new}
\end{figure}

Now that we have obtained the desired three-punctured sphere theory we are ready to derive A$\Delta$O. For this purpose we should close the remaining
minimal $\SU(2)_z$ puncture by giving nontrivial vev $\langle\partial_{12}^M\partial_{34}^KM \rangle\neq 0$  to a derivative of one of its moment maps.
In order to get zero total flux through the resulting two-punctured sphere we should choose the same moment map as we did closing
$\SU(2)_{\tz}$ minimal puncture previously. The difference is that now we have to choose non-trivial derivative that is at least one
of $K$ and $M$ numbers is not zero. Physically this corresponds to introducing codimension two defect into the tube theory with two maximal punctures as
shown in Figure \ref{pic:gluingmove2}. At the level of the index
we should give the following value to the $z$-fugacity:
\be
z=Z_{i;K,M}^*\equiv\pqm{1}{2}w^{-2\frac{(N+2)^2}{N+3}}\ta_i^{-1}
q^{-K}p^{-M}\,,
\label{z:closing}
\ee
Then, according to \eqref{diff:op:def}, capturing the corresponding pole and gluing the resulting tube with two maximal $\USp(2N)$ punctures to an arbitrary theory with
at least one puncture of this type as shown in Figure \ref{pic:gluingmove3}, we obtain an A$\Delta$O.
However, technically it sometimes appears not to be as straightforward. In particular if we perform these
operations with the expression \eqref{3pt:sphere:4thway} instead of A$\Delta$O we obtain some integral-finite difference operator. It is highly possible
that in fact this operator can be written in the form of A$\Delta$O. However technical issues make it too difficult task and we leave it for future
investigation.

\begin{figure}[h!]
    \centering
       \includegraphics[width=0.4\linewidth]{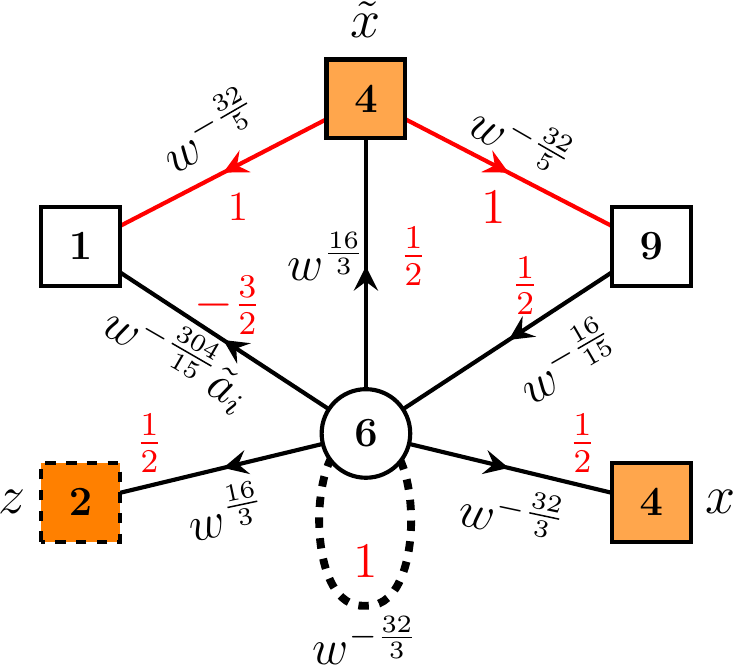}
       \caption{Three-punctured sphere theory with two maximal $\USp(4)$ and one minimal $\SU(2)$ punctures. Dashed line starting and ending on the gauge
       $\SU(6)$ node as previously corresponds to the matter in the $\overline{AS}$ representation.}
       \label{pic:c2:trinion}
\end{figure}

\begin{figure}[!btp]
          \centering
	  \subfigure[Gluing two trinions.]{\label{pic:gluingmove1}\includegraphics[width=0.7\linewidth]{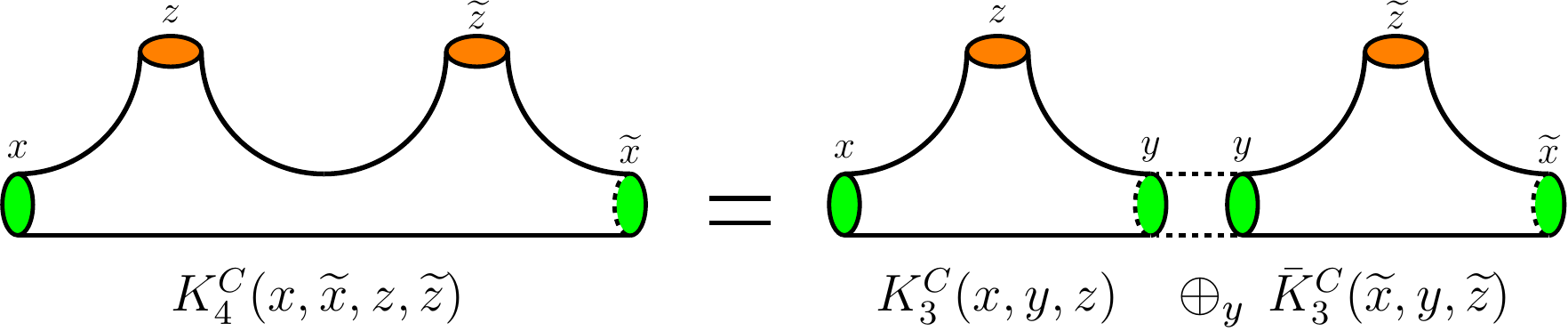}}
           \subfigure[Closing minimal punctures of the four-punctured sphere.]{\label{pic:gluingmove2}\includegraphics[width=0.7\linewidth]{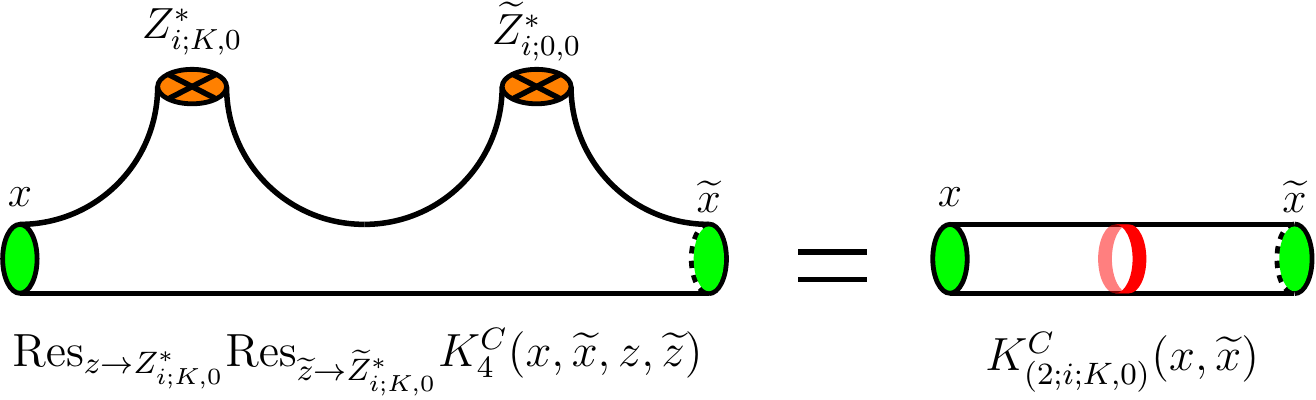}}
           \subfigure[Gluing tube with the defect to an arbitrary Riemann surface.]{\label{pic:gluingmove3}\includegraphics[width=0.85\linewidth]{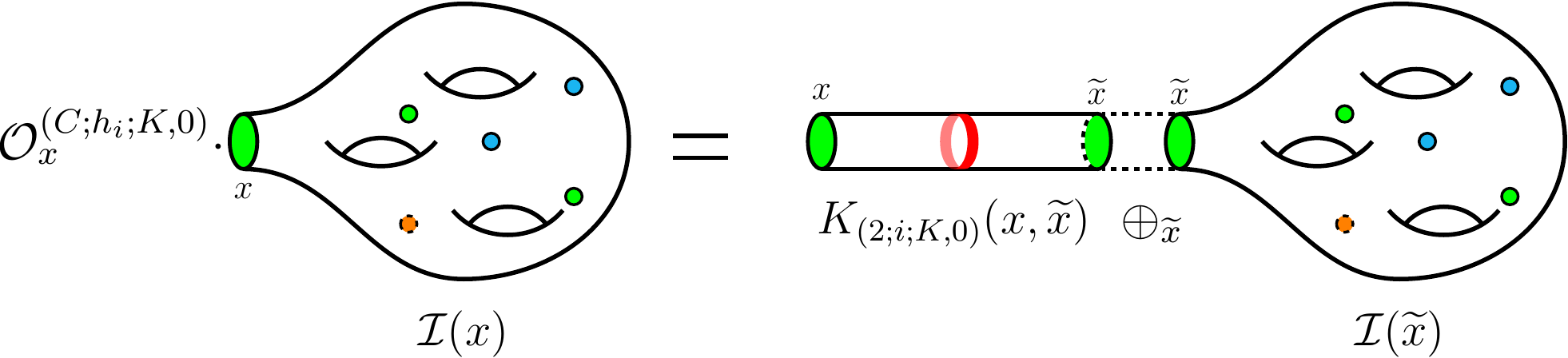}}
	   \caption{On the figures above we summarize all the steps we go through in order to derive $C_2$ A$\Delta$O given in \eqref{c2:operator:general:K}. First
		   as shown on the Figure (a) we glue two trinions with two maximal $\USp(4)$ (shown with \textit{green}) and one minimal $\SU(2)$ (shown with \textit{orange})
		   punctures each. For one of the trinions we conjugate all the
   charges in order to perform S gluing. At the level of the superconformal index this operation is expressed in \eqref{4puncture:gluing}. Next, as shown on the Figure (b),
   we close two minimal punctures of the four-punctured sphere. This operation is performed by giving vev $\langle\partial_+^K\widehat{M}_i\rangle\neq 0$
   to holomorphic derivative of one of the moment map operators with the $U(1)$ charge $h_i$. As the result we obtain tube theory with
   two maximal $\USp(4)$ punctures and codimension two defect introduced. On the Figure we denote this defect with the red ring. Finally, as the
   last step of our algorithm shown on the Figure (c), we glue this tube with the defect to an arbitrary surface with at least one maximal $\USp(4)$ puncture. This results
   in the action of the set of certain A$\Delta$Os on the index of the original theory.}
\end{figure}

Here we will instead concentrate on the derivation of next to the lowest rank $C_2$ operator. This case is simpler to analyze. For example, in our trinion theory
shown in Figure \ref{pic:cn:3pt:new}, when $N=2$ the $\overline{AS}$ multiplet of the right  $\SU(3)$ gauge node becomes just antifundamental, and a chain of duality transformations
can be used to simplify the theory. These calculations are summarized in Appendix \ref{app:punctured:spheres}. As a result we obtain single-node
$\SU(6)$ gauge theory shown in Figure \ref{pic:c2:trinion} with the corresponding index specified in \eqref{c2:3pt:sphere}.

Finally, we can close $\SU(2)_z$ puncture. In particular we give the following weight consistent with \eqref{tz:closing} to the fugacity $z$:
\be
z=Z^*_{i;K,0}=\pqm{1}{2}w^{-\frac{32}{5}}a_i^{-1}q^{-K}\,.
\label{z:closing:0}
\ee

Performing this closure we obtain the index of the theory for two punctured sphere with two $\USp(4)$ maximal punctures
\be
K_{(2;i;K,0)}^C(x\,,\tx)\sim\mathrm{Res}_{z\to Z_{i;K,0 }^*}K_{(3;i,0)}^C(x\,,\tx\,,z)
\label{c2:tube:def}
\ee
The index itself is specified in \eqref{tube:C2:general}. This time it
does not have natural gauge theory interpretation in case of general $K$ due to the presence of the codimension two defect.



As a final step of our derivation, we glue the obtained tube with the defect to an arbitrary $\NN=1$ theory with at least one maximal $\USp(4)$ puncture
and obtain A$\Delta$O as follows:
\be
\cO^{(C_2;h_{k};K,0)}_x\cdot\cI(x)\sim\frac{\qPoc{q}{q}^2\qPoc{p}{p}^2}{2^2\cdot 2!}\oint\frac{d\tx_{1,2}}{2\pi i\tx_{1,2}}\frac{1}{\GF{\tx_{1,2}^{\pm 2}}\GF{\tx_1^{\pm 1}\tx_2^{\pm 1}}}
\times\nn\\
K_{(2;i;K,0)}^C(x,\,\tx)\cI(\tx)
\label{c2:final:gluing}
\ee
Details of this calculation can be found in Appendix \ref{app:punctured:spheres}. It leads to the following expression for A$\Delta$O:
\begin{shaded}
\be
\cO^{(C_2;h_{k};K,0)}_x &= \sum\limits_{\vec{K}}\sum\limits_{m_1=-k_{1,+}}^{k_{1,-}}\sum\limits_{m_2=-k_{2,+}}^{k_{2,-}}\tilde{C}^k_{\vec{K}}\prod\limits_{l\neq k}^{10}
\prod\limits_{i=1}^2\prod\limits_{s_i=\pm 1}\prod\limits_{l_1=-k_{i,s_i}+s_im_i}^{2s_im_i-1}\thf{q^{l_1}x_i^{-2s_i}}^{-1}
\times\nn\\
&\prod\limits_{l_2=-k_{i,s_i}}^{s_i(k_{i,-}-k_{i,+})-1}\thf{q^{l_2}x_i^{-2s_i	}}^{-1}
\prod\limits_{l_3=-k_{2,s_2}-s_1m_1}^{s_2m_2-s_1m_1-1}\thf{q^{l_3}x_1^{s_1}x_2^{-s_2}}^{-1}
\times\nn\\
&\prod\limits_{j\neq i}^{2}\prod\limits_{l_4=-k_{j,s_j}}^{k_{i,s_i}- k_{j,s_j}-1}\thf{q^{l_4}x_i^{s_i}x_j^{-s_j}}^{-1}
\prod\limits_{l_5=-k_{1,s_1}-k_{2,s_2}}^{-k_{1,s_1}+s_2m_2-1}\thf{q^{l_5}x_1^{-s_1}x_2^{-s_2}}
\times\nn\\
&\prod\limits_{l_6=K-k_{i,s_i}-k_5}^{K-k_{i,s_i}-1}\thf{\pq{1}{2}h_kx_i^{-s_i}q^{l_6}}
\prod\limits_{l_7=K-k_5}^{K-k_5+k_{i,s_i}-1}\thf{\pq{1}{2}h_kx_i^{s_i}q^{l_7}}^{-1}
\times\nn\\
&\prod\limits_{l_8=-K-k_{i,s_i}}^{-K-k_{i,s_i}+k_5-1}\thf{\pqm{1}{2}h_k^{-1}x_i^{-s_i}q^{l_8}}
\prod\limits_{l_9=-k_{i,s_i}}^{-k_{i,s_i}+k_6-1}\thf{p^{-1}q^{l_9-1}h^{-\frac{1}{2}}h_kx_i^{-s_i} }
\times\nn\\
&\prod\limits_{l_{10}=-k_6}^{k_{i,s_i}-k_6-1}\thf{q^{-1-l_{10}}h^{-\frac{1}{2}}h_kx_i^{-s_i}}^{-1}
\prod\limits_{l_{11}=-k_{1,s_1}-k_{2,s_2}}^{-k_{1,s_1}+s_2m_2-1}\thf{q^{l_{11}}x_1^{-s_1}x_2^{-s_2} }
\times\nn\\
&\prod\limits_{l_{12}=-s_im_i}^{k_{i,s_i}-1 }\thf{\pq{1}{2}h_lx_i^{s_i}q^{l_{12}}}
\prod\limits_{l_{13}=k_6-s_im_i}^{k_{i,s_i}+k_6-1}\thf{h^{-\frac{1}{2}}h_kx_i^{s_i}q^{l_{13}}}
\times\nn\\
&\hspace{2cm}\prod\limits_{l_{14}=-s_im_i}^{K-k_5-s_im_i-1}\thf{\pq{1}{2}h_kx_i^{s_i}q^{l_{14}}}
\Delta_q^{-m_1}(x_1)\Delta_q^{-m_2}(x_2)\,,
\label{c2:operator:general:K}
\ee
\end{shaded}
where the constant $\tilde{C}^k_{\vec{K}}$ is given by
\begin{shaded}
\be
\tilde{C}^k_{\vec{K}}&=\prod\limits_{l\neq k}\prod\limits_{l_1=-K}^{k_5-K-1}\thf{h_lh_k^{-1}q^{l_1}}\prod\limits_{l_2=0}^{k_6-1}\thf{\pqm{1}{2}h^{-\frac{1}{2}}h_lh_kq^{l_2} }
\prod\limits_{l_3=-2K}^{-2K+k_5-1}\thf{h_k^{-2}q^{l_7}}
\times\nn\\
&\prod\limits_{l_4=K-k_5}^{K-k_5+k_6-1}\thf{\pqm{1}{2} h^{-\frac{1}{2}}h_k^2q^{l_5}}^{-1}\prod\limits_{l_4=K-k_6}^{K-1}\thf{\pq{3}{2}h^{\frac{1}{2}}q^{l_4}}
\times\nn\\
&\prod\limits_{l_6=-K-k_6}^{-K+k_5-k_6-1}\thf{\pq{1}{2}h^{\frac{1}{2}}h_k^{-2}q^{l_5}}
\prod\limits_{l_7=-K}^{-K+k_5+k_6-1}\thf{\pqm{1}{2}h^{-\frac{1}{2}}q^{l_6} }\,.
\label{c2:operator:const:gen:K}
\ee\end{shaded}

We have also introduced notation $h_i$, $i=1,\dots,10$ for  $U(1)$ charges of the moment maps we act on. Notice that
according to \eqref{cn:moment:maps} moment maps of both minimal and maximal punctures have the very same charges,
so that in fact
\be
h_i=w^{\frac{32}{5}}\ta_i\,,\qquad h\equiv \prod\limits_{j=1}^{10} h_j=w^{64}\,,
\label{c2:moment:maps}
\ee
for both charges of the maximal punctures we act on and minimal punctures we close. We have also introduced here an overall
$U(1)$ charge $h$. The operator itself is labelled by an index $k$ according to the choice of $U(1)$ charge $h_k$ of the moment map
we give vev to in order to close the minimal puncture and to introduce defects into the theory.

The first sum in the expression is performed over all possible partitions $\vec{K}=(k_1,\cdots,k_6)$ of the integer $K$.
Finally the products in \eqref{c2:operator:general:K} should be understood as ordered ones, i.e., one should assume  $\prod_{i=n_1}^{n_2}$
is just $1$ if $n_2<n_1$ since there are no terms in the product. The coefficient $\tilde{C}^k_{\vec{K}}$ is chosen so that the
$x$-independent factors of the  highest order shift terms $\Delta_q^{\pm K}(y_{1,2})$ are just one.

 As expected obtained operators  differ from the canonical $BC_n$ vD. 
In order to obtain the latter one we should instead consider compactifications of $6d$ 
rank-Q E string theory using the very same approach described in details on our paper. 
One of the most important difference from the $BC_n$ vD model  is the number of parameters it depends on.
Higher-rank (i.e. rank higher than one) vD model depends on $p$, $q$ and another $9$ parameters. 
This number of parameters is independent of the rank of the $BC$ type root 
system. Meanwhile our $C_2$ operators depend on $p$, $q$ and another $10$ 
parameters. Moreover in the present paper due to technical difficulties we failed 
to derive higher-rank $C_N$ generalizations. But from our construction we can expect 
them to depend on $2N+6$ parameters corresponding to $U(1)$ charges of the moment maps.\footnote{
Similar comments hold for the previously discussed $A_N$ operators \cite{Nazzal:2021tiu}, which depend on 
$p$, $q$ and another $2N+4$ parameters.}

Notice that the expression above is not the most general since we could have closed $\SU(2)_z$ minimal puncture using general $K$ and $M$
integers in \eqref{z:closing}, but we will concentrate here only on the case of non-trivial $K$ while putting $M=0$. As we see the operator appears to be
quite complicated and it is hard to analyze it. Instead it is useful to write down \textit{basic} operator corresponding to the choice of $K=1$,
i.e. the simplest non-trivial case. Taking into account six possible partitions $\vec{K}$ we can write down explicit form of A$\Delta$O:

\begin{shaded}
\be
\cO^{(C_2;h_k;1,0)}_x =\sum\limits_{i=1}^2\left(A^{(C_2;h_k;1,0)}_i(x,h)\Delta_q(x_i)+A^{(C_2;h_k;1,0)}_i(x_i\to x_i^{-1},h)\Delta_q^{-1}(x_i)\right)
\nn\\
\hspace{3cm}+W^{(C_2;h_k;1,0) }(x,h)\,,
\nn\\
A^{(C_2;h_k;1,0)}_i(x,h)=\prod\limits_{j\neq i}^2\frac{1}{\thf{x_i^2}\thf{qx_i^{2}}\thf{x_ix_j^{\pm 1}}}
\thf{\pq{1}{2}h_kx_j^{\pm 1}}
\prod\limits_{l=1}^{10}\thf{\pq{1}{2}h_l^{-1}x_i}\,,
\nn\\
W^{(C_2;h_k;1,0)}(x,h)=\left[\sum\limits_{i=1}^2\prod\limits_{j\neq i}^2\frac{1}{\thf{x_i^2}\thf{q^{-1}x_i^{-2}}\thf{x_ix_j^{\pm 1}}}
	\frac{\thf{h^{-\frac{1}{2}}h_kx_i}}{\thf{q^{-1}h^{-\frac{1}{2}}h_kx_i^{-1}}}
\times\right.\nn
\ee
\be
\left.
	\frac{\prod_{l\neq k}^{10}\thf{\pq{1}{2}h_lx_i}}{\thf{\pq{1}{2}h_kqx_i}}\prod\limits_{m=1}^2\thf{\pq{1}{2}h_kx_m^{\pm 1}}
	+\left(x_i\to x_i^{-1}\right)
\right]
+\nn\\
\prod\limits_{i=1}^2\prod\limits_{l\neq k}^{10}\frac{\thf{h_lh_k^{-1}q^{-1}}\thf{\pq{3}{2}h^{\frac{1}{2}}}}{\thf{q^{-2}h_k^{-2}}\thf{\pq{1}{2}q^{-1}h^{\frac{1}{2}}h_k^{-2}}}
\frac{\thf{\pq{1}{2}h_kx_i^{\pm 1}}}{\thf{\pqm{1}{2}h_k^{-1}q^{-1}  x_i^{\pm 1}}}
+\nn\\
\prod\limits_{i=1}^2\prod\limits_{l\neq k}^{10}\frac{\thf{\pqm{1}{2}h^{-\frac{1}{2}}h_kh_l}}{\thf{\pqm{1}{2}h^{-\frac{1}{2}}h_k^2q }}
\frac{\thf{\pq{1}{2}h_kx_i^{\pm 1}}}{\thf{p^{-1}q^{-1}h^{-\frac{1}{2}} h_k x_i^{\pm1}}}\,,
\label{c2:operator:final}
\ee
\end{shaded}

One thing to be noticed is that just as in the case of vD model, the constant term choice is not unique. What really defines this
constant term are the following properties. First of all it can be noticed that $W^{(h_k;1,0)}(x,h)$ is an elliptic function in both
$x_1$ and $x_2$ with periods $1$ and $p$. In the fundamental domain this elliptic function has poles located at:
\be
x_i=sq^{\pm \frac{1}{2}}\,,\qquad x_i=sq^{\pm \frac{1}{2}}p^{\frac{1}{2}}\,, \quad s=\pm 1\,,
\label{c2:poles}
\ee
which are in fact exactly the same as the poles of vD model. Corresponding residues are given by:
\begin{align}
&\mathrm{Res}_{x_i=sq^{\frac{1}{2}}}W^{(C_2;h_l;1,0)}(x,h)=-s\frac{q^{\frac{1}{2}}\prod\limits_{k=1}^{10}\thf{\pq{1}{2}sq^{-\frac{1}{2}}h_k}}
{2\qPoc{p}{p}^2\thf{q^{-1}}}\prod\limits_{j\neq i}^2 \frac{\thf{\pq{1}{2}h_lx_j^{\pm 1}}}{\thf{sq^{-\frac{1}{2}}x_j^{\pm 1} }}\,,
\nn\\
&\mathrm{Res}_{x_i=sq^{-\frac{1}{2}}}W^{(C_2;h_l;1,0)}(x,h)=s\frac{q^{-\frac{1}{2}}\prod\limits_{k=1}^{10}\thf{\pq{1}{2}sq^{-\frac{1}{2}}h_k}}
{2\qPoc{p}{p}^2\thf{q^{-1}}}\prod\limits_{j\neq i}^2 \frac{\thf{\pq{1}{2}h_lx_j^{\pm 1}}}{\thf{sq^{-\frac{1}{2}}x_j^{\pm 1} }}\,,
\nn\\
&\mathrm{Res}_{x_i=sp^{\frac{1}{2}}q^{\frac{1}{2}}}W^{(C_2;h_l;1,0)}(x,h)=-s\frac{ph^{-\frac{1}{2}}\prod\limits_{k=1}^{10}\thf{sh_k}}
{2\qPoc{p}{p}^2\thf{q^{-1}}}\prod\limits_{j\neq i}^2 \frac{\thf{\pq{1}{2}h_lx_j^{\pm 1}}}{\thf{\pqm{1}{2}sx_j^{\pm 1} }}\,,
\nn\\
&\mathrm{Res}_{x_i=sp^{\frac{1}{2}}q^{-\frac{1}{2}}}W^{(C_2;h_l;1,0)}(x,h)=s\frac{pq^{-1}h^{-\frac{1}{2}}\prod\limits_{k=1}^{10}\thf{sh_k}}
{2\qPoc{p}{p}^2\thf{q^{-1}}}\prod\limits_{j\neq i}^2 \frac{\thf{\pq{1}{2}h_lx_j^{\pm 1}}}{\thf{\pqm{1}{2}sx_j^{\pm 1} }}\,.
\label{c2:residues}
\end{align}

This concludes our derivation of the $C_2$ A$\Delta$O. All possible details of these derivations can be found in the Appendix \ref{app:punctured:spheres}.

\section{Properties of the $C_2$ operators.}
\label{sec:c2:properties}

There is a number of interesting properties that the operators we derived should posses by construction. In this section
we discuss checks and, where it is possible, give proofs of these properties using explicit expressions we have derived.

\begin{figure}[!btp]
          \centering
	  \includegraphics[width=\linewidth]{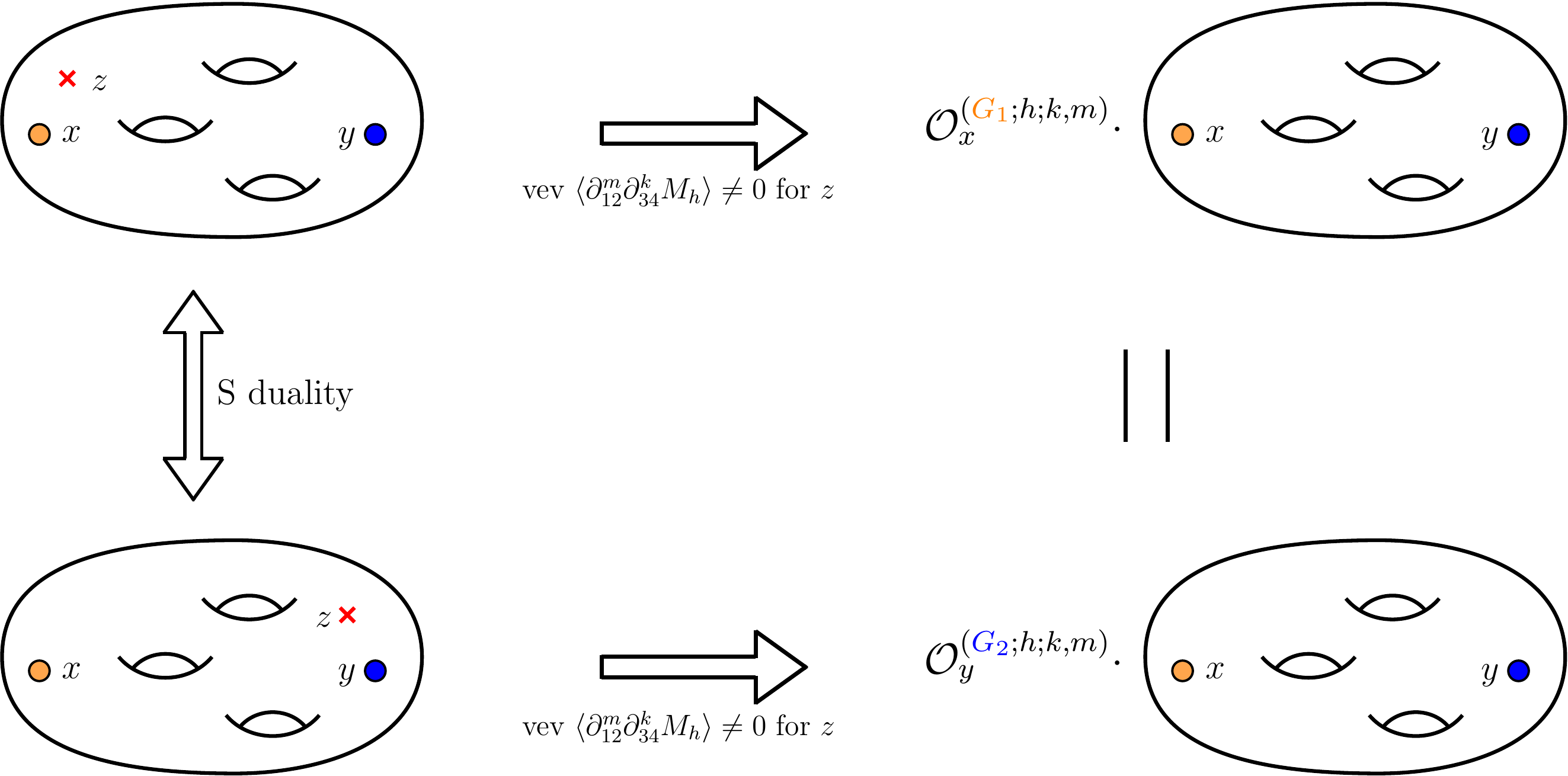}
	  \caption{On this Figure we represent the argument in favor of the kernel property of $\NN=1$ indices. For this we consider
	  an index of a theory obtained in the compactification of the minimal $(D_5,D_5)$ conformal matter theory on a generic Riemann
	  surface with at least two maximal punctures \textit{(shown as colored discs on the Figure)} of any types with fugacities $x$ and $y$  as well as one minimal
  $\SU(2)$ puncture \textit{(shown as red crosses)} with the fugacity $z$. For example in case of the $A_2C_2$ tube and kernel property \eqref{a2c2:kernel:property}
  one maximal puncture has $\USp(4)$ global symmetry while the second one has $\SU(3)$ symmetry. Then we close the minimal puncture by giving space dependent vev
  $\langle \partial_{12}^m\partial_{34}^k M \rangle\neq 0$ to one of the moment maps of this puncture. As discussed in the paper this corresponds to the introduction
  of the codimension-two defect into $4d$ effective description and results in the action of the A$\Delta$O on one of the maximal punctures of the index of
  the theory corresponding to the Riemann surface with only two maximal punctures. This operation can be performed in different duality frames. In each frame we
  obtain operators acting on one of the two punctures. Since the result of the calculation should not depend on the duality frame we choose we conclude that
  the action of two, possibly different operators, on two different punctures leads to two expressions equal to each other. Hence we arrive to the kernel property equation
  \eqref{kernel:property:definition}. }
  \label{pic:kernel}
\end{figure}

The most important but also most complicated property to discuss is the so-called \textit{kernel property} of our operators.
The main idea behind it is that the superconformal index of any $\NN=1$ theory obtained in the compactifications of $6d$
minimal $(D_5,D_5)$ conformal matter theory is  a \textit{kernel function} of our operators. The kernel function is defined by the following
mathematical identity:
\be
\cO_z^{(G_1;h_i;r,m)}\cdot\cI\left[\CC_{g,s}[z,u]\right]=\cO_u^{(G_2;h_i;r,m)}\cdot\cI\left[\CC_{g,s}[z,u]\right]\,.
\label{kernel:property:definition}
\ee
Physically we consider the superconformal index of the theory obtained in the $6d$ compactification on the Riemmann
surface $\CC_{g,s}[z,u]$ with at least two maximal punctures parametrized by the fugacities $z$ and $u$ correspondingly.
In general there can be as many punctures as we want. The claim here is that any such superconformal index
play the role of the \textit{kernel function} for the derived A$\Delta$Os according to \eqref{kernel:property:definition}.
Namely we can act with our operators on different punctures and we should always obtain the same result.
This property is expected to hold due to an argument coming from the geometry of compactification shown in Figure \ref{pic:kernel}. Equivalently
it can be understood from the invariance of the superconformal indices under $S$ duality transformations.

In principle we can choose any sphere with multiple punctures to check the kernel property.
Natural simplest candidate would be WZW model for the two-punctured sphere with two maximal $\USp(2N)$ punctures for which the index is given in \eqref{tube:C2:general} with $K=0$.
However, we will check a more interesting and
trickier case where two punctures are of different types. Namely we take the tube theory shown in Figure
\ref{pic:cn:tube} which has one $\SU(N+1)$ and one $\USp(2N)$ maximal puncture. Then for $N=2$ the kernel property \eqref{kernel:property:definition}
in this case reads:
\be
\cO^{(A_2;\tth_i^{-1};r,m)}_{y}K_2^{AC}(x,y)=\cO^{(C_2;h_i;r,m)}_{x}K_2^{AC}(x,y)\,,
\label{a2c2:kernel:property}
\ee
where $K_2^{AC}(x,y)$ is the index of the tube specified in \eqref{ancn:tube:index} with $x$ and $y$ being fugacities of
$\USp(4)$ and $\SU(3)$ punctures correspondingly. Operators $\cO^{(A_2;\tth_i;1,0)}_{y}$ and $\cO^{(C_2;h_i;1,0)}_{x}$ are given in
\eqref{sun:op:noflip:gen} and \eqref{c2:operator:final} correspondingly. Notice that in order for the kernel property to work we need
to close  minimal punctures in the same way on two sides of the equation, i.e. we have to give vev to the same moment map of the minimal puncture.
 In particular in $A_2$ case we should give vevs
to the minimal puncture moment map of the charge $\tth_i^{-1}\tth^{\frac{1}{4}}$ where $\tth_i$ are given in \eqref{a2:moment:maps}.
At the same time, in $C_2$ case minimal puncture is closed by giving vev to the moment map of charge $h_i$ specified in
\eqref{c2:moment:maps}. Using the map \eqref{ancn:dict} we can see that the only map that matches between the two is
\be
\tth_{10}^{-1}\tth^{\frac{1}{4}}=v^{-24}w^{-8}=w^{-\frac{32}{5}}\ta_{10}^{-1}=h_{10}\,.
\label{kernel:property:charge}
\ee
If we would like to check the kernel property for other ways of closing minimal punctures
we should consider $A_2$ operators \eqref{AN-HT} with the non-flipped moment maps so that in this case
we close the minimal puncture with the vevs of the moment map with charge $\tth_i\tth^{-\frac{1}{4}}$. Because of this
complication here we consider only closing with the vev given to the moment map of charge \eqref{kernel:property:charge}. The proofs for
other operators should work identically.
Also in our proof we restrict ourselves to only basic operators  \eqref{c2:operator:final} and \eqref{sun:op:noflip:gen}, i.e. we fix
$r=1\,,m=0$ (or equivalently $m=1\,,r=0$ on both sides of the kernel equation \eqref{a2c2:kernel:property}. We have to do it since
higher operators specified in \eqref{c2:operator:general:K} and \eqref{AN-HT} are too complicated for the analysis. However the kernel
property \eqref{a2c2:kernel:property} should also work for these higher operators as well.

Since we precisely know explicit expressions for both operators \eqref{sun:op:noflip:gen} and \eqref{c2:operator:final} as well as
supposed kernel function \eqref{ancn:tube:index} it is straightforward to check this kernel identity. Acting with finite difference
operators of two kinds on the kernel we get two algebraic functions. Equality of these functions implies validity of the
kernel property \eqref{a2c2:kernel:property}. In Appendix \ref{app:kernel} we give details of the analytic proof of this identity.

\begin{figure}[!btp]
          \centering
	  \includegraphics[width=\linewidth]{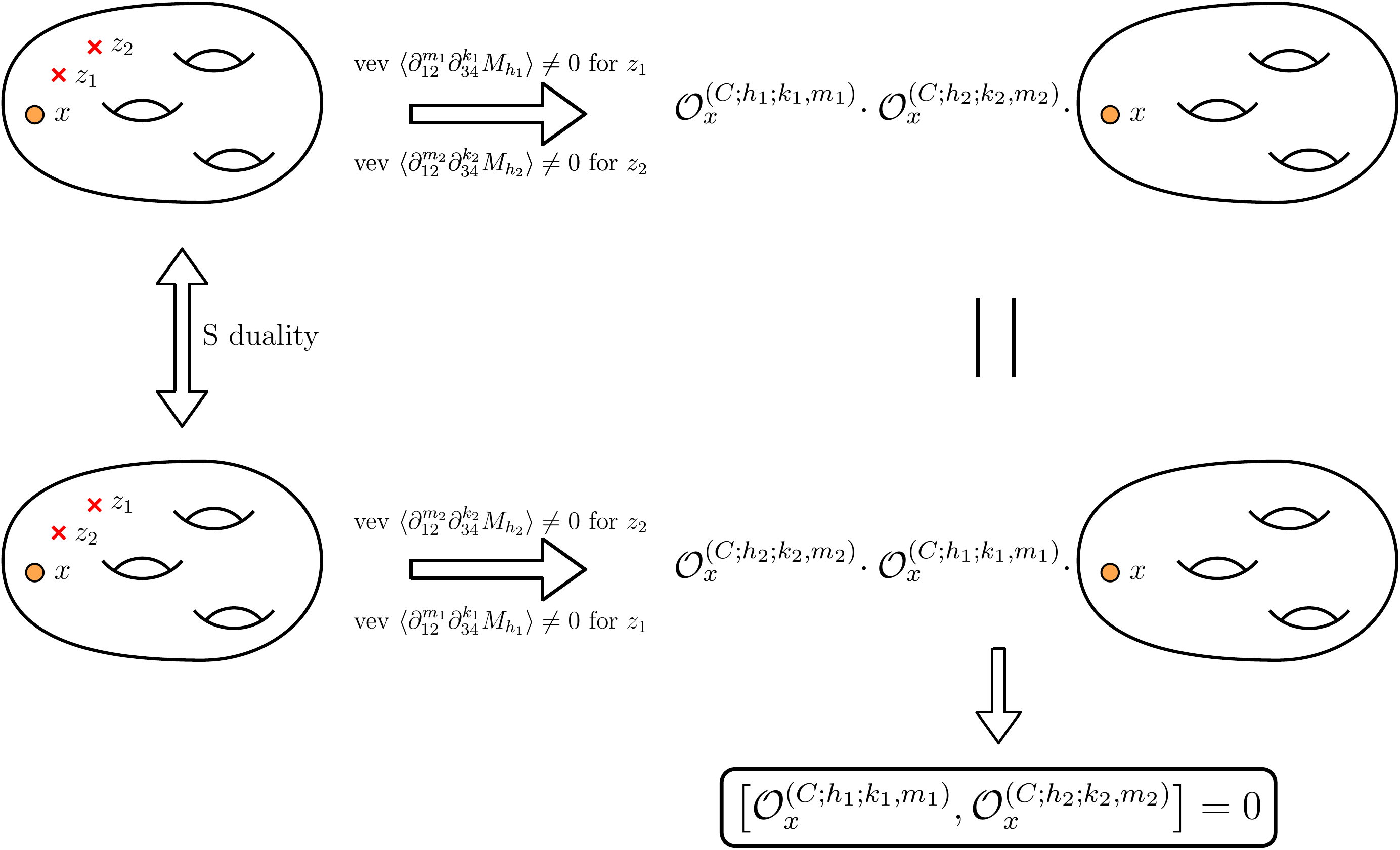}
           \caption{On this Figure we represent the argument in favor of commutation of A$\Delta$Os derived in the present paper. For this we start with a
		   theory obtained in compactification of $6d$ minimal $\left(D_5,D_5\right)$ conformal matter on an arbitrary Riemann surface. This surface
	   has one maximal $\USp(4)$ puncture \textit{(orange disc)} with fugacity $x$  and two minimal $\SU(2)$ punctures \textit{(red crosses)} with fugacities $z_1$ and
   $z_2$ correspondingly. Then we introduce codimension two defects into this theory giving space-dependent vevs to some of the moment maps $M_{h_1}$ and $M_{h_2}$ correspondingly.
   Giving this vev is equivalent to closing puncture and at the level of the index computations leads to an action of two different A$\Delta$Os.
   We can perform this operation in any duality frame and the result should not depend on the choice of a particular frame. Hence the action
   of operators on the index does not depend on their order. Since the compactification surface, and hence the effective $4d$ descriptions were chosen arbitrarily
   we conclude that the operators themselves also commute.}
   \label{pic:commutator}
\end{figure}

Another important property of the derived A$\Delta$Os is their commutation with each other which also directly
follows from the S duality of our compactification construction as shown in Figure \ref{pic:commutator}.
Since it does not depend which duality frame we close the minimal punctures in, it also does not matter in which order two
different A$\Delta$Os act on the index. Hence we conclude that all of operators we derived should commute with each other:
\be
\left[\cO_x^{(C_2;h_a;K_1,M_1)},\,\cO_x^{(C_2;h_b;K_2,M_2)}\right]=0\,,\quad \forall ~ ~a,b=1,\ldots,10;~ K_{1,2}\geq 0;~M_{1,2}\geq 0;
\label{c2:comm}
\ee

Here $h_a$ and $h_b$ are moment maps we give vev to in order to close two minimal punctures and $K_{1,2}, M_{1,2}$ are
numbers of holomorphic/antiholomorphic derivatives. Unfortunately it is very hard to prove or even to check these
relations for general operators \eqref{c2:operator:general:K} of the full tower. However we can perform checks of
the commutation relations \eqref{c2:comm} for the basic operators, i.e.
for the case when $K_{1,2}$ and $M_{1,2}$ are taking values $0$ and $1$. In particular we show that:
\be
&&\left[\cO_x^{(C_2;h_a;1,0)},\,\cO_x^{(C_2;h_b;1,0)}\right]=
\left[\cO_x^{(C_2;h_a;0,1)},\,\cO_x^{(C_2;h_b;0,1)}\right]=
\nn\\
&&\hspace{2cm}\left[\cO_x^{(C_2;h_a;1,0)},\,\cO_x^{(C_2;h_b;0,1)}\right]=
0\,,\qquad \forall ~ a,b =1,\dots,10\,.
\label{c2:basic:comm}
\ee
In Appendix \ref{app:commutators} we check these relations. In particular, using ellipticity properties of $\thf{x}$ function we
prove that the action of the third type of the commutators on an arbitrary trial function is zero. Hence these commutators are zero themselves.
For the first and second type of commutators it is hard to prove analytically that their action on a trial function is
zero. Instead we check these identities perturbatively in $p$ and $q$ expansion. These checks performed up to a sufficiently high order
suggest that both of the commutators are indeed zero.

\section{Discussion and outlook.}
\label{sec:discussion}

In this paper we have considered $4d$ description of the compactification of the minimal $(D_5,D_5)$ conformal matter theory
on a Riemann surface with $\USp(4)$ maximal punctures. Using this $4d$ description and introducing codimension-two defect in corresponding
theory we derive $C_2$ generalization of the van Diejen model. In particular we obtain an infinite set of analytic difference operators
acting on the maximal puncture of $C_2$ type. Different operators of our set correspond to different ways of closing minimal punctures
on the compactification surfaces. They  are organized in a decuplet of basic operators and an infinite tower of operators on
top of each of the basic ones.

The operators we obtain are supposed to satisfy interesting and important properties following
directly form the geometry of compactifications. First such property is the commutation of all operators. In our paper,
using combination of residue computations and perturbative expansions we show that at least all basic operators indeed commute with each other.
Second important property is that the superconformal index of any compactification of the minimal $(D_5,D_5)$ conformal matter theory on
a surface with several $\USp(4)$ punctures is the kernel function of our difference operators. As an example of such kernel function we considered
a tube theory corresponding to the compactification on a sphere with one $\USp(4)$ and one $\SU(3)$ puncture. In order to prove
the kernel property we should act on two punctures with two different operators: $A_2$ operator derived in \cite{Nazzal:2021tiu} from one side  and $C_2$ operator derived
in the present paper from the other. This fact makes corresponding kernel function very interesting for the study. In the paper we managed to prove
this kernel property fully analytically for the basic $A_2$ and $C_2$ operators.

Another result of our paper is derivation of the full tower of $A_N$ generalization of van Diejen operator orginating from the
compactification of the minimal $(D_{N+3},D_{N+3})$ conformal matter theory on a Riemann surface with $\SU(N+1)$ maximal punctures.
We have already derived basic operators of this kind in our previous paper \cite{Nazzal:2021tiu}. However the tower of operators
was missing and we filled this gap in the present paper.

The present paper is another brick in our program of establishing dictionary between compactifications of $6d$ SCFTs and integrable analytic
difference operators and studying this dictionary in details. There are plenty of ways this research can be continued in.
In particular there should be separate difference operator for each possible $5d$ compactification of $6d$ theory, or equivalently for
each possible puncture type from $4d$ prospective.
So far several results have been observed in this frame. First of all $BC_1$ van Diejen model was observed in E-string compactifications in
several ways \cite{Nazzal:2018brc,Nazzal:2021tiu}. Also previously unknown $A_2$ and $A_3$ operators were derived using compactifications of
the minimal conformal matter theories of types $\SU(3)$ and $\SO(8)$ \cite{Razamat:2018zel}. Finally we have considered compactifications
of the minimal $(D_{N+3},D_{N+3})$ conformal matter theories on a Riemann surfaces with $A_N$ and $C_2$ (in case of $(D_5,D_5)$ $6d$ theory) type punctures.
However there are plenty of examples of $6d$ SCFT compactificatoins that are known up to date
\cite{Razamat:2022gpm,Heckman:2015bfa,DelZotto:2014hpa,Bhardwaj:2015xxa,Hayashi:2015vhy,Hayashi:2015zka,Kim:2017toz,Razamat:2016dpl,Kim:2018lfo,Apruzzi:2019vpe}. Our main goal is to
extend our results to more examples of compactifications and ideally establish a dictionary between them and integrable models.
Such dictionary can from one point of view shed light on physics of $6d$, $5d$ and $4d$ supersymmetric gauge theories. From the other point of view
this research program can lead to important results in the field of integrable systems.

First of all as a continuation of the present paper it is natural to consider compactifications of the minimal $(D_{N+3},D_{N+3})$ conformal matter
on Riemann surfaces with other types of punctures. In first place these are of course punctures of $C_N$ type for general $N$. In the present work
we failed to derive explicit form of the corresponding A$\Delta$O  because of the technical difficulties but it is worth trying to overcome them.
 Second candidate for the study is $\left(A_1\right)^N$ puncture. Corresponding $4d$ description was derived in \cite{Razamat:2019ukg}. It would be interesting to derive explicit form
of the corresponding difference operators and study their properties. For example we immediately know examples of the kernel functions
of these operators even before deriving them explicitly.

Second candidate for the studies are non-minimal $(D_{N+3},D_{N+3})$ conformal matter theories, which are obtained as worldvolume theories of the
stack of $k$ M5 branes probing $D_{N+3}$ singularity \cite{Heckman:2015bfa}. In this case the known $5d$ gauge theory, and hence the puncture type
on Riemann surface, is a direct generalization of $\left(A_1 \right)^N$ case of minimal conformal matter compactification discussed above. It corresponds to the
$\left(A_{k-1}\right)^2\times \left(A_{2k-1}\right)^N\times\left(A_{k-1}\right)^2$ gauge theory in $5d$ and hence the puncture type with the same
global symmetry. Trinion theories with two maximal and one minimal punctures,
which are building blocks in the construction of A$\Delta$Os, are also known \cite{Sabag:2020elc}. So construction of the correspodning difference operators
should be straightforward though can happen to be technically complicated.

Finally one more thing to be made in this direction is the study of the rank Q E-string theory compactifictaions along the same lines of research.
Unfortunately so far corresponding trinion theories were not obtained but the tube theories were already derived in \cite{Hwang:2021xyw,Pasquetti:2019hxf}.
Despite we can not derive corresponding A$\Delta$O using only tube theories, in this case we can naturally conjecture it to be $BC_n$ van Diejen model.
We can at least test this conjecture by checking corresponding kernel property.

Another possible direction of the research is related to the study of the properties of the operators we derived. Most interesting questions here are
related to the eigenfunctions of these operators. Deriving full eigenfunctions is of course difficult and most probably impossible. Realistically
we can try to derive certain limits of these eigenfunction. In case we succeed these eigenfunctions can play the same role for
indices of $\NN=1$ theories as Macdonald and Schur polynomials played for the indices of $\NN=2$ theories \cite{Gadde:2011uv,Gadde:2009kb,Gadde:2011ik}.

It would also be interesting to establish connections of our research program with other integrable models emerging in supersymmetric gauge theories.
For example integrable systems derived using compactifications of the worldvolume theory of $k$ M5 branes probing $A_{N-1}$ singularity were shown
to be related to the set of transfer matrices \cite{Gaiotto:2015usa,Maruyoshi:2016caf,Ito:2016fpl,Yagi:2017hmj}. It would be interesting to understand if
similar connection emerges in case of $D_{N+3}$ singularity.

Also, a relation of E-string theory to the van Diejen model was observed in \cite{Chen:2021ivd} using quantization of the corresponding
Seiberg-Witten curve. Later, authors of this paper generalized their result to $6d$ $(1,0)$ SCFTs with $\SO(N)$ gauge group and $(N-8)$ fundamental
flavors obtaining yet another set of difference operators \cite{Chen:2021rek}. It would be interesting to clarify precise relation of our
construction with the methods and results of these papers.

\section*{Acknowledgments}

We would like to especially thank Shlomo Razamat for numerous discussion, collaboration on different stages of this projects and
for reading the draft of the paper. AN is also thankful for the hospitality of the Theoretical Physics Division of Uppsala University
where parts of this work were done. The work of BN is supported in part by Israel Science Foundation under grant no. 2289/18,
by a Grant No. I-1515-303./2019 from the GIF, the German-Israeli Foundation for Scientific Research and Development,  by BSF grant no. 2018204.
Research of AN is supported by the Swiss National Science Foundation, Grant No. 185723.

\section*{Conflict of interest statement}

On behalf of all authors, the corresponding author states that there is no conflict of interest.

\section*{Data availability statement}

Data sharing not applicable to this article as no datasets were generated or analysed during the current study.

\appendix

\section{Special functions}
\label{app:spec:func}

Here we summarize some definitions and properties of special functions used in the paper.

Elliptic Gamma function is defined through the following infinite product:
\be\label{gamma:def}
\GF{z}\equiv \prod\limits_{k,m=0}^\infty\frac{1-p^{k+1}q^{m+1}/z}{1-p^kq^m z}\,.
\ee
It can be easily seen that the poles of this function are located at the following values of the argument:
\be\label{gamma:poles}
z=p^{-k}q^{-m}\,,\qquad k,m\in\mathbb{Z}_{\geq 0}\,.
\ee
The following relation will be useful in our calculations:
\be\label{gamma:rels}
\GF{\frac{pq}{z}}\GF{z}=1\,.
\ee
Also we will often deal with the elliptic beta integral formula
\be
\kappa\oint\frac{dz}{4\pi i z}\frac{1}{\GF{z^{\pm2}}}\prod\limits_{j=1}^6\GF{t_iz^{\pm 1}}=\prod\limits_{i<j}\GF{t_it_j}\,.
\label{elliptic:beta}
\ee
Here $\kappa$ is defined to be
\be\kappa=(q;q)_{\infty}(p;p)_{\infty}=\prod_{\ell=0}^\infty (1-q^{1+\ell})(1-p^{1+\ell}).\ee
$A_N$ generalization of this formula is
\be
&&\frac{\kappa^N}{N!}\oint\prod\limits_{i=1}^N\frac{dz_i}{2\pi i z_i}\prod\limits_{i\neq j}^{N+1}\GF{\frac{z_i}{z_j}}^{-1}\prod\limits_{i=1}^{N+2}\prod\limits_{j=1}^{N+1}
\GF{s_iz_j}\GF{t_iz_j^{-1}}=\nn\\
&&\prod\limits_{i=1}^{N+2}\GF{Ss_i^{-1}}\GF{Tt_i^{-1}}\prod\limits_{i,j=1}^{N+2}\GF{s_it_j}\,,\qquad\quad \left(T=\prod\limits_{i=1}^{N+2}t_i\,,~S=\prod\limits_{i=1}^{N+2}s_i\right)\,.
\label{elliptic:beta:An}
\ee
The Theta function is defined as follows:
\be
\thf{x}\equiv\qPoc{x}{p}\qPoc{x^{-1}p}{p}\,,
\label{theta:def}
\ee
where $\qPoc{z}{p}$ is the usual q-Pochhammer symbol defined as follows:
\be
\qPoc{x}{p}=\prod\limits_{k=0}^\infty\left( 1-xp^k \right)\,.
\label{qpoc:def}
\ee
Following properties of theta function will be useful to us
\be
&&\thf{x}=\frac{\GF{qx}}{\GF{x}}\,,\quad \thf{x^{-1}}=-x^{-1}\thf{x}\,,\quad \thf{xp^m}=(-1)^mx^{-m}p^{-\frac{1}{2}m(m-1)}\thf{x}\,,
\nn\\
&&\hspace{4cm}\frac{ \GF{ p^L q^K x } } { \GF{x} } = \prod_{j=0}^{K-1} \thp{ q^j x} \prod_{j=0}^{L-1} \thq{ q^K p^j x}
\label{theta:properties}
\ee
We will also use the following duality identity from \cite{Spiridonov:2008}:
\be
V(\underline{t})=\prod\limits_{1\leq j<k\leq 4}^8\GF{t_jt_k}\GF{t_{j+4}t_{k+4}}V(\underline{s})\,,
\label{spiridonov:id}
\ee
where
\be
V(\underline{t})\equiv \kappa\oint\frac{dz}{2\pi iz}\frac{\prod\limits_{j=1}^8\GF{t_jz^{\pm 1}}}{\GF{z^{\pm 2} }}\,,
\qquad \prod\limits_{j=1}^8t_i=pq\,,\quad |t_j|,|s_j|<1\,.
\nn\\
s_j=\rho^{-1}t_j\,,\quad j=1,2,3,4;\quad s_j=\rho t_j\,,\quad j=5,6,7,8;\quad \rho\equiv \sqrt{\frac{t_1t_2t_3t_4}{pq}}
\ee

\section{Derivation of higher tower $A_N$ operators }
\label{app:an:operator}

In this appendix we will give details of the derivation of $A_N$ A$\Delta$Os
that we summarize in Section \ref{sec:an:oper}.
We start with the four-punctured sphere theory whose quiver is shown in Figure \ref{pic:an:trinion}.
This theory and its superconformal index were derived in \cite{Nazzal:2021tiu}. The index is
given by the following expression:
\be
K_4(x,\tx,z,\tz)=\kappa_{N+2}\oint\prod\limits_{i=1}^{N+2}\frac{dt_i}{2\pi i t_i}
\prod\limits_{i\neq j}^{N+3}\frac{1}{\GF{\frac{t_i}{t_j}}}\times
\nn\\
\prod\limits_{i=1}^{N+3}\prod\limits_{j=1}^{N+1}\prod\limits_{k=1}^{2N+4}
\GF{\pq{1}{N+3}\left( u^{-1}w \right)^{4\frac{N+2}{N+3}}x_j^{-1}t_i}
\GF{\pq{1}{N+3}\left( uw^{-1} \right)^{2\frac{(N+2)(N+1)}{N+3}}z^{\pm 1}t_i}
\nn\\
\GF{\pq{N+1}{2(N+3)}u^{-\frac{(N+1)^2}{N+3}}v^{N+1}w^{-2\frac{N+1}{N+3}}a_k^{-1}t_i^{-1}}\GF{\pq{1}{2}u^{N+3}v^{-N-1}w^{-2}a_kx_j}
\times\nn\\
\GF{\pq{N+1}{2(N+3)}u^{-2\frac{(N+1)(N+2)}{N+3}}v^{-2(N+1)(N+2)}w^{-4\frac{N+2}{N+3}}t_i^{-1}}
\times\nn\\
\GF{\pq{1}{2}u^{2N+4}v^{2(N+1)(N+2)}x_j}\GF{\pq{1}{2}\left( uv w^{-2}\right)^{-N-1}a_lz^{\pm 1} }
\times\nn\\
\GF{\pq{1}{2}v^{2(N+1)(N+2)}w^{2N+4}z^{\pm 1}}\GF{\pq{N+1}{2(N+3)}u^{2\frac{(N+1)(N+2)^2}{N+3}}w^{4\frac{(N+2)^2}{N+3}}t_i^{-1}}
\times\nn\\
\GF{\pq{1}{2}u^{-2N(N+2)}w^{-4N-8}\tx_j}\GF{\pq{1}{2}u^{-2(N+1)(N+2)}w^{-2N-4}\tz^{\pm 1}}
\times\nn\\
\GF{\pq{1}{N+3} \left( u^{-1}w \right)^{4\frac{N+2}{N+3}}t_i\tx_j^{-1}}\GF{\pq{1}{N+3}\left( uw^{-1} \right)^{2\frac{(N+1)(N+2)}{N+3}}t_i\tz^{\pm 1}}\,.
\label{sun:4pt:sphere}
\ee
This is an index of an $\SU(N+3)$ SQCD with $2N+6$ flavors and a superpotential. Variables $x,\tx,z,\tz$ are
fugacities of the global symmetry of two maximal $\SU(N+1)_{x,\tx}$ and two minimal $\SU(2)_{z,\tz}$ punctures
correspondingly.
\\
We close the two $\SU(2)$ punctures by setting $z$ and $\tz$ to the values in \eqref{an:pole}.
Performing calculations for general values of the charges is complicated. So we will perform here the calculation
only for one of the charges fixing $\tth_i=\tth_{2N+6}=\left(u^Nw^2\right)^{-2N-4}$. Corresponding positions of the poles are then given by
\be \label{ztz}
z=(pq)^{-\frac12} \left( w u^{N+1} \right)^{-2N-4} q^{-M} p^{-L} ,
\;\;\;\;\;\;\; \tilde{z}=(pq)^{-\frac12} \left( w u^{N+1} \right)^{2N+4} q^{-\tilde{M}} p^{-\tilde{L}}\,.
\ee
Also as discussed in Section \ref{sec:an:oper} we will further consider $\tilde{L}=\tilde{M}=0$.

The pole in $\tz$ is an explicit simple pole so we can just take the residue. The pole in $z$ however, is due to contour pinching. To see this clearly, it is useful to first perform a Seiberg duality leading to an index of $SU(N+2)$ gauge thoery with $2N+5$ flavors, with the followng index,
\be
&&K_{(2;2N+6;L,M)}^{A_N}\left(x,\tx \right)=\kappa_{N+1}\oint \prod\limits_{i=1}^{N+1}\frac{dt_i}{2\pi it_i}
\prod\limits_{i\neq j}^{N+2}\frac{1}{\GF{\frac{t_i}{t_j}}}
\times\nn\\
&&\prod\limits_{i=1}^{N+2}\prod\limits_{j=1}^{N+1}\prod\limits_{k=1}^{2N+4}\GF{\pq{1}{2(N+2)}u^{2N+4}x_jt_i}
\GF{\pq{1}{2}u^{-N-3}v^{N+1}w^2a_k^{-1}\tx_j^{-1}}
\times\nn\\
&&\GF{\pq{1}{2}\left(uv^{N+1}\right)^{-2N-4}\tx^{- 1}}
\GF{\pq{1}{2(N+2)}u^{2N+4}\tx_jt_i}
\times\nn\\
&&\GF{\pq{N+3}{2N+4} u^{(2N+4)(N+1)} w^{4N+8}  t_i  q^M p^L}
\GF{(pq)^{-\frac{N+1}{2N+4}} u^{-(2N+4)(N+1)}   t_i  q^{-M} p^{-L} }
\times\nn\\
&&\GF{\pq{N+3}{2(N+2)}u^{-2(N+1)(N+2)}t_i}\GF{\pq{N+1}{2(N+2)}\left(uv\right)^{-N-1}w^{-2}a_kt_i^{-1}}
\times\nn\\
&&\GF{\pq{N+1}{2(N+2)}v^{2(N+1)(N+2)}t_i^{-1}}\GF{\pq{1}{2}\left( u^Nw^2 \right)^{-2N-4}\tx_j}
\nn\\
\ee
where we have already dropped irrelevant overall factors and substituted values of $z$ from \eqref{ztz}.
The subscript of the index as usually contains all the information of the underlying theory. In particular
$2$ stands for the number of punctures. Index $2N+6$ stands for the index of the moment map we give vev to in order to close
the puncture (moment map of charge $\tth_{2N+6}\tth^{-\frac{1}{4}}$ in this case). Finally $(L,M)$ denotes derivative
powers in the vev and are the same as $L$ and $M$ integers in \eqref{ztz}.
As can be seen from the expression above the contour pinching is due to collision
of poles at the following values:
\be
&&t_i=\pqm{1}{2(N+2)}u^{-2(N+2)}x_i^{-1}q^{-m_i}p^{-l_i}\,,\quad t_{N+2}=\pq{N+1}{2(N+2)}u^{2(N+1)(N+2)}q^{M-m_{N+2}}p^{L-l_{N+2}} \,,
\nn\\
&&\mathrm{or}
\nn\\
&&t_i=\pqm{1}{2(N+2)}u^{-2(N+2)}\tx_i^{-1}q^{-m_i}p^{-l_i}\,,\quad t_{N+2}=\pq{N+1}{2(N+2)}u^{2(N+1)(N+2)}q^{M-m_{N+2}}p^{L-l_{N+2}}\,,
\nn\\
\label{an:pinching:1}
\ee
where $m_i$ and $l_i$ are partitions of $M$ and $L$ respectively. There are $(N+2)!$ such poles coming from permutations of $t_i$ but they all give
the same result and we are ignoring overall factors. It can be checked that the two lines in \eqref{an:pinching:1} give the same result so we will
consider only the first set of poles. After computing the residue we get the following expression for particular partitions $\vec l,\vec m$,
\be
&&K_{\left(2;2N+6;\vec{m},\vec{l}\right)}^{A_N}(x,\tx)=\GF{pq}
 \prod_{j=1}^{N+2}
\frac{
\prod_{n=1}^{M-m_{N+2}} \thp{ q^{-n} } \prod_{n=1}^{L-l_{N+2} } \thq{ q^{m_{N+2}-M} p^{-n} }
}
{
\prod_{n=1}^{m_j} \thp{ q^{-n} p^{-l_j} } \prod_{n=1}^{l_j } \thq{   p^{-n} }
}
\nn\\
&&\prod\limits_{i=1}^{N+1}  \prod\limits_{k=1}^{2N+4}\prod\limits_{j\neq i}^{N+1}
\GF{\frac{\tx_i}{x_i}q^{-m_i} p^{-l_i} }
\frac{\GF{\pq{1}{2}u^{2(N+2)^2}q^{M-m_{N+2}}p^{L-l_{N+2}} x_i}}{\GF{\pq{1}{2}u^{2(N+2)^2}q^{M+m_i-m_{N+2}}p^{L+l_i-l_{N+2}}x_i}}
\times\nn\\
&&\frac{\GF{\pqm{1}{2}u^{-2(N+2)^2}x_i^{-1}q^{-M-m_i} p^{-L-l_i} }}{\GF{\pqm{1}{2}u^{-2(N+2)^2}x_i^{-1}q^{m_{N+2}-M-m_i} p^{l_{N+2}-L-l_i} }}
\GF{\frac{\tx_j}{x_i}q^{-m_i} p^{-l_i} }
\times\nn\\
&&\GF{\pq{1}{2}u^{N+3}v^{-N-1}w^{-2}a_kx_iq^{m_i} p^{l_i} }\GF{\pq{1}{2}\left(u^Nw^2\right)^{2N+4}x_i^{-1}q^{M-m_i} p^{L-l_i} }
\times\nn\\
&&\GF{\pq{1}{2}u^{-2(N+2)^2}x_i^{-1}q^{-m_i} p^{-l_i} }\GF{\pq{1}{2}\left(uv^{N+1}\right)^{2N+4}x_iq^{m_i} p^{l_i} }
\times\nn\\
&&\GF{\pq{1}{2}u^{-N-3}v^{N+1}w^2a_k^{-1}\tx_i^{-1}}
\GF{q^{m_{N+2}-M}p^{l_{N+2}-L} \left(vu^{-1}\right)^{2(N+1)(N+2)}}
\times\nn\\
&&\frac{\GF{\frac{x_j}{x_i}q^{-m_i} p^{-l_i} }}{\GF{\frac{x_j}{x_i}q^{m_j-m_i} p^{l_j-l_i} }}
\GF{\pq{1}{2}u^{2(N+2)^2}q^{M-m_{N+2}} p^{L-l_{N+2}} \tx_i}
\times\nn\\
&&\GF{\pq{1}{2}\left(u^Nw^2\right)^{-2N-4}\tx_i}\GF{pq \left(u^{N+1}w\right)^{4N+8} q^{2M-m_{N+2}} p^{2L-l_{N+2}}  }
\times\nn\\
&&\GF{\pq{1}{2}\left(uv^{N+1}\right)^{-2N-4}\tx_i^{-1}}
\GF{\left(u^{2N+5}v\right)^{-N-1}w^{-2}a_kq^{m_{N+2}-M} p^{l_{N+2}-L} }\,.
\label{an:tube4pt:k}
\ee
To get the $A\Delta O$ acting on $\SU(N+1)_x$ puncture, we should S glue this tube to the index
of an arbitrary theory $\cI(\tx)$ with $SU(N+1)_{\tx}$ puncture and sum over all partitions,
\be
\cI _1(x)=\kappa_{N}\sum_{\{m_i\},\{l_i\}}\oint\prod\limits_{i=1}^{N}\frac{d\tx_i}{2\pi i \tx_i}\prod\limits_{i\neq j}^{N+1}\frac{1}{\GF{\frac{\tx_i}{\tx_j}}}
K_{\left(2;2N+6;\vec{m},\vec{l}\right)}^{A_N}(\tx,x)\cI_0(\tx)\,,
\label{an:gluing:bos1}
\ee
Note that there is a zero in \eqref{an:tube4pt:k} coming from $\GF{pq}$ but this is canceled against the pinching of $\tx$ at the following values,
\be
\tx_i=x_{i}q^{m_i-s_i}p^{l_i-r_i} \,,\quad \sum\limits_{i=1}^{N+1}s_i=M-m_{N+2}\,,\quad \sum\limits_{i=1}^{N+1}r_i=L-l_{N+2},
\ee
up to permutations of $x_i$ which give the same result due to the Weyl symmetry of $A_N$ root system.
If we specify $\tx_i$ to the values written above we obtain double pole due to the contour pinching
of \eqref{an:gluing:bos1} but as we mentioned one of them is cancelled by the zero of $\GF{pq}$.
Computing the residue of the remaining pole we obtain the following
contribution of each individual partition:
\be
&&\cI^{\boldsymbol{m,l,s,r}}_1(x)=
\prod_{i=1}^{N+1} \prod_{n=1}^{N+2}\prod\limits_{k=1}^{2N+4}\prod\limits_{j\neq i}^{N+1}
\frac{\GF{\pq{1}{2}u^{2(N+2)^2}q^{M-m_{N+2}}p^{L-l_{N+2}} x_i}}{\GF{\pq{1}{2}u^{2(N+2)^2}q^{M+m_i-m_{N+2}}p^{L+l_i-l_{N+2}}x_i}}
\times
\nn\\
&&
\frac{
\prod_{n=1}^{M-m_{N+2}} \thp{ q^{-n} } \prod_{n=1}^{L-l_{N+2} } \thq{ q^{m_{N+2}-M} p^{-n} }
}
{
\prod_{n=1}^{m_j} \thp{ q^{-n} p^{-l_j} } \prod_{n=1}^{l_j } \thq{   p^{-n} }
\prod_{n=1}^{s_i} \thp{ q^{-n} p^{-r_i} } \prod_{n=1}^{r_i } \thq{   p^{-n} }
}
\times\nn\\
&&\frac{\GF{\pqm{1}{2}u^{-2(N+2)^2}x_i^{-1}q^{-M-m_i} p^{-L-l_i} }}{\GF{\pqm{1}{2}u^{-2(N+2)^2}x_i^{-1}q^{m_{N+2}-M-m_i} p^{l_{N+2}-L-l_i} }}
\GF{\frac{x_j}{x_i}q^{m_j-s_j-m_i} p^{l_j-r_j-l_i} }
\times\nn\\
&&\GF{\pq{1}{2}u^{N+3}v^{-N-1}w^{-2}a_kx_iq^{m_i} p^{l_i} }\GF{\pq{1}{2}\left(u^Nw^2\right)^{2N+4}x_i^{-1}q^{M-m_i} p^{L-l_i} }
\times\nn\\
&&\GF{\pq{1}{2}u^{-2(N+2)^2}x_i^{-1}q^{-m_i} p^{-l_i} }
\GF{\pq{1}{2}u^{-N-3}v^{N+1}w^2a_k^{-1}x_i^{-1} q^{s_i-m_i}p^{r_i-l_i} }
\times\nn\\
&&\GF{\pq{1}{2}\left(uv^{N+1}\right)^{-2N-4}x_i^{-1} q^{s_i-m_i}p^{r_i-l_i}}
\frac{\GF{\frac{x_j}{x_i}q^{-m_i} p^{-l_i} }   }{\GF{\frac{x_j}{x_i}q^{m_j-m_i} p^{l_j-l_i} }}
\times\nn\\
&&
\frac{\GF{\pq{1}{2}\left(u^Nw^2\right)^{-2N-4}x_i q^{m_i-s_i}p^{l_i-r_i} }}{ \GF{ q^{m_i-m_j+s_j-s_i}p^{l_i-l_j+r_j-r_i} x_i/x_j } }
\GF{\pq{1}{2}\left(uv^{N+1}\right)^{2N+4}x_iq^{m_i} p^{l_i} }
\times\nn\\
&&\GF{pq \left(u^{N+1}w\right)^{4N+8} q^{2M-m_{N+2}} p^{2L-l_{N+2}}  }
\times\nn\\&&
\GF{q^{m_{N+2}-M}p^{l_{N+2}-L} \left(vu^{-1}\right)^{2(N+1)(N+2)}}
\times\nn\\
&&\GF{\pq{1}{2}u^{2(N+2)^2}q^{M-m_{N+2} + m_i-s_i} p^{L-l_{N+2}+l_i-r_i} x_i  }
\times\nn\\
&&\GF{\left(u^{2N+5}v\right)^{-N-1}w^{-2}a_kq^{m_{N+2}-M} p^{l_{N+2}-L} }
\cI_0(q^{m_i-s_i} p^{l_i-r_i}x_i)\,.
\nn\\
\label{an:tube4pt:k1}
\ee
Summing over all partitions and using properties of $\Gamma$ and $\theta$ functions \eqref{theta:properties}
we get the difference operator given in \eqref{AN-HT} with the $\tth_k=\tth_{10}$. Operators obtained using another
ways to close minimal punctures can be derived in absolutely identical manner and lead to the same A$\Delta$O \eqref{AN-HT}.

\section{Derivation of $C_2$ operator}
\label{app:punctured:spheres}

In this section we give details of the derivations of results summarized in Section \ref{sec:c2:operator} and show in all details how to derive
$C_2$ type generalization of the van Diejen model.

Let's start with the derivation of the trinion that has one maximal
$\USp(2N)$ puncture, one maximal $\SU(N+1)$ puncture and
one minimal $\SU(2)$ puncture. It is important since it
will be later used by us to derive four punctured sphere with
two maximal $\USp(2N)$ and two minimal $\SU(2)$ punctures as specified in
\eqref{4puncture:gluing}.


To obtain this kind of three punctured sphere theory we start with the trinion
$\cT^A_{x,y,z}$ with two maximal $\SU(N+1)_{x,y}$ and one minimal $\SU(2)_z$
punctures. This trinion was derived in \cite{Razamat:2020bix} and is shown
on the Figure \ref{pic:an:trinion}. Corresponding superconformal index is given by:
\be
K_3^{A}(x,y,z)=\kappa_{N+1}\oint\prod\limits_{i=1}^{N+1}\frac{dt_i}{2\pi i t_i}\prod\limits_{i\neq j}^{N+2}\frac{1}{\GF{\frac{t_i}{t_j}}}
\prod\limits_{i=1}^{N+2}\prod\limits_{j=1}^{N+1}\GF{(pq)^{\frac{1}{2(N+2)}}u^{2N+4}t_ix_j}\times
\nn\\
\GF{(pq)^{\frac{1}{2(N+2)}}v^{2N+4}t_iy_j}\GF{(pq)^{\frac{1}{2(N+2)}}w^{2N+4}t_iz^{\pm1}}
\times\nn\\
\prod\limits_{l=1}^{2N+4}\GF{(pq)^{\frac{N+1}{2(N+2)}}\left(uv\right)^{-N-1}w^{-2} t_i^{-1}a_l}\,,
\label{index:trinion:sun}
\ee
where $u$, $v$, $w$ and $a_i$ parametrize Cartans of the $6d$ $\SO(4N+8)$ global symmetry. $\SU(N+2)$ gauge symmetry is parametrized by $t_i$'s with the relation
\be
\prod\limits_{i=1}^{N+2}t_i=1\,.
\label{sun:constr}
\ee
Global $\SU(N+1)$ symmetries of the maximal punctures are parametrized by $x_i$ and $y_i$ satisfying
\be
\prod\limits_{j=1}^{N+1}x_j=\prod\limits_{j=1}^{N+1}y_j=1\,.
\label{sun:constr:gl}
\ee
Each puncture has $(2N+6)$ moment map operators with the charges specified in \eqref{an:moment:maps}.

Now in order to obtain desired trinion we should turn one of the $\SU(N+1)$ punctures
into $\USp(2N)$ type puncture. For this purpose we use an $A_NC_N$ tube theory introduced in \cite{Kim:2018bpg}
and shown in Figure \ref{pic:cn:tube}.
Chirals of these theory correspond to the moment maps of punctures. For $\USp(2N)$ puncture these are
just moment maps specified in \eqref{cn:moment:maps}. For $\SU(N+1)$ puncture as specified in \eqref{an:moment:maps} but with the last moment map
flipped so that all of them transform in the same fundamental representation of $\SU(N+1)$:
\be
M_u={\bf N+1}^x\otimes \left( {\bf 2N+4}_{u^{N+3}v^{-N-1}w^{-2}}\oplus {\bf 1}_{\left( uv^{N+1} \right)^{2N+4}}\oplus {\bf 1}_{\left(u^N w^2\right)^{-2N-4}}\right)
\label{moment:map:x:flipped}
\ee
The index of this tube theory is given by:
\be
K_2^{AC}(x,\tx)=\prod\limits_{i=1}^{N+1}\prod\limits_{j=1}^N\prod\limits_{l=1}^{2N+4}\GF{\left(u^{-1}w\right)^{2N+4}\tx_i^{-1}x_j^{\pm 1}}
\GF{\pq{1}{2}u^{N+3}v^{-N-1}w^{-2}a_l\tx_i}
\times\nn\\
\GF{\pq{1}{2}\left( uv^{N+1} \right)^{2N+4}\tx_i}
\GF{\pq{1}{2}\left(u^Nw^2\right)^{-2N-4}\tx_i}
\times\nn\\
\GF{\pq{1}{2}\left(uvw^{-2}\right)^{N+1}a_l^{-1}x_j^{\pm 1}}
\GF{\pq{1}{2}\left(v^{N+1}w \right)^{-2N-4}x_j^{\pm 1}}
\times\nn\\
\GF{\pq{1}{2}\left(u^{N+1}w\right)^{2N+4}x_j^{\pm 1}}
\prod\limits_{i<j}^{N+1}\GF{pq \left( uw^{-1}\right)^{4(N+2)}\tx_i\tx_j}\,,
\nn\\
\label{ancn:tube:index}
\ee

Now we take this theory and glue it along the  $\SU(N+1)$ puncture to the $A_N$ trinion shown in Figure \ref{pic:an:trinion}.
Consistent gluing in this case is the mixture of $\Phi$ and S gluings.
 In particular we $\Phi$ glue all moment maps except
${\bf 1}_{\left(u^N w^2\right)^{2N+4}}$ which is S glued. At the level of the index this operation corresponds to
\begin{align}
&K_3^{AC}(x,y,z)=\kappa_N\oint\prod\limits_{i=1}^N\frac{d\tx_i}{2\pi i \tx_i}\prod\limits_{i\neq j}^{N+1}\frac{1}{\GF{\frac{\tx_i}{\tx_j}}}K_3^A(\tx,y,z)K_2^{AC}(\tx,x)
\times\nn\\
&\prod\limits_{i=1}^{N+1}\prod\limits_{l=1}^{2N+4}\GF{\pq{1}{2} u^{-N-3}v^{N+1}w^2a_l^{-1}\tx_i^{-1}}\GF{\pq{1}{2}\left(uv^{N+1}\right)^{-2N-4}\tx_i^{-1}}
\label{gluing}
\end{align}
Substituting expressions \eqref{index:trinion:sun} for $A_N$ trinion and \eqref{ancn:tube:index} for the $A_NC_N$ tube we immediately see that
the $\SU(N+1)_{\tx}$ gauge theory appears to be S-confining. Corresponding S-confining theory is depicted in Figure \ref{pic:svw:duality}. This duality
was first introduced by Spiridonov and Warnaar \cite{MR2264067} as a mathematical identity for superconformal indices of the corresponding theories:

\begin{figure}[!btp]
          \centering
           \includegraphics[width=0.7\linewidth]{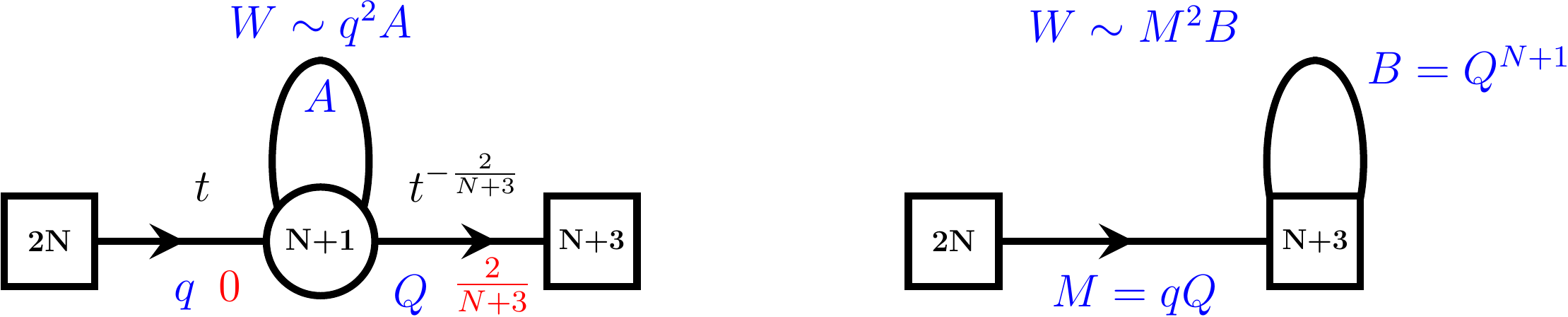}
            \caption{Spiridonov-Warnaar-Vartanov (SWV) duality. On the picture we also specify superpotentials, UV on the left and dynamically generated on the right sides
            of dualitites. The solid line starting and ending on the same node denotes multiplet in rank-two antisymmetric representation (AS).}
            \label{pic:svw:duality}
\end{figure}

\be
\kappa_N\oint\prod\limits_{i=1}^N\frac{d\tx_i}{2\pi i \tx_i}\prod\limits_{i\neq j}^{N+1}\frac{1}{\GF{\frac{\tx_i}{\tx_j}}}\prod\limits_{i<j}^{N+1}\GF{S\tx_i^{-1}\tx_j^{-1}}
\prod\limits_{j=1}^{N+1}\prod\limits_{k=1}^N\GF{\tx_j t_k}\GF{pq S^{-1}t_k^{-1}\tx_j}
\times\nn\\
\prod\limits_{m=1}^{N+3}\GF{s_m\tx_j^{-1}}=\prod\limits_{k=1}^N\prod\limits_{m=1}^{N+3}\frac{\GF{t_ks_m}}{\GF{St_ks_m^{-1}}}\prod\limits_{l<m}^{N+3}\GF{Ss_l^{-1}s_m^{-1}}\,,
\label{svw:duality:index}
\ee
where $S=\prod\limits_{i=1}^{N+3}s_i$. Later Spiridonov and Vartanov \cite{Spiridonov:2009za} discussed physical implications of the relation written above.
We refer to this duality as Spiridonov-Warnaar-Vartanov (SWV) duality. More detailed discussion about it can also been found in the Appendix B.5 of our previous
paper \cite{Nazzal:2021tiu}. Using this duality for the $\SU(N+1)_{\tx}$ node in \eqref{gluing} we can finally write down the index of the desired three-punctured sphere
as follows:
\begin{align}
&K_3^{AC}(x,y,z)=\kappa_{N+1}\oint\prod\limits_{i=1}^{N+1}\frac{dt_i}{2\pi it_i}\prod\limits_{i\neq j}^{N+2}\frac{1}{\GF{\frac{t_i}{t_j}}}\prod\limits_{k=1}^N
\prod\limits_{i=1}^{N+2}\prod\limits_{j=1}^{N+1}\GF{\pq{1}{2(N+2)}w^{2N+4}x_k^{\pm 1}t_i}
\times\nn\\
&\GF{\pq{N+1}{2(N+2)}u^{2(N+1)(N+2)}t_i^{-1}}\GF{\pq{1}{2(N+2)}v^{2N+4}y_jt_i}\GF{\pq{1}{2(N+2)}w^{2N+4}t_iz^{\pm 1}}
\times\nn\\
&\prod\limits_{l=1}^{2N+4}\GF{\pq{N+1}{2(N+2)}\left(uv \right)^{-N-1}w^{-2}t_i^{-1}a_l}\prod\limits_{i<j}^{N+2}\GF{\pq{N+1}{N+2}w^{-4N-8}t_i^{-1}t_j^{-1}}\,,
\label{AN:CN:trinion}
\end{align}
where we have omitted some of the $\USp(2N)$ singlets thus redefining type of the puncture. Corresponding trinion theory is shown in Figure \ref{pic:ancntrinion}.

\begin{figure}[!btp]
          \centering
           \includegraphics[width=0.4\linewidth]{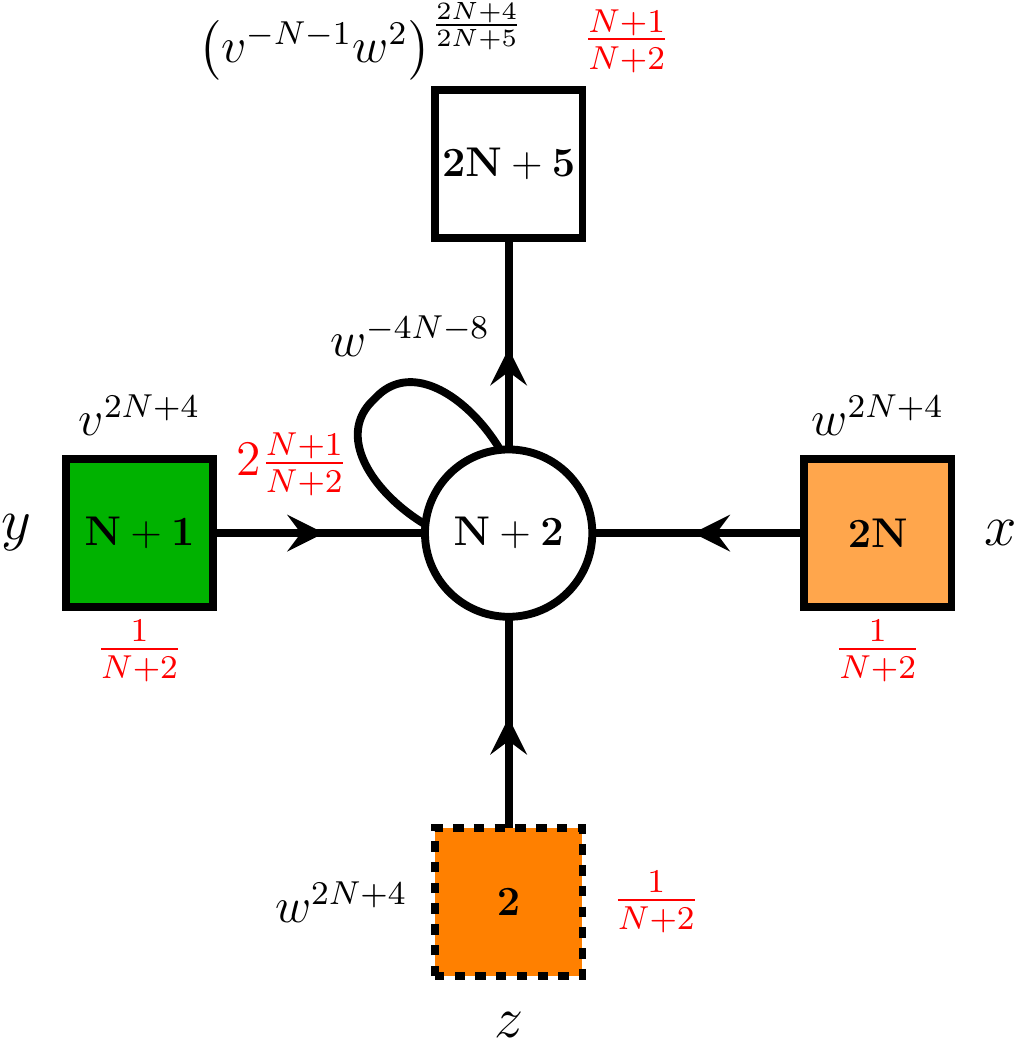}
            \caption{Trinion theory with $\USp(2N)_x$ maximal puncture \textit{(light orange)}, $\SU(N+1)_y$ maximal puncture \textit{(green)} and
            $\SU(2)_z$ minimal puncture \textit{(orange)}. $U(1)$ global symmetries are parametrized using the parametrization \eqref{an:parametrization} more natural for the
        $\SU(N+1)$ type punctures. The index of the theory is given in \eqref{AN:CN:trinion}. }
            \label{pic:ancntrinion}
\end{figure}

\begin{figure}[!btp]
          \centering
           \includegraphics[width=0.65\linewidth]{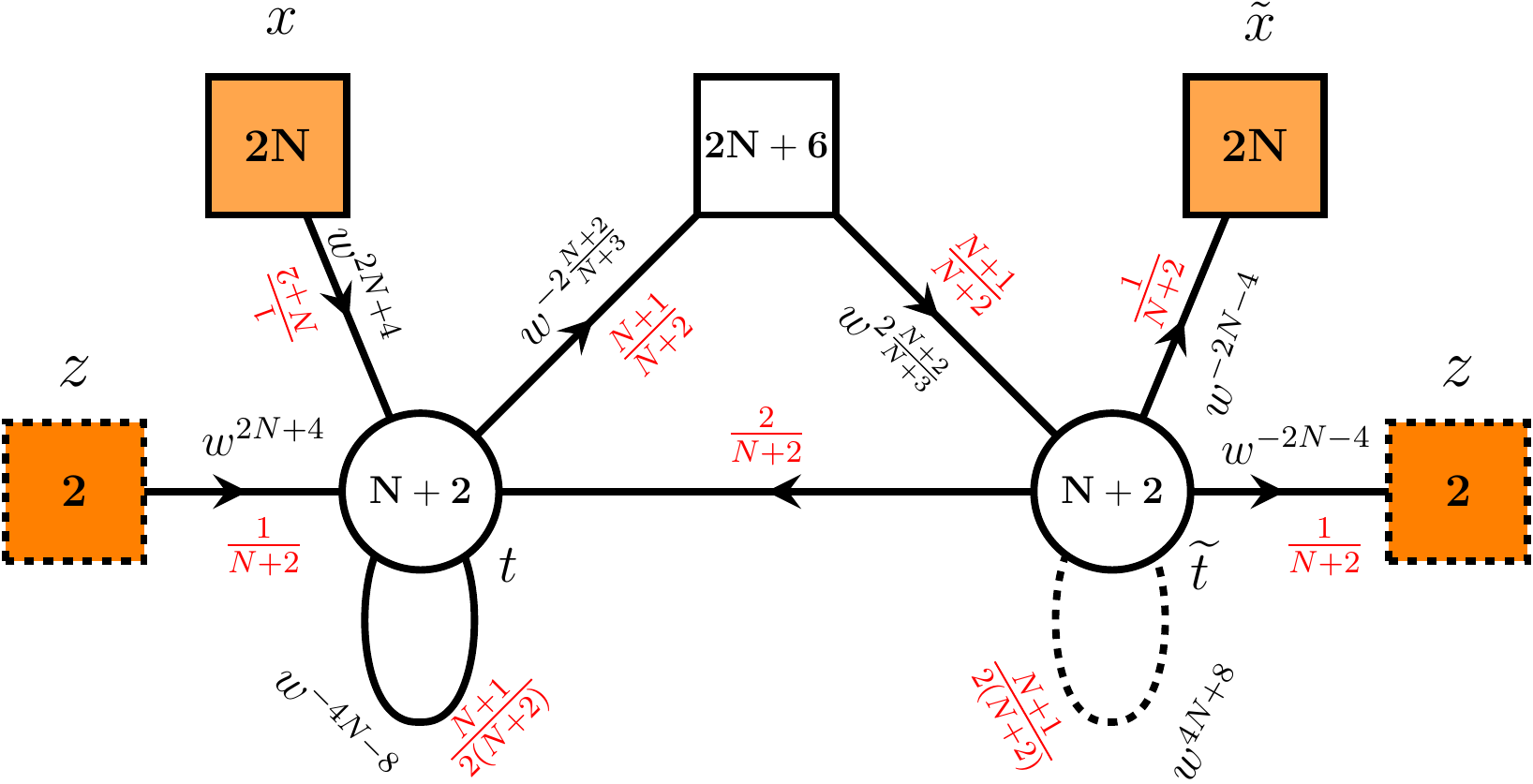}
            \caption{Four-punctured sphere theory with two maximal $\USp(2N)$ and two minimal $\SU(2)$ punctures. The four-punctured sphere is derived by gluing two trinion theories
            shown on the Figure \ref{pic:ancntrinion} and performing some duality transformations specified in the text.}
            \label{pic:cn4pt}
\end{figure}

Now we can use this $A_NC_N$ type trinion theory in order to derive four-punctured sphere theory with two $\USp(2N)$ maximal punctures. For this purpose we S glue two copies of
this trinion along $A_N$ type puncture. At the level of the index this corresponds to the operation specified in \eqref{4puncture:gluing}. Substituting trinion index
\eqref{AN:CN:trinion} into this expression we obtain the following four-punctured sphere index:
\begin{align}
K_4(x,\tx,z,\tz)\equiv \kappa_N\oint\prod\limits_{i=1}^N\frac{dy}{2\pi ix}\prod\limits_{i\neq j}\frac{1}{\GF{y_i/y_j}}
K_3^{AC}(x,y,z)\bar{K}_3^{AC}(\tx,y,\tz)=
\nn\\
\kappa_{N+1}^2\kappa_N\oint\prod\limits_{i=1}^{N+1}\frac{dt_i}{2\pi it_i}\frac{d\ttt_i}{2\pi i\ttt_i}
\prod\limits_{i=1}^N\frac{dy_i}{2\pi iy_i}\prod\limits_{i=1}^{N+2}\frac{1}{\GF{t_i/t_j}\GF{\ttt_i/\ttt_j}}\prod\limits_{i=1}^{N+1}\frac{1}{\GF{y_i/y_j}}
\times\nn\\
\prod\limits_{i=1}^{N+2}\prod\limits_{j=1}^{N+1}\prod\limits_{k=1}^N\prod\limits_{l=1}^{2N+4}\GF{\pq{1}{2N+4}w^{2N+4}x_k^{\pm 1}t_i}
\GF{\pq{N+1}{2N+4}u^{2(N+1)(N+2)}t_i^{-1}}
\times\nn\\
\GF{\pq{1}{2N+4}v^{2N+4}y_jt_i}\GF{\pq{1}{2N+4}w^{2N+4}t_iz^{\pm 1}}\GF{\pq{1}{2N+4}w^{-2N-4}\tx_k^{\pm 1}\ttt_i^{-1}}
\times\nn\\
\GF{\pq{N+1}{2N+4}\left(uv\right)^{-N-1}w^{-2}t_i^{-1}a_l}\GF{\pq{N+1}{2N+4}\left(uv\right)^{N+1}w^2\ttt_ia_l^{-1}}
\times\nn\\
\GF{\pq{N+1}{2N+4}u^{2(N+1)(N+2)}\ttt_i}\GF{\pq{1}{2N+4}v^{-2N-4}y_j^{-1}\ttt_i^{-1}}
\times\nn\\
\GF{\pq{1}{2N+4}w^{-2N-4}\ttt_i^{-1}\tz^{\pm 1}}
\prod\limits_{i<j}^{N+2}\GF{\pq{N+1}{N+2}w^{-4N-8}t_i^{-1}t_j^{-1}}\GF{\pq{N+1}{N+2}w^{4N+8}\ttt_i\ttt_j}
\end{align}
Now if we look on the $\SU(N+1)_y$ gauged node we see that it corresponds to the theory with $N+2$ hypermmultiplets, meaning
the node is S-confining and can be integrated out using the standard Seiberg duality. Performing this simple operation we land on the
theory shown on  the Figure \ref{pic:cn4pt} with the following superconformal index:
\begin{align}
K_4^C(x,\tx,z,\tz)=\kappa^2_{N+1}\oint \prod\limits_{i=1}^{N+1}\frac{dt_i}{2\pi it_i}\frac{d\ttt_i}{2\pi i\ttt_i}
\prod\limits_{i\neq j}^{N+2}\frac{1}{\GF{t_i/t_j}\GF{\ttt_i/\ttt_j}}
\times\nn\\
\prod\limits_{i=1}^{N+2}\prod\limits_{k=1}^N\prod\limits_{l=1}^{2N+6}\GF{\pq{1}{2N+4}w^{2N+4}x_k^{\pm 1}t_i}
\GF{\pq{N+1}{2N+4}w^{-2\frac{N+2}{N+3}}\ta_lt_i^{-1}}
\times\nn\\
\GF{\pq{1}{2N+4}w^{-2N-4}\tx_k^{\pm 1}\ttt_i^{-1}}\GF{\pq{1}{2N+4}w^{-2N-4}\ttt_i^{-1}\tz^{\pm 1}}
\times\nn\\
\GF{\pq{1}{2N+4}w^{2\frac{N+2}{N+3}}\ta_l^{-1}\ttt_i}\prod\limits_{i,j=1}^{N+2}\GF{\pq{1}{N+2}t_i\ttt_j^{-1}}
\GF{\pq{1}{2N+4}w^{2N+4}t_iz^{\pm 1}}
\times\nn\\
\prod\limits_{i<j}^{N+2}\GF{\pq{N+1}{N+2}w^{-4N-8}t_i^{-1}t_j^{-1}}\GF{\pq{N+1}{N+2}w^{4N+8}\ttt_i\ttt_j}\,,
\label{c2:4pt:index}
\end{align}
where we have also used dictionary \eqref{ancn:dict} in order to express everything in terms of $C_N$ parametrzation \eqref{cn:parametrization} since now we have only
this type of maximal punctures and hence latter parametrization is more natural.

Now we finally close one of the two $\SU(2)$ minimal punctures without introducing the defect. As discussed in Section \ref{sec:c2:operator}
this amounts to computing the residue of \eqref{c2:4pt:index} at the point  $\tz=\pqm{1}{2}w^{2\frac{(N+2)^2}{N+3}}\ta_i$.
Indeed it can be seen that at this value of $\tz$ there is pole coming from the contour pinching at
\be
\ttt_{N+2}=\pqm{N+1}{2N+4}w^{-2\frac{N+2}{N+3}}\ta_i\,,
\ee
where $\ttt_{N+2}$ variable is chosen without loss of generality. In fact we should compute such pinchings for each $\ttt_i$ variables
and sum the results. But due to the Weyl symmetry of the $\SU(N+1)$ root system  all such contributions are the same and
summing them results in an overall factor of $(N+1)!$. We omit such overall factors since they are irrelevant for the structure of
the A$\Delta$O we get in the end. Computation of the corresponding residue at the pinching point leads to the following expression
for the three-punctured sphere theory index:
\begin{align}
&K_{(3;i;0)}^C(x,\tx,z)=\kappa_{N+1}\kappa_N\oint\prod\limits_{j=1}^{N+1}\frac{dt_j}{2\pi it_j}\prod\limits_{j=1}^N\frac{d\ttt_j}{2\pi i\ttt_j}
\prod\limits_{k\neq j}^{N+2}\frac{1}{\GF{t_k/t_j}}\prod\limits_{i\neq j}^{N+1}\frac{1}{\GF{\ttt_k/\ttt_j}}
\times\nn\\
&\prod\limits_{m=1}^{N+2}\prod\limits_{j=1}^{N+1}\prod\limits_{k=1}^N\prod\limits_{l\neq i}^{2N+6}
\GF{\pq{1}{2N+4}w^{2N+4}x_k^{\pm 1}t_m}\GF{\pq{N+1}{2N+4}w^{-2\frac{N+2}{N+3}}\ta_lt_m^{-1}}
\times\nn\\
&\GF{\pq{1}{2N+4}w^{2N+4}t_mz^{\pm 1}}\GF{w^{-2\frac{(N+2)^3}{(N+1)(N+3)}}\ta_i^{\frac{1}{N+1}}\tx_k^{\pm 1}\ttt_j^{-1}}
\GF{\pq{1}{2}w^{-2\frac{(N+2)^2}{N+3}}\ta_i^{-1}\tz_k^{\pm 1}}
\times\nn\\
&\GF{\pq{1}{2}w^{2\frac{(N+2)^2}{(N+1)(N+3)}}\ta_l^{-1}\ta_i^{-\frac{1}{N+1}}\ttt_j}
\GF{\pq{1}{2N+4}w^{-2\frac{N+2}{(N+1)(N+3)}}\ta_i^{\frac{1}{N+1}}t_i/\ttt_j}
\times\nn\\
&\prod\limits_{i=m<j}^{N+2}\GF{\pq{N+1}{N+2}w^{-4N-8}t_m^{-1}t_j^{-1}}
\prod\limits_{m<j}^{N+1}\GF{pq w^{4\frac{(N+2)^3}{(N+1)(N+3)}}\ta_i^{-\frac{2}{N+1}}\ttt_m\ttt_j}\,.
\label{3pt:sphere:4thway}
\end{align}
 Corresponding quiver diagram is shown in Figure \ref{pic:cn:3pt:new}.

\begin{figure}[!btp]
          \centering
           \includegraphics[width=\linewidth]{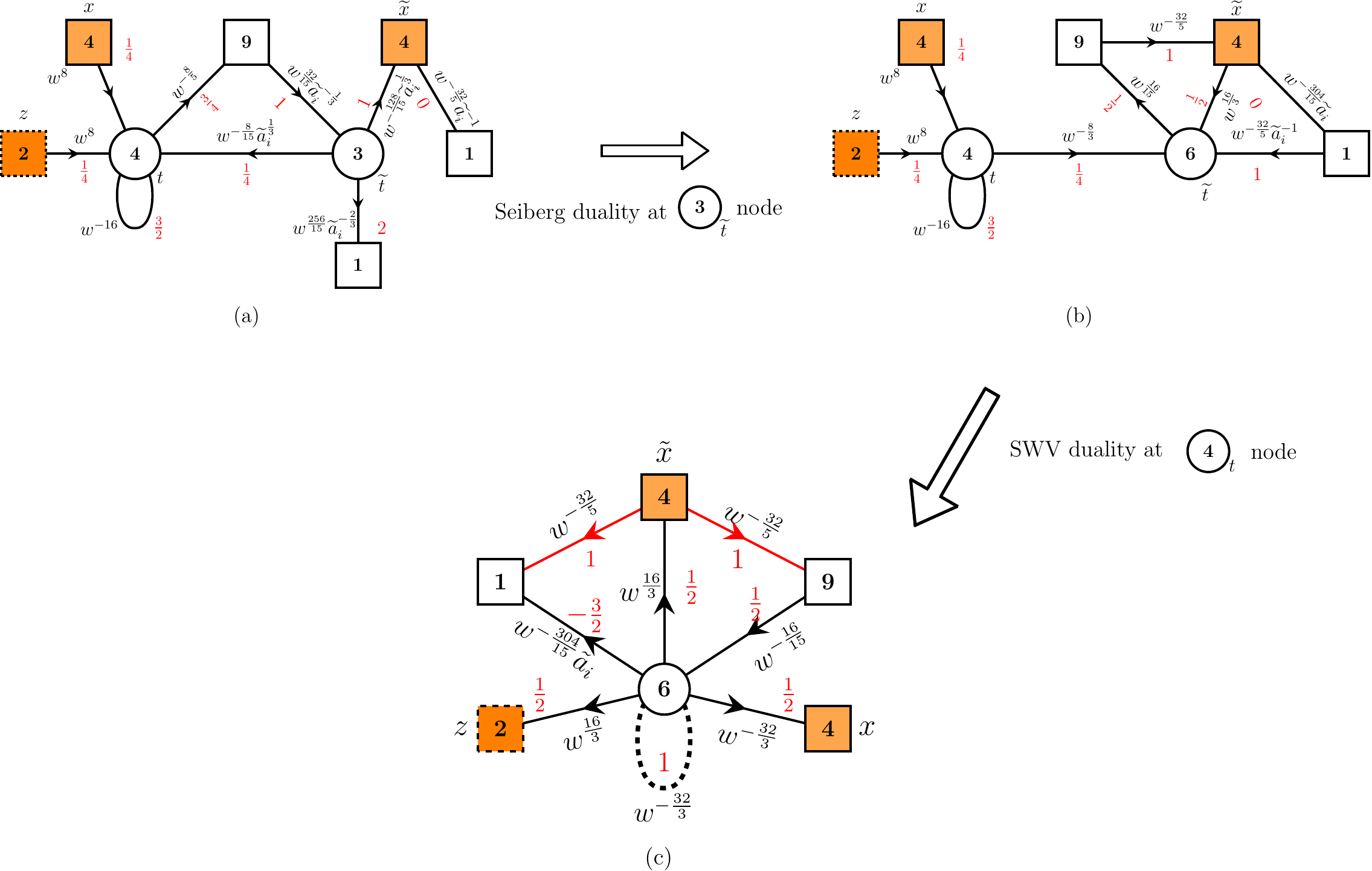}
            \caption{Chain of duality transformations of three-punctured sphere theory in $C_2$ case. We start with the theory derived by gluing two three-punctured
            sphere theories $\cT^{AC}$ along $A$-type puncture, integrating out $A_2$ node and closing $\SU(2)_{\tz}$ minimal puncture. This theory is shown on the
            Figure (a) above and corresponding index is given by \eqref{c2:3pt:sphere}. Then performing Seiberg duality on the right node of this theory
	    we obtain the quiver shown on the Figure (b). Finally we notice that right $\SU(4)$ gauge node of theory (b) is confining due to SWV duality.
            Integrating it out we obtain theory shown in the Figure (c). }
            \label{pic:c2:dualities}
\end{figure}

In case $N=1$ it can be shown that the theory we derived reduces to the one shown in Figure 22 (b) in our previous paper \cite{Nazzal:2021tiu}.
There we used it to derive $C_1$ A$\Delta$O
which appeared to be van Diejen operator. Studying general $N$ case appears to be too complicated since there is no clear
way to further simplify theory \eqref{3pt:sphere:4thway}. Instead we concentrate here on $C_2$ case. When $N=2$ $\overline{AS}$ representation
of $\SU(N+1)$ node becomes anti-fundamental representation and the index of the theory reads:
\begin{align}
K_{(3;i,0)}^C(x,\tx,z)=\kappa_{3}\kappa_2\oint\prod\limits_{m=1}^{3}\frac{dt_m}{2\pi it_m}\prod\limits_{i=1}^2\frac{d\ttt_m}{2\pi i\ttt_m}
\prod\limits_{m\neq j}^{4}\frac{1}{\GF{t_m/t_j}}\prod\limits_{m\neq j}^{3}\frac{1}{\GF{\ttt_m/\ttt_j}}
\times\nn\\
\prod\limits_{m=1}^{4}\prod\limits_{j=1}^{3}\prod\limits_{k=1}^2\prod\limits_{l\neq i}^{10}
\GF{\pq{1}{8}w^{8}x_k^{\pm 1}t_m}\GF{\pq{3}{8}w^{-\frac{8}{5}}\ta_lt_m^{-1}}
\GF{\pq{1}{8}w^{8}t_mz^{\pm 1}}
\times\nn\\
\GF{w^{-\frac{128}{15}}\ta_i^{\frac{1}{3}}\tx_k^{\pm 1}\ttt_j^{-1}}
\GF{\pq{1}{2}w^{-\frac{32}{5}}\ta_i^{-1}\tx_k^{\pm 1}}
\GF{\pq{1}{2}w^{\frac{32}{5}}\ta_l^{-1}\ta_i^{-\frac{1}{3}}\ttt_j}
\times\nn\\
\GF{\pq{1}{8}w^{-\frac{8}{15}}\ta_i^{\frac{1}{3}}t_m/\ttt_j}
\GF{pqw^{\frac{256}{15}}\ta_i^{-\frac{2}{3}}\ttt_j^{-1}}
\prod\limits_{m<j}^{N+2}\GF{\pq{3}{4}w^{-16}t_m^{-1}t_j^{-1}}
\,,
\label{c2:3pt:sphere}
\end{align}
with the corresponding quiver of the theory shown in Figure \ref{pic:c2:dualities} (a).

At the next step we perform Seiberg duality on the $\SU(3)$ gauge node of the theory above. This node has $9$ falvors so after dualising it we obtain
$\SU(6)$ node with the superconformal index of the resulting theory given by
\be
K_{(3;i,0)}^C(x,\tx,z)=\kappa_3 \kappa_5\oint\prod\limits_{m=1}^3\frac{dt_m}{2\pi it_m}\prod\limits_{m=1}^5\frac{d\ttt_m}{2\pi i\ttt_m}
\prod\limits_{m\neq j}^4\frac{1}{\GF{t_m/t_j}}\prod\limits_{m\neq j}^6\frac{1}{\GF{\ttt_m/\ttt_j}}
\times\nn\\
\prod\limits_{m=1}^4\prod\limits_{j=1}^6\prod\limits_{k=1}^2\prod\limits_{l=2}^{10}\GF{\pq{1}{8}w^8y_k^{\pm 1}t_m}
\GF{\pq{1}{8}w^8t_mz^{\pm 1}}\GF{\pq{1}{2}w^{-\frac{32}{5}}\ta_i^{-1}\tx_k^{\pm 1}}
\times\nn\\
\GF{\pq{1}{2}w^{-\frac{32}{5}}\ta_l^{-1}\tx_k^{\pm 1}}\GF{\pq{1}{4}w^{\frac{16}{15}}\ta_l\ttt_j}
\GF{\pq{1}{4}w^{\frac{16}{3}}\tx_k^{\pm 1}\ttt_j^{-1}}
\times\nn\\
\GF{\pqm{3}{4}w^{-\frac{304}{15}}\ta_i\ttt_j^{-1}}
\GF{\pq{1}{8}w^{-\frac{8}{3}}t_m^{-1}\ttt_j^{-1}}\prod_{m<n}^4\GF{\pq{3}{4}w^{-16}t_m^{-1}t_n^{-1}}\,,
\ee
and corresponding quiver is shown in Figure \ref{pic:c2:dualities} (b).
We see that $\SU(4)$ gauge node of this theory appears to be S-confining due to SWV duality.
Hence using SWV index identity \eqref{svw:duality:index} we arrive to the gauge theory with the following
superconformal index:
\be
K_{(3;i,0)}^C(x,\tx,z)=\kappa_5\oint\prod\limits_{i=1}^5\frac{d\ttt_i}{2\pi i\ttt_i}\prod\limits_{i\neq j}^6\frac{1}{\GF{\ttt_i/\ttt_j}}
\prod\limits_{i=1}^6\prod\limits_{k=1}^2\GF{\pq{1}{4}w^{\frac{16}{3}}x_k^{\pm 1}\ttt_i^{-1}}
\times\nn\\
\GF{\pq{1}{4}w^{\frac{16}{3}}z^{\pm 1}\ttt_i^{-1}}\GF{\pq{1}{4}w^{\frac{16}{3}}\tx_k^{\pm 1}\ttt_i^{-1}}
\GF{\pqm{3}{4}w^{-\frac{304}{15}}\ta_1\ttt_i^{-1}}
\times\nn\\
\GF{\pq{1}{2}w^{-\frac{32}{5}}\ta_1^{-1}\tx_k^{\pm 1}}
\prod\limits_{l=2}^{10}\GF{\pq{1}{4}w^{\frac{16}{15}}\ta_l\ttt_i}\GF{\pq{1}{2}w^{-\frac{32}{5}}\ta_l^{-1}\tx_k^{\pm 1}}
\times\nn\\
\prod\limits_{i<j}^6\GF{\pq{1}{2}w^{-\frac{32}{3}}\ttt_i\ttt_j}\,,
\label{3pt:sphere:c2:final:2}
\ee
with the quiver shown in Figure \ref{pic:c2:dualities} (c).
Now we are ready to close second minimal puncture leaving us with the sphere with
only two maximal $\USp(2N)$ punctures left. As discussed in the Section \ref{sec:c2:operator} for this purpose we fix
$z$ fugacity to be
\be
z=Z^*_{i;K,0}=\pqm{1}{2}w^{-\frac{32}{5}}a_i^{-1}q^{-K}\,.
\ee
Carefully studying expression \eqref{3pt:sphere:c2:final:2} we see that at this value of $z$  superconformal index indeed has
pole coming from the contour pinching at:
\be
\ttt_{1,2}=\pq{1}{4}w^{\frac{16}{3}}y_1^{\pm 1}q^{k_{1,\pm}}\,,\quad\ttt_{3,4}=\pq{1}{4}w^{\frac{16}{3}}y_1^{\pm 1}q^{k_{2,\pm}}\,,
\nn\\
\ttt_5=\pqm{1}{4}w^{-\frac{16}{5}}a_i^{-1}q^{k_5-K}\,,\quad \ttt_6=\pqm{3}{4}w^{-\frac{304}{15}}a_i q^{k_6}\,,
\label{pinching:c2}
\ee
where $\{k_{1,\pm},\,k_{2,\pm},\,k_5,\,k_6\}\equiv\vec{K}$ is partition of the integer $K$. Of course any permutation of \eqref{pinching:c2} would also result in the
same contour pinching and hence the pole of the superconformal index. Due  to the  Weyl group symmetry contributions of all these
permutations into our operator is the same and taking them into account just results in an overall factor which we omit.
Substituting values \eqref{pinching:c2} into three punctured sphere index \eqref{3pt:sphere:c2:final:2} we get the following
result for the index of the two-punctured sphere:
\begin{align}
&K_{(2;i;K,0)}(x,\tx)=\sum\limits_{\vec{K}}C_{(\vec{K};i)}\prod\limits_{j=1}^2\prod\limits_{l\neq i}^{10}\frac{\GF{x_j^{\pm 2}q^{-k_{j,\mp}}}}{\GF{x_j^{\pm 2}q^{\pm(k_{j,+}-k_{j,-})}}}
\times\nn\\
&\frac{\GF{x_1 x_2^{\pm 1}q^{-k_{2,\mp}}}\GF{x_1^{-1}x_2^{\pm 1}q^{-k_{2,\mp}}}}{\GF{\left(x_1x_2^{\pm 1}q^{\left(k_{1,+}-k_{2,\mp} \right)} \right)^{\pm 1}}}
\frac{\GF{x_1^{\pm 1} x_2q^{-k_{1,\mp}}}\GF{x_1^{\pm 1}x_2^{-1}q^{-k_{1,\mp}}}}{\GF{\left(x_1^{-1}x_2^{\pm 1}q^{\left(k_{1,-}-k_{2,\mp} \right)} \right)^{\pm 1}}}
\times\nn\\
&\frac{\GF{\pq{1}{2}w^{\frac{32}{5}}\ta_ix_j^{\pm 1}q^{K-k_{j,\mp}}} }
{\GF{\left(\pq{1}{2}w^{\frac{32}{5}}\ta_ix_j^{\pm 1}q^{K+k_{j,\pm}-k_5}\right)^{\pm 1}}}
\frac{\GF{p^{-1}q^{-1}w^{-\frac{128}{5}}\ta_ix_j^{\pm 1}q^{-k_{j,\mp}}}}
{\GF{\left(pqw^{\frac{128}{5}}\ta_i^{-1}x_j^{\pm 1}q^{k_{j,\pm}-k_6}\right)^{\pm 1}}}
\times\nn\\
&\GF{\pqm{1}{2}w^{-\frac{32}{5}}\ta_i^{-1}x_j^{\pm 1}q^{-K-k_{j,\mp}}}
\GF{w^{-\frac{128}{5}}\ta_ix_j^{\pm 1}q^{k_{j,\pm}+k_6}}\GF{pqx_1x_2^{\pm 1}q^{k_{1,+}+k_{2,\pm}}}
\times\nn\\
&\GF{\pq{1}{2}w^{-\frac{32}{5}}\ta_i^{-1}x_j^{\pm 1}q^{k_{j,\pm}+k_5-K}}\GF{pq w^{\frac{128}{5}}\ta_i^{-1}\tx_j^{\pm 1}q^{-k_6}}
\GF{\pq{1}{2}w^{\frac{32}{5}}\ta_lx_j^{\pm 1}q^{k_{j,\pm}}}
\times\nn\\
&\GF{\pq{1}{2}w^{\frac{32}{5}}x_j^{\pm 1}\ta_iq^{K-k_5}}\GF{\pq{1}{2}w^{\frac{32}{5}}\ta_i\tx_j^{\pm 1}q^{K-k_5}}
\GF{\pq{1}{2}w^{-\frac{32}{5}}\ta_l^{-1}\tx_j^{\pm 1}}
\times\nn\\
&\GF{\pq{1}{2}w^{-\frac{32}{5}}\ta_i^{-1}\tx_j^{\pm 1}}\prod\limits_{k=1}^2\GF{\tx_k^{\pm 1} x_jq^{-k_{j,-}}}\GF{\tx_k^{\pm 1} x_j^{-1}q^{-k_{j,+}}}
\GF{pq^{1+k_{1,+}+k_{1,-}}}
\times\nn\\
&\hspace{2cm}\GF{pqw^{\frac{128}{5}}\ta_i^{-1}x_j^{\pm 1}q^{-k_6}}\GF{pqx_1^{-1}x_2^{\pm 1}q^{k_{1,-}+k_{2,\pm}}}
\GF{pq^{1+k_{2,+}+k_{2,-}}}\,,
\label{tube:C2:general}
\end{align}
where $C_{(\vec{K};i)}$ is the following constant:
\begin{align}
C_{(\vec{K};i)}&=\prod\limits_{l\neq i}^{10}\GF{\pq{3}{2}w^{32}q^{K-k_6}}\GF{\ta_l\ta_i^{-1}q^{k_5-K}}\GF{\pqm{1}{2}w^{-\frac{96}{5}}\ta_l\ta_iq^{k_6}}
\times\nn\\
&\frac{\GF{\pqm{1}{2}w^{-\frac{96}{5}}\ta_i^2q^{K-k_5}} \GF{\pq{1}{2}w^{\frac{96}{5}}\ta_i^{-2}q^{-K-k_6}}}
{\GF{\pq{1}{2}w^{\frac{96}{5}}\ta_i^{-2}q^{-K+k_5-k_6}} \GF{\pqm{1}{2}w^{-\frac{96}{5}}\ta_i^{2}q^{K+k_6-k_5}}}
\times\nn\\
&\hspace{2cm}\GF{\pqm{1}{2}w^{-32}q^{k_5+k_6-K}}\GF{pqw^{\frac{64}{5}}\ta_i^2 q^{2K-k_5}}\,,
\label{C2:ck:constant}
\end{align}
here subscript contains all the information on how the tube was obtained. Inside this subscript $2$ means that we obtain two-punctured sphere (the tube)
in the end. Index $i$ stands for the choice
of the moment map we give vev to according to \eqref{tz:closing} and \eqref{z:closing}. Finally $K$ denotes the power of holomorphic derivative of the moment map we give vev to.
Summation in \eqref{tube:C2:general} goes over all possible partitions of $K$ integer. Due to the presence of the  defect this index expression unlike our previous
four- and three-punctured sphere does not have nice gauge theory interpretation.

Now gluing tube \eqref{tube:C2:general} to an arbitrary $\NN=1$ theory with one maximal $\USp(2N)$
puncture we can obtain desired A$\Delta$O. At the level of the superconformal index the gluing is
performed as specified in \eqref{c2:final:gluing}:
\be
\cO^{(C_2;h_{k};K,0)}_x\cdot\cI(x)=\frac{\kappa_N}{2^N}\oint\frac{d\tx_{1,2}}{2\pi i\tx_{1,2}}\frac{1}{\GF{\tx_{1,2}^{\pm 2}}\GF{\tx_1^{\pm 1}\tx_2^{\pm 1}}}K_{(2;i;K,0)}^C(x,\,\tx)\cI(\tx)
\label{c2:final:gluing:app}
\ee
Notice that expression \eqref{tube:C2:general}
for the tube index has zeroes coming from the last two $\Gamma$-functions. Due to these zeroes only the poles of the integral in \eqref{c2:final:gluing:app}
coming from other $\Gamma$ functions will contribute to the final expression. In particular these poles are coming from the
$\GF{\tx_k^{\pm 1}x_jq^{-k_{j,-}}}$ and $\GF{\tx_k^{\pm 1}x_j^{-1}q^{-k_{j,+}}}$ terms and are located at
\be
\tx_i=\left(x_{\sigma(i)}q^{-m_i}\right)^{\pm 1}\,,\qquad -k_{\sigma(i),+}\leq m_i \leq k_{\sigma(i),-}\,.
\ee
where $\sigma(i)$ is permutation of $i$. In the end we should sum over such permutations as well as over all combinations of $\pm 1$ powers in the expression
above. Since the tube expression \eqref{tube:C2:general} is symmetric w.r.t. permutations of $\tx_1\leftrightarrow \tx_2$ and $\tx_i\to \tx_i^{-1}$ every contribution
of the pinchings specified above is the same and summation just gives an irrelevant overall factor. So for simplicity we can consider only one of the combinations
\be
\tx_i=x_i q^{-m_i}\,,\qquad -k_{i,+}\leq m_i \leq k_{i,-}\,.
\label{c2:pinchings:final}
\ee
The condition on $m_i$ comes from the fact that both types of specified $\Gamma$ functions should have poles. Half of these terms should lead to integration  contour pinching
in \eqref{c2:final:gluing:app} while other half will cancel zeroes coming from the last two $\Gamma$ functions in the tube index expression \eqref{tube:C2:general}.

Now computing the contribution of pinchings \eqref{c2:pinchings:final} into gluing \eqref{c2:final:gluing:app}
and using $\Gamma$ functions identity \eqref{theta:properties} we can directly obtain the full tower of operators \eqref{c2:operator:general:K}.

\section{Proof of the kernel property of $A_2C_2$ tube theory}
\label{app:kernel}

Here we give a proof of the kernel property \eqref{a2c2:kernel:property} where we act with $A_2$ operator
\eqref{sun:op:noflip:gen} and $C_2$ operator \eqref{c2:operator:final} on the index \eqref{ancn:tube:index}
of the tube with $\SU(3)$ and $\USp(4)$ maximal punctures.

Let's start the proof of the kernel property by explicitly calculating action of our A$\Delta$Os on
the tube index and write two sides of the equation in terms of algebraic expressions.
First of all we should fix the parametrization of  the Cartans of global  $6d$ symmetry. For this it
will be more convenient to use $C_2$ parametrization given in \eqref{cn:parametrization}.
We should also use the dictionary \eqref{ancn:dict} between
$A_2$ and $C_2$ parametrizations in order to rewrite everything in terms of the latter one.
Fixing $N=2$ in the tube index \eqref{ancn:tube:index} and writing everything in terms
of $C_2$ parametrization we obtain:
\be
K_2^{A_2C_2}(x,y)=\prod\limits_{i=1}^3\prod\limits_{j=1}^2\prod\limits_{l=1}^9\GF{w^{\frac{128}{15}}\ta_{10}^{-\frac{1}{3}}y_i^{-1}x_j^{\pm 1}}
\GF{\pq{1}{2}w^{-\frac{32}{15}}\ta_{10}^{\frac{1}{3}}\ta_ly_i}
\times\nn\\
\GF{\pq{1}{2}w^{-\frac{224}{15}}\ta_{10}^{-\frac{2}{3}}y_i}\GF{\pq{1}{2}w^{\frac{32}{5}}\ta_{10}x_j^{\pm 1 }}\GF{pq w^{-\frac{256}{15}}\ta_{10}^{\frac{2}{3}}y_i^{-1}}\,,
\label{a2c2:tube:index}
\ee
where as previously $x_{1,2}$ are fugacities for the global symmetry of the $\USp(4)$ puncture, while
$y_{1,2,3}$ are fugacities of the $\SU(3)$ puncture. Latter ones are as usually constrained by the
identity $\prod_{i=1}^3y_i=1$.

Now we can start with the r.h.s. of  \eqref{a2c2:kernel:property} describing the
action of $C_2$ operator. Main ingredient we need to find the action of the full
operator is the action of the shift $\Delta_q(x_i)$ on the
tube index. Studying this action on \eqref{a2c2:tube:index} we obtain:
\be
\Delta_q(x_l)K_2^{A_2C_2}(x,y)=D_l(x,y)K_2^{A_2C_2}(x,y)\,,
\ee
where the function $D_l(x,y)$ is given by:
\be
D_l(x,y)=\prod\limits_{i=1}^3\frac{\thf{\pq{1}{2}\ta_{10}w^{\frac{32}{5}}x_l}}{\thf{\pq{1}{2}q^{-1}\ta_{10}w^{\frac{32}{5}}x_l^{-1}}}
\frac{\thf{\ta_{10}^{-\frac{1}{3}}w^{\frac{128}{15}}x_ly_i^{-1}}}{\thf{q^{-1}\ta_{10}^{-\frac{1}{3}}w^{\frac{128}{15}}x_l^{-1}y_i^{-1}}}
\ee
Then on the r.h.s. of \eqref{a2c2:kernel:property} we obtain:
\be
\cO^{(C_2;h_{10};1,0)}_{x}K_2^{A_2C_2}(x,y)=\left[F^{(C_2;h_{10};1,0)}(x,y)+W^{(C_2;h_{10};1,0)}(x)\right]K_2^{A_2C_2}(x,y)
\nn\\
F^{(C_2;h_{10};1,0)}(x,y)\equiv\sum_{i=1}^2\left(A^{(C_2;h_{10};1,0)}_i(x)D_i(x,y)+\left(x_i\to x_i^{-1}\right)\right)=
\nn\\
\sum_{i=1}^2\left(\prod\limits_{j\neq i}^2 \frac{\thf{\pq{1}{2}w^{\frac{32}{5}}\ta_{10}x_j^{\pm 1}}}{\thf{x_i}\thf{qx_i^2}\thf{x_ix_j^{\pm 1}}}
\prod\limits_{k=1}^{10}\thf{\pq{1}{2}w^{-\frac{32}{5}}\ta_k^{-1}x_i}\right.
\times\nn\\
\left.\frac{\thf{\pq{1}{2}w^{\frac{32}{5}}\ta_{10}x_i}}{\thf{\pq{1}{2}w^{\frac{32}{5}}\ta_{10}q^{-1}x_i^{-1}}}
\prod\limits_{l=1}^3\frac{\thf{w^{\frac{128}{15}}\ta_{10}^{-\frac{1}{3}}y_l^{-1}x_i}}
{\thf{w^{\frac{128}{15}}\ta_{10}^{-\frac{1}{3}}y_l^{-1}q^{-1}x_i^{-1}}}+\left(x_i\to x_i^{-1} \right)\right)\,.
\label{C2:action}
\ee

Now we consider l.h.s. of the equation above. Similarly we need to understand action of $\Delta_{lm}(y)$ operator on the
$A_2C_2$ tube \eqref{a2c2:kernel:property}:
\be
\Delta_{lm}(y)K_2^{A_2C_2}(x,y)=D_{lm}(x,y)K_2^{A_2C_2}(x,y)\,,
\ee
where
\be
D_{lm}(x,y)=\prod\limits_{j=1}^2\prod\limits_{k=1}^9
\frac{\thf{\ta_{10}^{-\frac{1}{3}}w^{\frac{128}{15}}x_j^{\pm 1}y_l^{-1}}}{\thf{q^{-1}\ta_{10}^{-\frac{1}{3}}w^{\frac{128}{15}}x_j^{\pm 1}y_m^{-1}}}
\frac{ \thf{\pq{1}{2}\ta_{10}^{\frac{1}{3}}w^{-\frac{32}{15}}\ta_k y_m}}{  \thf{\pq{1}{2}\ta_{10}^{\frac{1}{3}}w^{-\frac{32}{15}}\ta_k q^{-1} y_l}}
\times\nn\\
\frac{ \thf{\pq{1}{2}\ta_{10}^{-\frac{2}{3}}w^{-\frac{224}{15}} y_m}}{  \thf{\pq{1}{2}\ta_{10}^{-\frac{2}{3}}w^{-\frac{224}{15}} q^{-1}y_l}}
\frac{\thf{\ta_{10}^{-\frac{2}{3}}w^{\frac{256}{15}}q^{-1}y_l}}{\thf{\ta_{10}^{-\frac{2}{3}}w^{\frac{256}{15}}y_m}}
\ee
So the l.h.s. of \eqref{a2c2:kernel:property} takes the following form:
\be
&&\cO^{(A_2;\tth_{10}^{-1};1,0)}_{y}K_2^{A_2C_2}(x,y)=\left[F^{(A_2;\tth_{10};1,0)}(x,y)+W^{(A_2;\tth_{10}^{-1};1,0)}(y)\right]K_2^{A_2C_2}(x,y)\,,
\nn\\
&&F^{(A_2;\tth_{10};1,0)}(x,y)\equiv \sum\limits_{l\neq m}^3D_{lm}(x,y)A^{(A_2;\tth_{10};1,0)}_{lm}(y)=
\nn\\
&&\sum\limits_{l\neq m}^3\prod\limits_{k=1}^9\prod\limits_{j=1}^2
\frac{\thf{\pq{1}{2}w^{-\frac{32}{15}}\ta_{10}^{\frac{1}{3}}\ta_ky_m}}{\thf{q\frac{y_m}{y_l}}\thf{\frac{y_m}{y_l}}}
\frac{\thf{w^{\frac{128}{15}}\ta_{10}^{-\frac{1}{3}}y_l^{-1}x_j^{\pm 1}}\thf{w^{\frac{256}{15}}\ta_{10}^{-\frac{2}{3}}q^{-1}y_l}}
{\thf{w^{\frac{128}{15}}\ta_{10}^{-\frac{1}{3}}y_m^{-1}q^{-1}x_j^{\pm 1}} \thf{w^{\frac{256}{15}}\ta_{10}^{-\frac{2}{3}}y_m}}
\times\nn\\
&&\thf{\pq{1}{2}w^{-\frac{224}{15}}\ta_{10}^{-\frac{2}{3}}y_m}
\prod\limits_{i\neq m\neq l}^3\frac{\thf{\pq{1}{2}w^{\frac{224}{15}}\ta_{10}^{\frac{2}{3}}y_i^{-1}}
\thf{\pq{1}{2}w^{-\frac{32}{15}}\ta_{10}^{\frac{4}{3}}y_i}}
{\thf{\frac{y_i}{y_l}}\thf{\frac{y_m}{y_i}}}\,.
\label{A2:action}
\ee

Hence kernel property \eqref{a2c2:kernel:property} can be reduced to the following algebraic identity:
\be
F^{(A_2;\tth_{10};1,0)}(x,y)+W^{(A_2;\tth_{10}^{-1};1,0)}(y)=F^{(C_2;h_{10};1,0)}(x,y)+W^{(C_2;h_{10};1,0)}(x)
\label{ancn:kernel:algebraic}
\ee

It can be checked that both $F^{(A_2;\tth_{10};1,0)}(x,y)$ and $F^{(C_2;h_{10};1,0)}(x,y)$ functions
defined above are elliptic w.r.t both $x_i$ and $y_i$ variables with periods $1$ and $p$
\footnote{As usually when we discuss ellipticity w.r.t. $A_2$ variables we should
keep in mind the constraint so when we shift one of the variables, say $y_l\to py_l$ we
are forced to also make another shift $y_m\to p^{-1}y_m$ for some other variable. }.
In order to prove the identity we now need to check poles and residues in the fundamental domain on two sides of equation \footnote{
We have complicated functions of many variables. Hence when we speak about poles and corresponding residues we always consider
our functions as functions of one chosen variable while all other variables are kept fixed. The poles and residues should be
checked in this way for all variables of functions one by one.}. We already know the poles and residues
of the constant parts $W^{(A_2;\tth_{10}^{-1};1,0)}(y)$ and $W^{(C_2;h_{10};1,0)}(x)$ summarized in sections \ref{sec:an:oper} and
\ref{sec:c2:operator} correspondingly. Now let's study functions
$F^{(A_2;\tth_{10};1,0)}(x,y)$ and $F^{(C_2;h_{10};1,0)}(x,y)$ coming from the action of the shift parts.

We start with $F^{(A_2;\tth_{10};1,0)}(x,y)$ function given in \eqref{A2:action}. This elliptic
function appears to have poles at the following positions:
\be
y_a=w^{\frac{128}{15}}\ta_{10}^{-\frac{1}{3}}q^{-1}x_b^s\,,\quad y_a=q y_b\,,\quad y_a=q^{-1}y_b\,,\quad y_a=sq^{\frac{1}{2}}P_a^{-\frac{1}{2}}\,,\quad y_a=sq^{-\frac{1}{2}}P_a^{-\frac{1}{2}}\,,
\nn\\
y_a=sq^{\frac{1}{2}}p^{\frac{1}{2}}P_a^{-\frac{1}{2}}\,,\quad y_a=sq^{-\frac{1}{2}}p^{\frac{1}{2}}P_a^{-\frac{1}{2}}\,,\quad x_j=\left(w^{\frac{128}{15}}\ta_{10}^{-\frac{1}{3}}q^{-1}y_a^{-1}\right)^{\pm 1}\,,
\label{FA:poles}
\ee
where $s=\pm $ and $P_a\equiv \prod\limits_{j\neq a}^3 y_j$. Seemingly there are more poles in expression \eqref{A2:action} but checking all residues
 shows that the only actual poles are the ones listed above.

 Residues of poles \eqref{FA:poles} are given by
\begin{align}
&\mathrm{Res}_{y_a=w^{\frac{128}{15}}\ta_{10}^{-\frac{1}{3}}q^{-1}x_b^s}F^{(A_2;\tth_{10};1,0)}(x,y)=\frac{w^{\frac{128}{15}}
\ta_{10}^{-\frac{1}{3}}q^{-1}x_b^s}{\qPoc{p}{p}^2}
\sum\limits_{l\neq a}^3\prod\limits_{k=1}^9\frac{\thf{\pq{1}{2}w^{\frac{32}{5}}\ta_kq^{-1}x_b^s}}{\thf{w^{\frac{128}{5}}q_{10}^{-1}q^{-1}x_b^s}}
\times\nn\\
&\frac{\thf{w^{\frac{256}{15}}\ta_{10}^{-\frac{2}{3}}q^{-1}y_l}}{\thf{w^{\frac{128}{15}}\ta_{10}^{-\frac{1}{3}}q^{-1}y_l^{-1}x_b^s}}
\prod\limits_{i\neq l\neq a}^3\frac{\thf{\pq{1}{2}w^{\frac{224}{15}}\ta_{10}^{\frac{2}{3}}y_i^{-1}} \thf{\pq{1}{2}w^{-\frac{32}{15}}\ta_{10}^{\frac{4}{3}}y_i}}
{\thf{\frac{y_i}{y_l}}\thf{w^{\frac{128}{15}}q^{-1}\ta_{10}^{-\frac{1}{3}}y_i^{-1}x_b^s}}
\times\nn\\
&\prod\limits_{j\neq b}^2\frac{\thf{w^{\frac{128}{15}}\ta_{10}^{-\frac{1}{3}}y_l^{-1}x_j^{\pm 1}}\thf{w^{\frac{128}{15}}\ta_{10}^{-\frac{1}{3}}y_l^{-1}x_b^{-s}}}
{\thf{x_b^{-s}x_j^{\pm 1} }\thf{x_b^{-2s}}}\thf{\pq{1}{2}w^{-\frac{32}{15}}\ta_{10}^{-1}q^{-1}x_b^s }\,,
\end{align}
\begin{align}
&\mathrm{Res}_{x_j=\left(w^{\frac{128}{15}}\ta_{10}^{-\frac{1}{3}}q^{-1}y_a^{-1}\right)^{s}}F^{(A_2;\tth_{10};1,0)}(x,y)=
s\frac{\left(w^{\frac{128}{15}}\ta_{10}^{-\frac{1}{3}}q^{-1}y_a^{-1}\right)^{s}}{\qPoc{p}{p}^2}
\times\nn\\
&\sum_{l\neq a}^3\prod\limits_{k=1}^9\prod\limits_{i\neq j}^2\frac{\thf{\pq{1}{2}w^{-\frac{32}{15}}\ta_{10}^{\frac{1}{3}}\ta_ky_a}}{\thf{q\frac{y_a}{y_l}}\thf{\frac{y_a}{y_l}}}
\frac{\thf{w^{\frac{128}{15}}\ta_{10}^{-\frac{1}{3}}y_l^{-1}x_i^{\pm 1}}}{\thf{w^{\frac{128}{15}}\ta_{10}^{-\frac{1}{3}}y_a^{-1}q^{-1}x_i^{\pm 1}}}
\times\nn\\
&\frac{\thf{w^{\frac{256}{15}}\ta_{10}^{-\frac{2}{3}}q^{-1}y_l^{-1}y_a^{-1}} \thf{w^{\frac{256}{15}}\ta_{10}^{-\frac{2}{3}}q^{-1}y_l}}
{\thf{w^{\frac{256}{15}}\ta_{10}^{-\frac{2}{3}}q^{-2}y_a^{-2}}\thf{w^{\frac{256}{15}}\ta_{10}^{-\frac{2}{3}}y_a}}
\thf{\pq{1}{2}w^{-\frac{224}{15}}\ta_{10}^{-\frac{2}{3}}y_a}
\times\nn\\
&\thf{qy_l^{-1}y_a}\prod\limits_{k\neq a\neq l}^3\frac{\thf{\pq{1}{2}w^{\frac{224}{15}}\ta_{10}^{\frac{2}{3}}y_k^{-1}}\thf{\pq{1}{2}w^{-\frac{32}{15}}\ta_{10}^{\frac{4}{3}}y_k}}
{\thf{\frac{y_k}{y_l}} \thf{\frac{y_a}{y_k}}}
\end{align}
\begin{align}
&\mathrm{Res}_{y_a=qy_b}F^{(A_2;\tth_{10};1,0)}(x,y)=\frac{qy_b}{\thf{q^{-1}}\qPoc{p}{p}^2}
\prod\limits_{k=1}^9\thf{\pq{1}{2}w^{-\frac{32}{15}}\ta_{10}^{\frac{1}{3}}\ta_ky_b}
\times\nn\\
&\thf{\pq{1}{2}w^{-\frac{224}{15}}\ta_{10}^{-\frac{2}{3}}y_b}\prod\limits_{j\neq a\neq b}^3\frac{\thf{\pq{1}{2}w^{\frac{224}{15}}\ta_{10}^{\frac{2}{3}}y_j^{-1}}\thf{\pq{1}{2}w^{-\frac{32}{15}}\ta_{10}^{\frac{4}{3}}y_j}}
{\thf{q^{-1}\frac{y_j}{y_b}}\thf{\frac{y_b}{y_j}}}\,,
\end{align}
\begin{align}
&\mathrm{Res}_{y_a=q^{-1}y_b}F^{(A_2;\tth_{10};1,0)}(x,y)=-\frac{q^{-1}y_b}{\thf{q^{-1}}\qPoc{p}{p}^2}\prod\limits_{k=1}^9
\thf{\pq{1}{2}w^{\frac{32}{15}}\ta_{10}^{-\frac{1}{3}}\ta_k^{-1}y_b^{-1}}
\times\nn\\
&\thf{\pq{1}{2}w^{\frac{224}{15}}\ta_{10}^{\frac{2}{3}}y_b^{-1}}\prod\limits_{j\neq a\neq b}^3\frac{\thf{\pq{1}{2}w^{\frac{224}{15}}\ta_{10}^{\frac{2}{3}}y_j^{-1}}\thf{\pq{1}{2}w^{-\frac{32}{15}}\ta_{10}^{\frac{4}{3}}y_j}}
{\thf{\frac{y_j}{y_b}}\thf{q^{-1}\frac{y_b}{y_j}}}\,,
\end{align}
\begin{align}
&\mathrm{Res}_{y_a=sq^{\frac{1}{2}}P_a^{-\frac{1}{2}}}F^{(A_2;\tth_{10};1,0)}(x,y)=
s\frac{q^{\frac{1}{2}}P_a^{-\frac{1}{2}}}{2\thf{q^{-1}}\qPoc{p}{p}^2}
\prod\limits_{k=1}^9\thf{sp^{\frac{1}{2}}w^{-\frac{32}{15}}\ta_{10}^{\frac{1}{3}}\ta_kP_a^{-\frac{1}{2}}}
\times\nn\\
&\thf{sp^{\frac{1}{2}}w^{-\frac{224}{15}}\ta_{10}^{-\frac{2}{3}}P_a^{-\frac{1}{2}}}
\frac{\thf{\pq{1}{2} w^{\frac{224}{15}}\ta_{10}^{\frac{2}{3}}P_a^{-1}}\thf{\pq{1}{2} w^{-\frac{32}{15}}\ta_{10}^{\frac{4}{3}}P_a}}{\thf{sq^{-\frac{1}{2}}P_a^{\frac{3}{2}}}
\thf{sq^{-\frac{1}{2}}P_a^{-\frac{3}{2}}}}\,,
\end{align}
\begin{align}
&\mathrm{Res}_{y_a=sq^{-\frac{1}{2}}P_a^{-\frac{1}{2}}}F^{(A_2;\tth_{10};1,0)}(x,y)=
-s\frac{q^{-\frac{1}{2}}P_a^{-\frac{1}{2}}}{2\thf{q^{-1}}\qPoc{p}{p}^2}
\prod\limits_{k=1}^9\thf{sp^{\frac{1}{2}}w^{-\frac{32}{15}}\ta_{10}^{\frac{1}{3}}\ta_kP_a^{-\frac{1}{2}}}
\times\nn\\
&\thf{sp^{\frac{1}{2}}w^{-\frac{224}{15}}\ta_{10}^{-\frac{2}{3}}P_a^{-\frac{1}{2}}}
\frac{\thf{\pq{1}{2} w^{\frac{224}{15}}\ta_{10}^{\frac{2}{3}}P_a^{-1}}\thf{\pq{1}{2} w^{-\frac{32}{15}}\ta_{10}^{\frac{4}{3}}P_a}}{\thf{sq^{-\frac{1}{2}}P_a^{\frac{3}{2}}}\thf{sq^{-\frac{1}{2}}P_a^{-\frac{3}{2}}}}\,,
\end{align}
\begin{align}
&\mathrm{Res}_{y_a=sq^{\frac{1}{2}}p^{\frac{1}{2}}P_a^{-\frac{1}{2}}}F^{(A_2;\tth_{10};1,0)}(x,y)=
s\frac{p^{\frac{3}{2}}q^{\frac{1}{2}}P_a^{\frac{1}{2}}w^{\frac{256}{15}}\ta_{10}^{-\frac{2}{3}}}{2\thf{q^{-1}}\qPoc{p}{p}^2}
\prod\limits_{k=1}^9\thf{sw^{-\frac{32}{15}}\ta_{10}^{\frac{1}{3}}\ta_kP_a^{-\frac{1}{2}}}
\times\nn\\
&\thf{sw^{-\frac{224}{15}}\ta_{10}^{-\frac{2}{3}}P_a^{-\frac{1}{2}}}
\frac{\thf{\pq{1}{2} w^{\frac{224}{15}}\ta_{10}^{\frac{2}{3}}P_a^{-1}}\thf{\pq{1}{2} w^{-\frac{32}{15}}\ta_{10}^{\frac{4}{3}}P_a}}
{\thf{sq^{-\frac{1}{2}}p^{\frac{1}{2}}P_a^{\frac{3}{2}}}\thf{sq^{-\frac{1}{2}}p^{-\frac{1}{2}}P_a^{-\frac{3}{2}}}}\,,
\end{align}
\begin{align}
&\mathrm{Res}_{y_a=sq^{-\frac{1}{2}}p^{\frac{1}{2}}P_a^{-\frac{1}{2}}}F^{(A_2;\tth_{10};1,0)}(x,y)=
-s\frac{p^{\frac{3}{2}}q^{-\frac{1}{2}}P_a^{\frac{1}{2}}w^{\frac{256}{15}}\ta_{10}^{-\frac{2}{3}}}{2\thf{q^{-1}}\qPoc{p}{p}^2}
\prod\limits_{k=1}^9\thf{sw^{-\frac{32}{15}}\ta_{10}^{\frac{1}{3}}\ta_kP_a^{-\frac{1}{2}}}
\times\nn\\
&\thf{sw^{-\frac{224}{15}}\ta_{10}^{-\frac{2}{3}}P_a^{-\frac{1}{2}}}
\frac{\thf{\pq{1}{2} w^{\frac{224}{15}}\ta_{10}^{\frac{2}{3}}P_a^{-1}}\thf{\pq{1}{2} w^{-\frac{32}{15}}\ta_{10}^{\frac{4}{3}}P_a}}
{\thf{sq^{-\frac{1}{2}}p^{\frac{1}{2}}P_a^{\frac{3}{2}}}\thf{sq^{-\frac{1}{2}}p^{-\frac{1}{2}}P_a^{-\frac{3}{2}}}}\,,
\end{align}

Looking on these residues we immediately see that on the l.h.s. of \eqref{ancn:kernel:algebraic} there are vast cancellation of residues between $F^{(A_2;\tth_{10};1,0)}(x,y)$ and constant part $W^{(A_2;\tth_{10}^{-1};1,0)}(y)$. Latter ones can be found from \eqref{an:residues} upon charges identification given in \eqref{ancn:dict}.
In particular the poles at  $y_a=q y_b\,, y_a=q^{-1}y_b\,,$ $y_a=sq^{\frac{1}{2}}P_a^{-\frac{1}{2}}\,,$ $y_a=sq^{-\frac{1}{2}}P_a^{-\frac{1}{2}}\,,$
$y_a=sq^{\frac{1}{2}}p^{\frac{1}{2}}P_a^{-\frac{1}{2}}\,,$ $y_a=sq^{-\frac{1}{2}}p^{\frac{1}{2}}P_a^{-\frac{1}{2}}\,,$ get cancelled. Hence we are left with the following poles and residues on the l.h.s. of  \eqref{ancn:kernel:algebraic}:
\begin{shaded}
\begin{align}
&\mathrm{Res}_{y_a=w^{\frac{128}{15}}\ta_{10}^{-\frac{1}{3}}q^{-1}x_b^s}
\left[F^{(A_2;\tth_{10};1,0)}(x,y)+W^{(A_2;\tth_{10}^{-1};1,0)}(y)\right]=\frac{w^{\frac{128}{15}}\ta_{10}^{-\frac{1}{3}}q^{-1}x_b^s}{\qPoc{p}{p}^2}
\times\nn\\
&\sum\limits_{l\neq a}^3\prod\limits_{k=1}^9\frac{\thf{\pq{1}{2}w^{\frac{32}{5}}\ta_kq^{-1}x_b^s}}{\thf{w^{\frac{128}{5}}\ta_{10}^{-1}q^{-1}x_b^s}}
\frac{\thf{w^{\frac{256}{15}}\ta_{10}^{-\frac{2}{3}}q^{-1}y_l}}{\thf{w^{\frac{128}{15}}\ta_{10}^{-\frac{1}{3}}q^{-1}y_l^{-1}x_b^s}}
\times\nn\\
&\prod\limits_{i\neq l\neq a}^3\frac{\thf{\pq{1}{2}w^{\frac{224}{15}}\ta_{10}^{\frac{2}{3}}y_i^{-1}} \thf{\pq{1}{2}w^{-\frac{32}{15}}\ta_{10}^{\frac{4}{3}}y_i}}
{\thf{\frac{y_i}{y_l}}\thf{w^{\frac{128}{15}}q^{-1}\ta_{10}^{-\frac{1}{3}}y_i^{-1}x_b^s}}
\times\nn\\
&\prod\limits_{j\neq b}^2\frac{\thf{w^{\frac{128}{15}}\ta_{10}^{-\frac{1}{3}}y_l^{-1}x_j^{\pm 1}}\thf{w^{\frac{128}{15}}\ta_{10}^{-\frac{1}{3}}y_l^{-1}x_b^{-s}}}
{\thf{x_b^{-s}x_j^{\pm 1} }\thf{x_b^{-2s}}}\thf{\pq{1}{2}w^{-\frac{32}{15}}\ta_{10}^{-1}q^{-1}x_b^s }\,,
\label{A2:res:1}
\end{align}
\begin{align}
\mathrm{Res}_{x_j=\left(w^{\frac{128}{15}}\ta_{10}^{-\frac{1}{3}}q^{-1}y_a^{-1}\right)^{s}}
\left[F^{(A_2;\tth_{10};1,0)}(x,y)+W^{(A_2;\tth_{10}^{-1};1,0)}(y)\right]
=s\frac{\left(w^{\frac{128}{15}}\ta_{10}^{-\frac{1}{3}}q^{-1}y_a^{-1}\right)^{s}}{\qPoc{p}{p}^2}
\times\nn\\
\sum_{l\neq a}^3\prod\limits_{k=1}^9\prod\limits_{i\neq j}^2\frac{\thf{\pq{1}{2}w^{-\frac{32}{15}}\ta_{10}^{\frac{1}{3}}\ta_ky_a}}{\thf{q\frac{y_a}{y_l}}\thf{\frac{y_a}{y_l}}}
\frac{\thf{w^{\frac{128}{15}}\ta_{10}^{-\frac{1}{3}}y_l^{-1}x_i^{\pm 1}}}{\thf{w^{\frac{128}{15}}\ta_{10}^{-\frac{1}{3}}y_a^{-1}q^{-1}x_i^{\pm 1}}}
\times\nn\\
\frac{\thf{w^{\frac{256}{15}}\ta_{10}^{-\frac{2}{3}}q^{-1}y_l^{-1}y_a^{-1}} \thf{w^{\frac{256}{15}}\ta_{10}^{-\frac{2}{3}}q^{-1}y_l}}
{\thf{w^{\frac{256}{15}}\ta_{10}^{-\frac{2}{3}}q^{-2}y_a^{-2}}\thf{w^{\frac{256}{15}}\ta_{10}^{-\frac{2}{3}}y_a}}
\thf{\pq{1}{2}w^{-\frac{224}{15}}\ta_{10}^{-\frac{2}{3}}y_a}
\times\nn\\
\thf{qy_l^{-1}y_a}\prod\limits_{k\neq a\neq l}^3\frac{\thf{\pq{1}{2}w^{\frac{224}{15}}\ta_{10}^{\frac{2}{3}}y_k^{-1}}\thf{\pq{1}{2}w^{-\frac{32}{15}}\ta_{10}^{\frac{4}{3}}y_k}}
{\thf{\frac{y_k}{y_l}} \thf{\frac{y_a}{y_k}}}
\label{A2:res:2}
\end{align}
\end{shaded}

Now let's move to the $C_2$ side of the kernel function equation \eqref{ancn:kernel:algebraic}.
As mentioned previously the function  $F^{(C_2;\tth_{10};1,0)}(x,y)$ defined in \eqref{C2:action}
 is elliptic with periods $1$ and $p$.
In the fundamental domain the poles of the function are located at

\be
y_a=w^{\frac{128}{15}}\ta_{10}^{-\frac{1}{3}}q^{-1}x_b^s\,,\quad x_i=sq^{\pm\frac{1}{2}}\,,\quad x_i=sq^{\pm\frac{1}{2}}p^{\frac{1}{2}}\,,
\quad  x_j=\left(w^{\frac{128}{15}}\ta_{10}^{-\frac{1}{3}}q^{-1}y_a^{-1}\right)^{\pm 1}\,,
\label{FA:poles2}
\ee
where as usually $s=\pm 1$. The residues at the poles $x_i=sq^{\pm\frac{1}{2}}$ and $x_i=sq^{\pm\frac{1}{2}}p^{\frac{1}{2}}$ are
given by
\begin{align}
&\mathrm{Res}_{x_i=sq^{\frac{1}{2}}}F^{(C_2;h_{10};1,0)}(x,y)=s\frac{q^{\frac{1}{2}}\prod\limits_{k=1}^{10}
\thf{\pq{1}{2}sq^{-\frac{1}{2}}w^{-\frac{32}{5}}\ta_k^{-1}}}
{2\qPoc{p}{p}^2\thf{q^{-1}}}\prod\limits_{j\neq i}^2
\frac{\thf{\pq{1}{2} w^{\frac{32}{5}}\ta_{10} x_j^{\pm 1}}}{\thf{sq^{-\frac{1}{2}}x_j^{\pm 1} }}\,,
\nn\\
&\mathrm{Res}_{x_i=sq^{-\frac{1}{2}}}F^{(C_2;h_{10};1,0)}(x,y)=-s\frac{q^{-\frac{1}{2}}
\prod\limits_{k=1}^{10}\thf{\pq{1}{2}sq^{-\frac{1}{2}}w^{-\frac{32}{5}}\ta_k^{-1}}}
{2\qPoc{p}{p}^2\thf{q^{-1}}}\prod\limits_{j\neq i}^2
\frac{\thf{\pq{1}{2} w^{\frac{32}{5}}\ta_{10} x_j^{\pm 1}}}{\thf{sq^{-\frac{1}{2}}x_j^{\pm 1} }}\,,
\nn\\
&\mathrm{Res}_{x_i=sp^{\frac{1}{2}}q^{\frac{1}{2}}}F^{(C_2;h_{10};1,0)}(x,y)=s\frac{pw^{-32}\prod\limits_{k=1}^{10}\thf{sw^{\frac{32}{5}}\ta_k} }
{2\qPoc{p}{p}^2\thf{q^{-1}}}\prod\limits_{j\neq i}^2 \frac{\thf{\pq{1}{2}w^{\frac{32}{5}}\ta_{10}x_j^{\pm 1}}}{\thf{\pqm{1}{2}sx_j^{\pm 1} }}\,,
\nn\\
&\mathrm{Res}_{x_i=sp^{\frac{1}{2}}q^{-\frac{1}{2}}}F_C^{(C_2;h_{10};1,0)}(x,y)=-s\frac{pq^{-1}w^{-32}
\prod\limits_{k=1}^{10}\thf{sw^{\frac{32}{5}}\ta_k} }
{2\qPoc{p}{p}^2\thf{q^{-1}}}\prod\limits_{j\neq i}^2 \frac{\thf{\pq{1}{2}w^{\frac{32}{5}}\ta_{10}x_j^{\pm 1}}}{\thf{\pqm{1}{2}sx_j^{\pm 1} }}\,.
\nn\\
\label{FC:residues1}
\end{align}
It can be easily seen that these residues are cancelled by the corresponding residues
\eqref{c2:residues} of the constant part $W^{(C_2;h_{10};1,0)}$. Hence
$C_2$ side of kernel function equation \eqref{ancn:kernel:algebraic} does not have poles in $x_i=sq^{\pm \frac{1}{2}}$
and $x_i=sq^{\pm \frac{1}{2}}p^{\frac{1}{2}}$.
The only poles are the same as on the $A_2$ side and are located in $y_a=w^{\frac{128}{15}}\ta_{10}^{-\frac{1}{3}}q^{-1}x_b^s$ and
$x_j=\left(w^{\frac{128}{15}}\ta_{10}^{-\frac{1}{3}}q^{-1}y_a^{-1}\right)^{\pm 1}$.
Corresponding residues of $F^{(C_2;h_{10};1,0)}$ and hence of the full function
$F^{(C_2;h_{10};1,0)}+W^{(C_2;h_{10};1,0)}$ are given by
\begin{shaded}
\begin{align}
&\mathrm{Res}_{y_a=w^{\frac{128}{15}}\ta_{10}^{-\frac{1}{3}}q^{-1}x_b^s}
\left[F^{(C_2;\ta_{10};1,0)}(x,y)+W^{(C_2;\ta_{10};1,0)}(x)\right]=\frac{w^{\frac{128}{15}}\ta_{10}^{-\frac{1}{3}}q^{-1}x_b^s}{\qPoc{p}{p}^2}
\times\nn\\
&\prod\limits_{j\neq b}^2\frac{\thf{\pq{1}{2}w^{\frac{32}{5}}\ta_{10}x_j^{\pm 1}}}{\thf{x_b^{-2s}}\thf{x_b^{-s}x_j^{\pm 1}}}
\prod\limits_{k=1}^9\thf{\pq{1}{2}w^{-\frac{32}{5}}\ta_k^{-1}x_b^{-s}}\thf{\pq{1}{2}w^{\frac{32}{5}}\ta_{10}x_b^{-s}}
\times\nn\\
&\hspace{8cm}\prod\limits_{l\neq a}^3\frac{\thf{w^{\frac{128}{15}}\ta_{10}^{-\frac{1}{3}}y_l^{-1}x_b^{-s}}}
{\thf{w^{\frac{128}{15}}\ta_{10}^{-\frac{1}{3}}q^{-1}y_l^{-1}x_b^{-s}} }\,,
\label{C2:res:1}
\end{align}
\begin{align}
&\mathrm{Res}_{x_j=\left(w^{\frac{128}{15}}\ta_{10}^{-\frac{1}{3}}q^{-1}y_a^{-1}\right)^{s}}
\left[F^{(C_2;\ta_{10};1,0)}(x,y)+W^{(C_2;\ta_{10};1,0)}(x)\right]
=s\frac{\left(w^{\frac{128}{15}}\ta_{10}^{-\frac{1}{3}}q^{-1}y_a^{-1}\right)^{s}}{\qPoc{p}{p}^2}
\times\nn\\
&\prod\limits_{j\neq a}^2 \frac{\thf{\pq{1}{2}w^{\frac{32}{5}}\ta_{10}x_j^{\pm 1} }}
{\thf{w^{\frac{256}{15}}\ta_{10}^{-\frac{2}{3}}q^{-2}y_b } \thf{w^{\frac{128}{15}}\ta_{10}^{-\frac{1}{3}}q^{-1}y_b^{-1}x_j^{\pm 1} }}
\prod\limits_{k=1}^{10}\thf{\pq{1}{2} w^{\frac{32}{15}}\ta_{10}^{-\frac{1}{3}}\ta_k^{-1}q^{-1}y_b^{-1}}
\times\nn\\
&\hspace{4cm}\prod\limits_{l\neq b}^3\frac{\thf{w^{\frac{256}{15}}\ta_{10}^{-\frac{2}{3}}q^{-1}y_l^{-1}y_b^{-1}}}{\thf{\frac{y_b}{y_l}}}
	\frac{\thf{\pq{1}{2}w^{\frac{224}{15}}\ta_{10}^{\frac{2}{3}}q^{-1}y_b^{-1}}}{\thf{\pq{1}{2}w^{-\frac{32}{15}}\ta_{10}^{\frac{4}{3}}y_b}}\,.
\label{C2:res:2}
\end{align}
\end{shaded}
As the next step we have to show that the residues in \eqref{C2:res:1} and \eqref{C2:res:2} coincide with the residues \eqref{A2:res:1} and \eqref{A2:res:2} in order for the kernel identity  \eqref{ancn:kernel:algebraic} to work.  Notice that in both pairs \eqref{C2:res:1},\eqref{C2:res:2} and \eqref{A2:res:1}, \eqref{A2:res:2} position of the poles are defined by a single equation
\be
w^{\frac{128}{15}}\ta_{10}^{-\frac{1}{3}}q^{-1}x_b^s y_a^{-1}=1
\ee
The two residues in each of the pairs differ by the choice of the variables with respect to which we compute this residue. It is either fixing $x_a$ variable and computing residue in  $y_a$ variable (like in \eqref{A2:res:1} and \eqref{C2:res:1}) or vice versa (like in \eqref{A2:res:2} and \eqref{C2:res:2}).
Hence it is sufficient to prove equality of residues only in one of the variables. It then authomatically works 
for another variable since it is the very same pole. 
Let's choose to study residues w.r.t. $x_b$
\footnote{As a crosscheck we have also checked the residue w.r.t. $y_a$}. Also since all expressions are
symmetric w.r.t. $x_i\to x_i^{-1}$ we can perform the check only for one of the signs $s$. The other
one will work automatically. So let's perform the check only for the residue at $x_a=w^{\frac{128}{15}}\ta_{10}^{-\frac{1}{3}}q^{-1}y_b^{-1}$
Subtracting residue \eqref{A2:res:2} from the residue \eqref{C2:res:2} we can obtain the following equation:
\begin{align}
&\mathrm{Res}_{x_a=w^{\frac{128}{15}}\ta_{10}^{-\frac{1}{3}}q^{-1}y_b^{-1}}\left[F^{(C_2;h_{10};1,0)}(x,y)+
W^{(C_2;h_{10};1,0)}(x)-F^{(A_2,\tth_{10};1,0)}(x,y)
\right.\nn\\&\left.
-W^{(A_2;\tth_{10}^{-1};1,0)}(y)\right]=
\frac{w^{\frac{128}{15}}\ta_{10}^{-\frac{1}{3}}y_b^{-1}q^{-1}}{\qPoc{p}{p}^2}\prod\limits_{l\neq b}^3\frac{\thf{w^{\frac{256}{15}}\ta_{10}^{-\frac{2}{3}}q^{-1}y_l}}{\thf{\frac{y_b}{y_l}}}
\frac{\thf{\pq{1}{2}w^{\frac{224}{15}}\ta_{10}^{\frac{2}{3}}y_b}}{\thf{w^{\frac{256}{15}}\ta_{10}^{-\frac{2}{3}}q^{-2}y_b^{-2}}}
\times\nn\\
&\hspace{5cm}\prod\limits_{j\neq a}^2\frac{\thf{\pq{1}{2}w^{\frac{32}{5}}\ta_{10}x_j^{\pm 1}}}
{\thf{w^{\frac{128}{15}}\ta_{10}^{-\frac{1}{3}}q^{-1}y_b^{-1}x_j^{\pm 1}}}\left[ 1-M_{ab}(x,y) \right]\,,
\label{residue:diff}
\end{align}
where we have introduced  an auxiliary function
\begin{align}
M_{ab}(x,y)&=\sum\limits_{l\neq b}^3\prod\limits_{j\neq a}^2\frac{\thf{w^{\frac{128}{15}}\ta_{10}^{-\frac{1}{3}}y_l^{-1}x_j^{\pm 1} } }
{\thf{\pq{1}{2}w^{\frac{32}{5}}\ta_{10}x_j^{\pm 1} } }\times
\nn\\
&\hspace{2cm}\prod\limits_{k\neq l\neq b}^3
\frac{\thf{\pq{1}{2}w^{\frac{224}{15}}\ta_{10}^{\frac{2}{3}}y_k^{-1} }\thf{\pq{1}{2}w^{-\frac{32}{15}}\ta_{10}^{\frac{4}{3}}y_k }}
{\thf{\frac{y_k}{y_l}}\thf{ w^{\frac{256}{15}}\ta_{10}^{-\frac{2}{3}}y_b}  }\,.
\label{M:function}
\end{align}
It can be checked that this function is elliptic with periods $1$ and $p$. Seemingly this function has poles at
$x_j=\left[\pq{1}{2}w^{\frac{32}{5}}\ta_{10}\right]^{\pm 1}$, $y_a=y_c$  with $a,c\neq b$ and $y_b=w^{-\frac{256}{15}}\ta_{10}^{\frac{2}{3}}$.
However as it often happens in our calculations accurate derivation of the corresponding residues results in zeros for all of them\footnote{To calculate these
residues it is crucial to use the $A_2$ constraint $\prod\limits_{j=1}^3y_j=1$ for $y$ variables.}. Hence these are not really
poles and our function $M(x,y)$ is an entire function for all its variables. Since function is entire elliptic function we
conclude that it is just a constant.

In order to define which particular constant it is we just need to put in some values for $x$ and $y$ variables. Here it is worth noticing that $M(x,y)$ function is the sum of
two products of $\theta$ function. Without loss of generality let's choose the following values of $y$ variables:
\be
y_c=\pqm{1}{2}w^{\frac{32}{15}}\ta_{10}^{-\frac{4}{3}}\,,\quad y_d=\pq{1}{2}w^{-\frac{32}{15}}\ta_{10}^{\frac{4}{3}}y_b^{-1}\,,
\label{y:special:M}
\ee
where the $c$ and $d$ indices are any indices not equal to $b$. The second equality, i.e. value of $y_d$
follows from the first one and
$A_2$ constraint for $y_j$ variables. Now if we look on the definition \eqref{M:function} of $M(x,y)$
we see the sum of two terms. First term is when $l=c\,, k=d$ and the second term
is vice versa when $l=d\,, k=c$. However we immediately see that if
$y_k=\pqm{1}{2}w^{\frac{32}{15}}\ta_{10}^{-\frac{4}{3}}$ in \eqref{M:function} we obtain zero of theta function. Hence
this second term in the sum is automatically zero and that was the idea behind  our choice \eqref{y:special:M}.
We are now left only with the first term. Substituting chosen values into this term we see
vast cancellations leading to
\be
\left.M_{ab}(x,y)\right|_{y_c=\pqm{1}{2}w^{\frac{32}{15}}\ta_{10}^{-\frac{4}{3}}}=1\,,
\ee
and hence the $M_{ab}(x,y)$ function given in \eqref{M:function} is just $1$, i.e. $M_{ab}(x,y)=1$.
Notice that to find the value of the constant $M_{ab}(x,y)$ we fixed only one of $y$ variables.
Other $y$'s as well as all of $x$ stayed random.
Using our finding  from \eqref{residue:diff} we immediately conclude that
\begin{align}
\mathrm{Res}_{x_a=w^{\frac{128}{15}}\ta_{10}^{-\frac{1}{3}}q^{-1}y_b^{-1}}
\left[F^{(C_2;h_{10};1,0)}(x,y)+W^{(C_2;h_{10};1,0)}(x)\right]=
\nn\\
\mathrm{Res}_{x_a=w^{\frac{128}{15}}\ta_{10}^{-\frac{1}{3}}q^{-1}y_b^{-1}}
\left[F^{(A_2;\tth_{10};1,0)}(x,y)+W^{(A_2;\tth_{10}^{-1};1,0)}(y)\right]\,.
\end{align}

Now since both functions $F^{(C_2;h_{10};1,0)}(x,y)+W^{(C_2;h_{10};1,0)}(x)$ and $F^{(A_2;\tth_{10};1,0)}(x,y)+W^{(A_2;\tth_{10}^{-1};1,0)}(y)$ are elliptic with the coinciding
poles and residues in the fundamental domain we conclude that they can differ at most by additive constant. And just as we did it with $M_{ab}(x,y)$ function
in order to find this constant we need to choose convenient values of $x$ and $y$ variables. Here it is a bit harder to do then in the case of $M_{ab}$ function.
However first thing we can notice is that constant part $W^{(C_2;h_{10};1,0)}(x)$ of $C_2$ operator
\eqref{c2:operator:final} has zero at the following value of variables:
\be
x_1=\pq{1}{2}w^{\frac{32}{5}}\ta_{10}\,,
\label{c2:zero}
\ee
where we have chosen $x_1$ variable without loss of generality (it can as well be $x_2$ variable) and the other $x_2$ variable is kept arbitrary.
At the same time constant part $W^{(A_2;\tth_{10}^{-1};1,0)}(y)$ of $A_2$ operator given in \eqref{sun:op:noflip:gen:const}
has zero at the following value of variables:
\be
y_1=\pq{1}{2}w^{\frac{224}{15}}\ta_{10}^{\frac{2}{3}}\,,\quad y_2=\pqm{1}{2}w^{\frac{32}{15}}\ta_{10}^{-\frac{4}{3}}\,,\quad y_3=w^{-\frac{256}{15}}\ta_{10}^{\frac{2}{3}}\,.
\label{a2:zero}
\ee
It is natural to try these values and check what happens with the shift parts $F^{(A_2;\tth_{10};1,0)}(y)$
and $F^{(C_2;h_{10};1,0)}(x)$ on two sides of the
kernel identity \eqref{ancn:kernel:algebraic}. On the $C_2$ side it is
straightforward to see from \eqref{C2:action}  that simultaneous substitution of values
specified in \eqref{a2:zero} and \eqref{c2:zero} leads to zero of the shift part $F^{(C_2;h_{10};1,0)}(x,y)$. With the $A_2$ side it is a bit more tricky since
the shift part $F^{(A_2;\tth_{10};1,0)}(x,y)$ seemingly has singularity at $y_3=w^{-\frac{256}{15}}\ta_{10}^{\frac{2}{3}}$ due to theta function
$\thf{w^{\frac{256}{15}}\ta_{10}^{-\frac{2}{3}}}$ in the denominator of the expression. However accurate analysis shows that this singularity is always cancelled by
zeros of other theta functions standing in the numerator of the expression. Taking this into account it is once again pretty easy to show that shift part on $A_2$ side of the
kernel property also has zero at values \eqref{a2:zero} and \eqref{c2:zero}.

Summarizing we have shown that expressions on two sides of the kernel property equation \eqref{ancn:kernel:algebraic} are elliptic functions with periods $1$ and $p$ and
same sets of poles and residues in the fundamental domain given in \eqref{A2:res:1}, \eqref{A2:res:2}, \eqref{C2:res:1} and \eqref{C2:res:2}.
This means that the two functions differ at most by constant. To fix this constant we also notice that
both function on two sides of the equation have zero at the same values of variables given in \eqref{a2:zero} and \eqref{c2:zero}. Hence the constant
two functions can differ by is just zero and the functions appears to be the same. This concludes the proof of the kernel property \eqref{a2c2:kernel:property}.

\section{Commutators of $C_2$ operators}
\label{app:commutators}

In this appendix we discuss commutation relations \eqref{c2:basic:comm} of the basic $C_2$ operators
\eqref{c2:operator:final}. Let's start with checking the following commutation relation:
\be
\left[\cO_x^{(C_2;h_a;1,0)},\,\cO_x^{(C_2;h_b;0,1)}\right]=0\,,\quad \forall ~ ~ a,b=1,\ldots,10\,.
\ee
Using explicit expression \eqref{c2:operator:final} we can write out all the terms of the commutator:
\be
&&\left[\cO_x^{(C_2;h_a;1,0)},\,\cO_x^{(C_2;h_b;0,1)}\right]\cI(x)=
\nn\\
&&\sum\limits_{i,j=1}^2\left[A_i^{\left(C_2;h_a;1,0\right)}(x_i)
\left(\Delta_q(x_i) A_j^{\left(C_2;h_b;0,1\right)}(x) \right)
\Delta_q(x_i)\Delta_p(x_j)\cI(x_i, x_j)+
\right.\nn\\
&&A_i^{\left(C_2;h_a;1,0\right)}(x)\left(\Delta_q(x_i) A_j^{\left(C_2;h_b;0,1\right)}\left(x^{-1}\right) \right)\Delta_q(x_i)\Delta_p^{-1}(x_j)\cI(x)+
\nn \\
&&A_i^{\left(C_2;h_a;1,0\right)}\left(x^{-1}\right)\left(\Delta^{-1}_q(x_i) A_j^{\left(C_2;h_b;0,1\right)}(x) \right)\Delta^{-1}_q(x_i)\Delta_p(x_j)\cI(x)+
\nn\\
&&\left.A_i^{\left(C_2;h_a;1,0\right)}\left(x^{-1}\right)\left(\Delta^{-1}_q(x_i) A_j^{\left(C_2;h_b;0,1\right)}\left(x^{-1}\right) \right)\Delta^{-1}_q(x_i)\Delta^{-1}_p(x_j)\cI(x)\right]+
\nn\\
&&\sum\limits_{i=1}^2\left[A_i^{\left(C_2;h_a;1,0\right)}(x)\left( \Delta_q(x_i)W^{\left(C_2;h_b;0,1\right)}(x)-W^{\left(C_2;h_b;0,1\right)}(x)\right)\Delta_q(x_i)\cI(x)+
 \right.\nn\\
 &&\left.
 A_i^{\left(C_2;h_b;0,1\right)}\left(x^{-1}\right)\left( \Delta^{-1}_p(x_i)W^{\left(C_2;h_a;1,0\right)}(x)-W^{\left(C_2;h_a;1,0\right)}(x)\right) \Delta^{-1}_q(x_i)\cI(x)\right]-
 \nn\\
&&
\hspace{11cm}\left(
\begin{array}{c}
p\leftrightarrow q \\
a \leftrightarrow b
\end{array}
\right)\,,
\label{commutator:1:prem}
\ee
where in the last line we substract all the terms written in the first six lines but with $p$ and $q$ parameters  as well as $a$ and $b$ indices exchanged.
Now in order to  compute this commutator we should find the action of the shift operators $\Delta_q(x_i)$ and $\Delta_p(x_i)$
on the shift part $A_i(x)$ and constant part $W(x)$. Using explicit expressions from \eqref{c2:operator:final} we obtain:
\be
&\Delta_p(x_i)W^{(C_2;h_a;1,0)}(x)=W^{(C_2;h_a;1,0)}(x)\,,
\nn\\
&\Delta_p(x_i)A_j^{(C_2;h_a;1,0)}(x)=\Delta_p(x_i)^{-1}A_j^{(C_2;h_a;1,0)}(x)=A_j^{(C_2;h_a;1,0)}(x)\,,
\nn\\
&\Delta_p(x_i)A_i^{(C_2;h_a;1,0)}(x)=(pq)^{-3}hA_i^{(C_2;h_a;1,0)}(x)\,,
\nn\\
&\Delta^{-1}_p(x_i)A_i^{(C_2;h_a;1,0)}(x)=(pq)^{3}h^{-1}A_i^{(C_2;h_a;1,0)}(x)\,,
\ee
where $h$ in the last expression is as usually total $U(1)$ charge defined in \eqref{c2:moment:maps}.
Completely identical expressions can be written for $p$ and $q$ exchanged. Using these expressions for the commutator action
\eqref{commutator:1:prem} it is straightforward to see that it is just zero:
\be
\left[\cO_x^{(C_2;h_a;1,0)},\,\cO_x^{(C_2;h_b;0,1)}\right]\cI(x)=0\,.
\ee
Since the test function $\cI(x)$ is arbitrary we can conclude that the commutator itself is also zero.

Now let's move to the computation of more complicated type of commutators
\be
\left[\cO_x^{(C_2;h_a;1,0)},\,\cO_x^{(C_2;h_b;1,0)}\right]=0\,, \quad \forall~ ~ a,b=1,\ldots,10\,.
\ee
In this case the proof is more complicated since the periodicity properties of $\theta$ functions can not be used
anymore. Instead we will be using expansion in $p$ and $q$ parameters to perform this check. Once again we will
use action of the commutator on an arbitrary test function $\cI(x)$:
\be
&&\left[\cO_x^{(C_2;h_a;1,0)},\,\cO_x^{(C_2;h_b;1,0)}\right]\cI(x)=
\nn\\
&&\sum\limits_{i,j=1}^2\left[A_i^{\left(C_2;h_a;1,0\right)}(x_i)
	\left(\Delta_q(x_i) A_j^{\left(C_2;h_b;1,0\right)}(x) \right)
\Delta_q(x_i)\Delta_p(x_j)\cI(x_i, x_j)+
\right.\nn\\
&&A_i^{\left(C_2;h_a;1,0\right)}(x)\left(\Delta_q(x_i) A_j^{\left(C_2;h_b;1,0\right)}\left(x^{-1}\right) \right)\Delta_q(x_i)\Delta_q^{-1}(x_j)\cI(x)+
\nn \\
&&A_i^{\left(C_2;h_a;1,0\right)}\left(x^{-1}\right)\left(\Delta^{-1}_q(x_i) A_j^{\left(C_2;h_b;1,0\right)}(x) \right)\Delta^{-1}_q(x_i)\Delta_q(x_j)\cI(x)+
\nn\\
&&\left.A_i^{\left(C_2;h_a;1,0\right)}\left(x^{-1}\right)\left(\Delta^{-1}_q(x_i) A_j^{\left(C_2;h_b;1,0\right)}\left(x^{-1}\right) \right)\Delta^{-1}_q(x_i)\Delta^{-1}_q(x_j)\cI(x)\right]+
\nn\\
&&\sum\limits_{i=1}^2\left[A_i^{\left(C_2;h_a;1,0\right)}(x)\left( \Delta_q(x_i)W^{\left(C_2;h_b;1,0\right)}(x)-W^{\left(C_2;h_b;1,0\right)}(x)\right)\Delta_q(x_i)\cI(x)+
 \right.\nn\\
 &&\left.
 A_i^{\left(C_2;h_b;1,0\right)}\left(x^{-1}\right)\left( \Delta^{-1}_q(x_i)W^{\left(C_2;h_a;1,0\right)}(x)-W^{\left(C_2;h_a;1,0\right)}(x)\right) \Delta^{-1}_q(x_i)\cI(x)\right]-
 \nn\\
&&
\hspace{11cm}\left(
\begin{array}{c}
a \leftrightarrow b
\end{array}
\right)\,,
\label{commutator:2:prem}
\ee
Since the test function $\cI(x)$ is arbitrary in order to perform our checks we have to consider contributions of all possible shifts of $\cI(x)$ separately.
Below we discuss such contributions one by one.

\textit{1) Terms with $\Delta_q(x_i)\Delta_q(x_j)\cI(x)$.} This kind of contributions in \eqref{commutator:2:prem} come from the following terms:
\be
&&\left[\cO_x^{(C_2;h_a;1,0)},\,\cO_x^{(C_2;h_b;1,0)}\right]\cI(x)\sim \left[A_i^{\left(C_2;h_a;1,0\right)}(x)\left(\Delta_q(x_i)A_j^{\left(C_2;h_b;1,0\right)}(x)\right)
+
\right.\nn\\
&&\left.
A_j^{\left(C_2;h_a;1,0\right)}(x)\left(\Delta_q(x_j)A_i^{\left(C_2;h_b;1,0\right)}(x)\right)
-\left(a\leftrightarrow b \right)
\right]\Delta_q(x_i)\Delta_q(x_j)\cI(x)\,.
\ee
In order for the prefactor to be zero the following algebraic identity has to be satisfied:
\be
&&F_1(x_i,x_j,h_a,h_b)+F_1(x_j,x_i,h_a,h_b)-F_1(x_i,x_j,h_b,h_a)-F_1(x_j,x_i,h_b,h_a)=0\,,
\nn\\
&&F_1(x_i,x_j,h_a,h_b)=
\nn\\
&&\hspace{2cm}\frac{\thf{x_jx_i^{\pm 1}}\thf{\pq{1}{2}h_bqx_i}\thf{\pq{1}{2}h_bq^{-1}x_i^{-1}}\thf{\pq{1}{2}h_ax_j^{\pm 1}}}{\thf{qx_jx_i}\thf{q^{-1}x_jx_i^{-1}}}\,.
\ee
In order to check this equation we expand all functions in $p$ and $q$ parameters. Since there are just two variables $x_1$ and $x_2$ in this case
we can fix $i=1$ and $j=2$ without loss of generality and check equality in expansion up to the order $O\left(p^3q^3\right)$. This check suggests that indeed these kind of terms
do not contribute to the commutator \eqref{commutator:2:prem}.

\textit{2) Terms with  $\Delta^2_q(x_i)\cI(x)$.} This contribution comes from the following terms:
\be
&&\left[\cO_x^{(C_2;h_a;1,0)},\,\cO_x^{(C_2;h_b;1,0)}\right]\cI(x)\sim \left[A_i^{\left(C_2;h_a;1,0\right)}(x)\left(\Delta_q(x_i)A_i^{\left(C_2;h_b;1,0\right)}(x)\right)
-
\right.\nn\\
&&\left.
\hspace{4.5cm} A_i^{\left(C_2;h_b;1,0\right)}(x)\left(\Delta_q(x_i)A_i^{\left(C_2;h_a;1,0\right)}(x)\right)
\right]\Delta^2_q(x_i)\cI(x)\,.
\ee
We can now notice that
\be
&&\Delta_q(x_i)A_i^{\left(C_2;h_a;1,0\right)}(x)=\prod\limits_{j\neq i}^2\frac{\thf{x_i^2}\thf{qx_i^2}\thf{x_ix_j^{\pm 1}}}{\thf{q^2x_i^2}\thf{q^3x_i^2}\thf{qx_ix_j^{\pm 1}}}
\times
\nn\\
&&\hspace{6cm}\prod\limits_{l=1}^{10}\frac{\thf{\pq{1}{2}h_l^{-1}qx_i}}{\thf{\pq{1}{2}h_l^{-1}x_i}}A_i^{\left(C_2;h_a;1,0\right)}(x)\,,
\ee
so that the overall factor we get after acting with the shift $\Delta_q(x_i)$ does not depend on the index $a$. Due to this independence
it is clear that corresponding contribution to the commutator \eqref{commutator:2:prem} is zero.

\textit{3) Terms with $\Delta_q(x_i)\Delta_q^{-1}(x_j)\cI(x)$.} These terms come from the contribution:
\be
&&\left[\cO_x^{(C_2;h_a;1,0)},\,\cO_x^{(C_2;h_b;1,0)}\right]\cI(x)\sim \left[A_i^{\left(C_2;h_a;1,0\right)}(x)\left(\Delta_q(x_i)A_j^{\left(C_2;h_b;1,0\right)}\left(x^{-1}\right)\right)
+
\right.\nn\\
&&\left.
	\hspace{0.5cm}A_j^{\left(C_2;h_a;1,0\right)}\left(x^{-1}\right)\left(\Delta^{-1}_q(x_j)A_i^{\left(C_2;h_b;1,0\right)}(x)\right)-(a\leftrightarrow b)
\right]\Delta_q(x_i)\Delta_q^{-1}(x_j)\cI(x)\,.
\ee
In order for this term to vanish we need the following equation to be satisfied:
\be
&&F_2(x_i,x_j,a,b)-F_2(x_i,x_j,b,a)-F_2(x_j^{-1},x_i^{-1},b,a)+F_2(x_j^{-1},x_i^{-1},a,b)=0\,,
\nn\\
&&F_2(x_i,x_j,a,b)=
\nn\\
&&\hspace{1cm}\frac{\thf{x_j^{-1}x_i^{\pm 1}}\thf{\pq{1}{2}h_bqx_i}\thf{\pq{1}{2}h_bq^{-1}x_i^{-1}}\thf{\pq{1}{2}h_ax_j^{\pm 1}}}{\thf{qx_ix_j^{-1}}\thf{q^{-1}x_i^{-1}x_j^{-1}}}
\ee
Just as previously we check this identity in $p$ and $q$ expansion up to an order of $O\left(p^3q^3\right)$ and thus show that corresponding contribution to the commutator
is zero.

\textit{4) Terms with $\Delta_q(x_i)\cI(x)$.} This kind of terms come from the contribution:
\be
&&\left[\cO_x^{(C_2;h_a;1,0)},\,\cO_x^{(C_2;h_b;1,0)}\right]\cI(x)\sim \left[A_i^{\left(C_2;h_a;1,0\right)}(x)\left(\Delta_q(x_i)W^{\left(C_2;h_b;1,0\right)}\left(x\right)\right)
+
\right.\nn\\
&&\left.
	\hspace{3cm}W^{\left(C_2;h_a;1,0\right)}\left(x\right)A_i^{\left(C_2;h_b;1,0\right)}\left(x\right)-(a\leftrightarrow b)
\right]\Delta_q(x_i)\cI(x)\,.
\ee
This expression is hard to simplify so we checked it directly fixing $i=1\,,j=2$ and expanding up to an order $O\left(p^2q^0 \right)$,
$O\left(p^0q^2 \right)$ and $O\left(pq\right)$ in $p$ and $q$ parameters. Up to these orders in expansion we confirmed that corresponding
contribution to the commutator is zero.

\textit{5) Constant terms $\cI(x)$.} This term comes from the contribution of
\be
&&\left[\cO_x^{(C_2;h_a;1,0)},\,\cO_x^{(C_2;h_b;1,0)}\right]\cI(x)\sim \left[W^{\left(C_2;h_a;1,0\right)}(x)W^{\left(C_2;h_b;1,0\right)}\left(x\right)
-
\right.\nn\\
&&\left.
\hspace{3cm}W^{\left(C_2;h_b;1,0\right)}(x)W^{\left(C_2;h_a;1,0\right)}\left(x\right)
\right]\cI(x)=0\,.
\ee
This term is obviously zero since it does not involve any shift.

\textit{6) Terms with $\Delta_q^{-1}(x_i)\Delta_q^{-1}(x_j)\cI(x)$.} These terms can be directly obtained from Type 1 terms containing  $\Delta_q(x_i)\Delta_q(x_j)\cI(x)$
by $x_i\to x_i^{-1}$ transformation of variables. Due to the $x_i\to x_i^{-1}$ symmetry of A$\Delta$O \eqref{c2:operator:final} we can immediately
conclude that corresponding contribution to the commutator action \eqref{commutator:2:prem} is zero.

\textit{7) Terms with $\Delta_q^{-2}(x_i)\cI(x)$.} These terms can be directly obtained from Type 2 terms containing  $\Delta^2_q(x_i)\cI(x)$
by $x_i\to x_i^{-1}$ transformation of variables. Hence by the same symmetry argument as in the previous case we can immediately
conclude that corresponding contribution to the commutator action \eqref{commutator:2:prem} is zero.

\textit{8) Terms with $\Delta_q^{-1}(x_i)\cI(x)$.} Contributions to the commutator action \eqref{commutator:2:prem} of this type are also zero which follows from the same
argument as previous two terms. In this case it can be obtained by using $x_i\to x_i^{-1}$ transformation in Type 4 terms.

Thus all eight types of terms present in the commutator \eqref{commutator:2:prem} give zero contribution and the full commutator action, and hence
the commutator itself, appears to be zero.
So far we have shown it only in $p$ and $q$ expansion. But in principle this can be also proven analytically by computing positions of the poles and
corresponding residues of all equalities we have checked in expansion. This way of proof is similar to the one we used in Appendix \ref{app:kernel} to
prove kernel property \eqref{a2c2:kernel:property}.

The last type of commutator identities we would like to prove is
\be
\left[\cO_x^{(C_2;h_a;0,1)},\,\cO_x^{(C_2;h_b;0,1)}\right]=0\,,\quad \forall ~ ~a,b=1,\ldots,10\,.
\ee
But this commutator can be directly obtained from the previous commutator
\\$\left[\cO_x^{(C_2;h_a;1,0)},\,\cO_x^{(C_2;h_b;1,0)}\right]$ by the exchange
$p\leftrightarrow q$. Hence all the arguments above work also for this commutator which as result is indeed equal to zero.

To summarize, in this Appendix we gave arguments in favor of all commutation relations \eqref{c2:basic:comm}. For the third commutator we gave full analytic proof
while for the first and second we checked commutation relations perturbatively in $p$ and $q$ expansion to sufficiently high order.

\bibliographystyle{ytphys.bst}
\bibliography{refs.bib}

\end{document}